\documentclass[11pt]{article}
\usepackage[left=1in,top=1in,right=1in,bottom=1in,head=.1in,nofoot]{geometry}

\usepackage{setspace,url,bm,amsmath} 

\usepackage{titlesec} 
\titlelabel{\thetitle.\quad} 

\titleformat*{\section}{\bf\Large\center}

\usepackage{graphicx}
\usepackage{longtable}
\usepackage{bbm}
\usepackage{latexsym}
\usepackage{caption}
\usepackage[margin=20pt]{subcaption}
\usepackage{hyperref}
\hypersetup{colorlinks,citecolor=blue,urlcolor=blue, linkcolor=blue}
\usepackage{booktabs}
\usepackage{multirow}
\usepackage{placeins}
\usepackage{enumitem}
\usepackage[table]{xcolor}
\usepackage{rotating}
\usepackage{amsthm}
\usepackage{amssymb}
\usepackage{amsmath}
\usepackage{color}
\usepackage{algorithm}
\usepackage{algorithmic}
\usepackage{tikz}
\usepackage{xcolor}
\usetikzlibrary{arrows.meta, positioning}
\usepackage{rotating}
\usepackage{comment}
\usepackage[sort]{natbib} 
\usepackage{etoolbox} 
\usepackage{listings}

\newcommand{\GG}[1]{}

\newcommand{\p}[1]{\left(#1\right)}
\newcommand{\sqb}[1]{\left[#1\right]}
\newcommand{\cb}[1]{\left\{#1\right\}}

\theoremstyle{definition}
\newtheorem{assumption}{Assumption}
\newtheorem*{theorem*}{Theorem}
\newtheorem{theorem}{Theorem}
\newtheorem*{rmk*}{Remark}

\newtheorem{proposition}{Proposition}
\newtheorem{lemma}{Lemma}
\newtheorem{example}{Example}

\newtheorem{estimator}{Estimator}
\newtheorem{varestimator}{Variance Estimator}
\newtheorem{coexample}{Covariate Construction}

\newtheorem{remark}{Remark}
\newtheorem{corollary}{Corollary}
\newtheorem*{corollary*}{Corollary}

\def\bX{\boldsymbol{X}}

\def\TTEhat{\hat{\tau}}

\def\rbhat{\hat{\bs \theta}_{\text{Reg}}}
\def\rb{\bs \theta_{\text{Reg}}}
\def\estchat{\hat{\bs \theta}_{\text{VIM}}}
\def\estc{\bs \theta_{\text{VIM}}}

\def\VIM{\TTEhat(\estchat)}
\def\Reg{\TTEhat(\rbhat)}
\def\SNIPE{\hat\tau_{\text{unadj}}}
\def\TTE{\tau}

\def\Var{\text{Var}}
\def\Cov{\text{Cov}}
\def\sumn{\sum_{i=1}^n}

\def\I{\mathbbm{1}}
\def\E{\mathbb{E}}
\def\bs{\boldsymbol}

\bibpunct{(}{)}{;}{a}{}{,} 
\apptocmd{\sloppy}{\hbadness 10000\relax}{}{}

\DeclareMathOperator*{\argmin}{arg\,min}

\renewcommand{\today}{\ifcase \month \or January\or February\or March\or %
April\or May \or June\or July\or August\or September\or October\or November\or %
December\fi, \number \year} 

\allowdisplaybreaks

\title{Covariate Adjustment Cannot Hurt: Treatment Effect Estimation under Interference with Low-Order Outcome Interactions}
\author{
	\makebox[45mm]{Xinyi Wang} \\ University of California, Berkeley \and
	\makebox[45mm]{Shuangning Li} \\  University of Chicago
}
\date{\today}

\begin{document}

\maketitle

\begin{abstract}
In randomized experiments, covariates are often used to reduce variance and improve the precision of treatment effect estimates. However, in many real-world settings, interference between units, where one unit’s treatment affects another’s outcome, complicates causal inference. This raises a key question: how can covariates be effectively used in the presence of interference? Addressing this challenge is nontrivial, as direct covariate adjustment, such as through regression, can increase variance due to dependencies across units. 
In this paper, we study covariate adjustment for estimating the total treatment effect under interference. We work under a neighborhood interference model with low-order interactions and build on the estimator of \citet{CortezRodriguezEichhornYu+2023}. We propose a class of covariate-adjusted estimators and show that, under sparsity conditions on the interference network, they are asymptotically unbiased and achieve a no-harm guarantee: their asymptotic variance is no larger than that of the unadjusted estimator. This parallels the classical result of \citet{Lin2013} under no interference, while allowing for arbitrary dependence in the covariates. 
We further develop a variance estimator for the proposed procedures and show that it is asymptotically conservative, enabling valid inference in the presence of interference. Compared with existing approaches, the proposed variance estimator is less conservative, leading to tighter confidence intervals in finite samples.
\end{abstract}

\section{Introduction}

Understanding the effects of treatments on outcomes of interest is a fundamental goal across many scientific fields, including medicine, economics, and education \citep{rubin1974estimating, holland1986statistics, imbens2015causal}. The field of causal inference seeks to develop methods for estimating these treatment effects, enabling researchers to address questions such as: How does a new medical intervention influence health outcomes? What is the impact of a job training program on labor market performance? How does an educational policy reform affect student achievement?

To answer such questions, a common approach is to conduct a randomized experiment, where units of interest are randomly assigned to either a treatment or a control group. The difference in average outcomes between treated and control units yields an unbiased estimator of the average treatment effect. In many such experiments, in addition to treatment assignment and outcome data, researchers also have access to auxiliary covariate information. For instance, in a randomized clinical trial evaluating the effects of hormone therapy on coronary heart disease, researchers recorded age, BMI, blood pressure, and hormone use history as covariates \citep{rossouw2002risks}; \citet{schochet2008does} studied the Job Corps training program and its effects on employment and earnings outcomes, incorporating covariates such as age, education, prior earnings, and employment history; and \citet{krueger1999experimental} analyzed the effect of being assigned to a small kindergarten class on student test scores, incorporating covariates including race, gender, and free-lunch eligibility.

Covariates can play a crucial role in improving the precision of causal effect estimates in experimental studies \citep{fisher1971design, freedman2008regression, Lin2013, negi2021revisiting, fogarty2018regression, su2021model, zhao2022reconciling, wang2023model}. While randomization ensures that treatment assignment is independent of both observed and unobserved confounders on average, in finite samples, there may still be chance imbalances in covariates that affect the outcome. Adjusting for these covariates can mitigate such imbalances and reduce the variance of the estimated treatment effect without introducing bias \citep{Lin2013}. Covariate adjustment can be implemented by regressing the outcome on the treatment indicator, (centered) covariates, and their interactions, with the adjusted treatment effect given by the fitted coefficient on the treatment indicator \citep{Lin2013}.

A key assumption underlying many causal inference methods is the Stable Unit Treatment Value Assumption (SUTVA), which posits that a unit’s outcome depends solely on the treatment it receives and is unaffected by the treatments assigned to others \citep{imbens2015causal}. While this assumption simplifies analysis and is reasonable in some settings, it is often violated in real-world contexts where units interact. For example, in a study examining the effect of information sessions about weather insurance on farmers’ financial decisions, farmers' choices may be influenced by the decisions and experiences of their peers \citep{cai2015social}. Similarly, in education, a pedagogical innovation may affect not only the treated students but also their classmates \citep{sacerdote2001peer}. These examples illustrate interference, where the treatment assigned to one unit influences the outcomes of others.

Interference complicates statistical analysis and presents significant challenges to causal inference. In the presence of interference, treatment–outcome pairs across units are no longer independent, invalidating many standard estimators. Overcoming these challenges requires methods that explicitly account for the interdependencies between units and the mechanisms of interference \citep{sobel2006randomized, hudgens2008toward, tchetgen2012causal, toulis2013estimation, eckles2017design, athey2018exact, aronow2017estimating, leung2020treatment, savje2021average,  li2022random, CortezRodriguezEichhornYu+2023}. 

In this paper, we study how to leverage covariate information to reduce the variance of treatment effect estimators under interference. Specifically, we focus on estimating the total treatment effect, defined as the difference in average outcomes when all units receive treatment versus when all receive control.

Our analysis builds on the low-order interaction outcome model introduced by \citet{CortezRodriguezEichhornYu+2023}, which offers a structured yet flexible framework for modeling interference. This model is built on the neighborhood interference model (also referred to as the network interference model in the literature), which assumes the existence of a known interference network such that each unit’s outcome depends only on its own treatment and the treatments of its neighbors \citep{hudgens2008toward, athey2018exact, leung2020treatment, li2022random}. The low-order interaction model imposes further structure by restricting the outcome to depend only on low-order interactions among neighbors' treatment assignments.
To estimate the total treatment effect, \citet{CortezRodriguezEichhornYu+2023} propose the Structured Neighborhood Interference Polynomial Estimator, which they show is unbiased under the low-order interaction model. Throughout the paper, we denote this estimator by $\SNIPE$. They also establish variance bounds and a central limit theorem under sparsity assumptions on the interference network. The construction of $\SNIPE$ explicitly incorporates information about treatment assignments, outcomes, and the interference network.

Building on $\SNIPE$, we propose a covariate-adjusted version of it. We show that under sparsity assumptions on the interference network, our estimator remains asymptotically unbiased and importantly has asymptotic variance no greater than that of the original unadjusted estimator $\SNIPE$. This parallels the well-known result in \citet{Lin2013} under SUTVA, where incorporating covariates through regression adjustment is shown to not hurt and often improve the precision of treatment effect estimators.

Achieving such variance improvement uniformly across all cases is nontrivial in the presence of interference. For instance, direct regression adjustment can inflate variance when interference exists \citep{gao2023causal}. While regression adjustment tends to reduce the variance of individual components, it can inadvertently increase the covariance across components due to interference, an effect that is absent under SUTVA but must be accounted for in interference settings. Our covariate-adjusted estimator avoids this pitfall by carefully accounting for the interference effects.

Our variance improvement result does not require strong assumptions on the covariates. The covariates can be arbitrarily dependent on each other, and unit $i$’s outcome may depend on their own covariates as well as the covariates of other units. More interestingly, the covariates used by our estimator can also be dependent on the interference network itself. In other words, our framework allows the use of both traditional covariates and network-derived features to reduce variance.

\subsection{Overview of results}
As an overview, we begin by considering a general approach to incorporating covariates into $\SNIPE$, inspired by the control variates method \citep{nelson1990control}. This approach is parameterized by a vector $\bs \theta$, which governs how covariate information is used.
The most straightforward choice of $\bs \theta$ is obtained via regression, leading to what we refer to as the regression-based covariate-adjusted estimator. Empirically, we find that this estimator often outperforms the unadjusted estimator $\SNIPE$ in terms of mean squared error (MSE). However, we also identify specific scenarios in which the regression-based version performs worse, motivating a more principled strategy for selecting $\bs \theta$.

To this end, we propose a new estimator, the variance improvement maximized covariate-adjusted estimator. This estimator is constructed by estimating the difference in variance between the unadjusted estimator $\SNIPE$ and a $\bs \theta$-adjusted estimator, and then choosing $\bs \theta$ to maximize this estimated variance reduction. Plugging this chosen $\bs \theta$ into the general adjustment form yields the variance improvement maximized covariate-adjusted estimator.

Theoretically, we show that under sparsity conditions on the interference network, our variance improvement maximized covariate-adjusted estimator is asymptotically unbiased and achieves asymptotic variance no greater than that of the original unadjusted estimator proposed by \cite{CortezRodriguezEichhornYu+2023}. Furthermore, we prove that it is asymptotically optimal in terms of mean squared error (MSE) within the class of estimators parameterized by $\bs \theta$. We also establish asymptotic normality, derive variance bounds for the general $\bs\theta$-adjusted estimator, following the analysis of \citet{CortezRodriguezEichhornYu+2023}; these results apply to both the Regression-based and the variance improvement maximized covariate-adjusted estimators.

Empirically, we conduct extensive simulation studies across a range of settings and consistently find that the variance improvement maximized covariate-adjusted estimator outperforms the original unadjusted estimator $\SNIPE$ in terms of MSE. The gains are especially large in scenarios where covariates explain a substantial portion of the outcome variance.

To support inference, we develop a variance estimator for the covariate-adjusted estimators.  The variance estimator applies to both the regression-based and VIM-based adjustments and remains valid under interference. We show that it is asymptotically conservative and, in empirical settings, far less conservative than existing approaches, leading to tighter confidence intervals in finite samples.

\subsection{Problem setup}

Suppose we have a finite population indexed $i = 1, \ldots, n$, where each unit is independently assigned a binary treatment $Z_i \in \{0, 1\}$, with $Z_i \sim \text{Bernoulli}(p_i)$ for some known $p_i \in [p, 1-p]$ with $p > 0$. We adopt the randomization-based framework, where the only source of randomness is the treatment assignment. Let $\bs Z = [Z_1, \ldots, Z_n]$ be the treatment vector of the population.

A network structure is observed among the population, represented by a directed graph with self-loops and edge set $\cb{E_{ij}}_{i,j = 1}^n$. For each unit $i$, let $\mathcal{N}_i = \cb{j \in [n] : (j, i) \in E}$ denote the set of in-neighbors of unit $i$.
We define the maximum in-degree and out-degree of the graph as
\[
d_{\text{in}} = \max_{i \in [n]} |\mathcal{N}_i|, \qquad
d_{\text{out}} = \max_{j \in [n]} \left| \{ i \in [n] : (j, i) \in E \} \right|,
\]
and let $d = \max(d_{\text{in}}, d_{\text{out}})$. 

Let $\bX_i$ be a $d_{\bs X}$-dimensional covariate vector of unit $i$, where $d_{\bs X}$ is a fixed constant independent of $n$. For simplicity, we assume $\bX_i$'s are mean-centered, i.e., $\bar{\bX} = \frac{1}{n} \sum_{i=1}^n \bX_i = 0$.
Let $Y_i$ be the observed outcome and $Y_i(\bs{z})$ the potential outcome of unit $i$ under treatment assignment $\bs{z}$. The potential outcome function satisfies $Y_i = Y_i(\bs{Z})$.
We impose the following assumption of neighborhood interference. 
\begin{assumption}[Neighborhood interference]\label{as:interference}
For any treatment assignment vectors $\bs z, \bs z' \in \{0,1\}^n$, if $\bs z_{\mathcal{N}_i} = \bs z_{\mathcal{N}_i}'$, then $Y_i(\bs z) = Y_i(\bs z')$.
\end{assumption}
Assumption \ref{as:interference} states that the outcome of unit $i$ depends only on the treatment assignments of units in $\mathcal{N}_i$. The neighborhood interference assumption (also known as the network interference assumption) is widely used in the interference literature \citep{toulis2013estimation, eckles2017design, leung2020treatment, li2022random, savje2021average, CortezRodriguezEichhornYu+2023}, both for its practical relevance and theoretical elegance.

In this paper, we focus on estimating the total treatment effect defined as follows
\begin{align*}
    \TTE:= \frac{1}{n}\sumn\left[Y_i(\bs 1) - Y_i(\bs 0)\right],
\end{align*}
where $\bs 1$ represents the all-ones vector and $\bs 0$ represents the all-zeros vector. $\TTE$ is a well-studied estimand in the literature, capturing the average treatment effect of assigning everyone to treatment versus everyone to control \citep{yu2022estimating, CortezRodriguezEichhornYu+2023, eckles2017design, ugander2023randomized, chin2019regression}. It is particularly relevant in settings where a decision-maker is considering whether to implement a new treatment for all units or maintain the existing standard (control). For example, an online platform may be evaluating whether to adopt a new recommendation algorithm or user interface for all users.

We use the following standard asymptotic and norm notations. For deterministic sequences, $a_n = o(b_n)$ means that $a_n / b_n \to 0$ as $n \to \infty$, and $a_n = O(b_n)$ means that there exists a constant $C > 0$ such that $|a_n| \le C |b_n|$ for all sufficiently large $n$. For random variables, $X_n = o_p(1)$ indicates convergence in probability to zero, i.e., $\mathbb{P}(|X_n| > \epsilon) \to 0$ for every $\epsilon > 0$. We write $\|\cdot\|$ to denote the Euclidean norm for vectors and the operator norm for matrices, and $\|\cdot\|_1$ to denote the $\ell_1$ norm.

\subsection{Related work}

A large body of literature has studied the role of covariates in randomized experiments under SUTVA. Covariate adjustment has long been recognized as a way to improve efficiency \citep[e.g.,][]{fisher1971design}, and regression-based approaches such as \citet{Lin2013} formally show that adjustment never reduces asymptotic precision \citep[see also][]{negi2021revisiting}. Related developments have extended regression adjustment to other experimental settings \citep{rosenbaum2002covariance, fogarty2018regression, su2021model, zhao2022reconciling, wang2023model, chang2024exact, wang2024model, zhao2024covariate}, further underscoring the central role of covariates in improving inference.  

Estimation of causal effects under network interference raises additional challenges compared to settings that satisfy SUTVA, since a unit’s outcome may depend not only on its own treatment but also on the treatments assigned to other units. Foundational contributions established frameworks for defining causal effects when interference is present \citep{sobel2006randomized, hudgens2008toward, tchetgen2012causal}, and a growing literature has proposed estimators under various assumptions on interference \citep{eckles2017design, aronow2017estimating, leung2020treatment, savje2021average, li2022random, CortezRodriguezEichhornYu+2023}. These works differ in the assumptions they impose, ranging from exposure mappings to random graphs to approximate neighborhood interference, but in most cases do not directly incorporate covariates into estimation.

More recent work has studied covariate adjustment under interference with the goal of improving the precision of treatment effect estimation. \citet{aronow2017estimating} noted this possibility, while \citet{basse2018analyzing} analyzed two-stage randomized experiments and showed that covariates can be leveraged to sharpen inference. Under the approximate neighborhood interference framework of \citet{leung2022causal}, \citet{lu2024adjusting} and \citet{gao2023causal} developed covariate-adjusted estimators. \citet{fan2025causal} considered adjustment when estimating the average direct effect defined by \citet{hu2022average} under a random graph model. \citet{chin2019regression} and \citet{han2023model} both focus on estimating $\TTE$ (or global average treatment effect), viewing regression adjustment primarily as a tool for debiasing, though in \citet{han2023model} it can also improve variance. This contrasts with work such as \citet{Lin2013}, where the central motivation for adjustment is variance reduction. Our paper contributes to this line of research but operates under a different modeling framework, namely the low-order interaction outcomes model.

Covariates also play important roles beyond direct adjustment for estimation in the presence of interference. In observational studies with network interference, covariates are critical confounders and are required to identify causal effects \citep{tchetgen2012causal, liu2019doubly, barkley2020causal, forastiere2021identification}. In experimental design, covariates have been used to optimize assignments and improve efficiency in the presence of interference \citep{basse2018model, viviano2020experimental}. Recent work has also considered policy design, learning, and targeting under interference, where covariates inform optimal assignment rules \citep{galeotti2020targeting, kitagawa2023should, zhang2023individualized, park2024minimum, viviano2024policy, viviano2025policy, hu2025optimal}. Finally, in the context of inference and testing, covariates can be incorporated to improve the power of tests and sharpen inference \citep{rosenbaum2007interference, athey2018exact, han2023detecting}.


\section{Adjusting for Covariates under Low-Order Outcome Interactions}
\label{sec:method}
\subsection{The low-order interaction model and the SNIPE estimator}\label{sec:low-order_SNIPE}

Following \cite{CortezRodriguezEichhornYu+2023}, we consider the low-order interaction model for the potential outcomes. For a fixed integer $\beta$, define $\mathcal S_i^\beta = \{\mathcal S \subseteq \mathcal N_i: |\mathcal S| \leq \beta\}$ for $i = 1, \ldots,n$ as the collection of all subsets of $\mathcal{N}_i$ of size at most $\beta$.

\begin{assumption}[Low-order interactions model \citep{CortezRodriguezEichhornYu+2023}]\label{as:model} 
For each unit $i$, there exists a vector $\bs \alpha_i$ such that the potential outcomes of unit $i$ can be expressed as
\begin{equation}
\label{eqn:low_order}
        Y_i(\bs z) = \sum_{\mathcal{S} \in \mathcal{S}_i^\beta}\alpha_{i, \mathcal S} \prod_{j \in \mathcal S} z_j.
\end{equation}
    This specification is referred to as a $\beta$-order interaction model.
\end{assumption}
Assumption \ref{as:model} posits that the potential outcome of unit $i$, under any treatment assignment vector $\bs z$, can be written as a sum of interaction effects up to degree $\beta$ from its neighbors. Each $\alpha_{i, \mathcal{S}}$ represents the additional effect on the outcome of unit $i$ when all units in $\mathcal{S}$ receive treatment.
When $\beta = 1$, the model reduces to a linear outcome model in the treatment indicators $z_j$:
\begin{align}\label{equ: Ydeg1}
 Y_i(\bs z) = \alpha_{i, \varnothing} + \sum_{j \in \mathcal{N}_i}\alpha_{i, \{j\}} z_j,   
\end{align}
which captures only the individual (additive) effects of each neighbor's treatment on the outcome of unit $i$. When $\beta = 2$, the model additionally includes pairwise interaction effects:
\begin{align}\label{equ: Ydeg2}
 Y_i(\bs z) = \alpha_{i, \varnothing} + \sum_{j \in \mathcal{N}_i}\alpha_{i, \{j\}} z_j + \sum_{j,k \in \mathcal{N}_i, j < k }\alpha_{i, \{j, k\}} z_j z_k.   
\end{align}
Including interaction terms in the potential outcomes model allows us to capture non-additive effects among treated neighbors, which often arise in real-world settings. Since $\bs{z}$ is a binary vector, any potential outcome function mapping $\bs{z}$ to $Y_i(\bs{z})$ can be expressed as a polynomial in $\bs{z}$ of degree at most $\lvert \mathcal{N}_i \rvert$. Consequently, the potential outcome function can always be represented by a $\lvert \mathcal{N}_i \rvert$-order interaction model. By restricting the order of interaction from $\lvert \mathcal{N}_i \rvert$ to a smaller integer $\beta$, the low-order interaction model reduces the complexity of the potential outcomes function class, enabling more efficient estimation of $\TTE$.

Under Assumption \ref{as:model}, we can rewrite $\TTE$ as
\begin{align}
\label{equ: defTTE}
    \TTE = \frac{1}{n}\sumn\left[Y_i(\bs 1) - Y_i(\bs 0)\right] = \frac{1}{n}\sumn\sum_{\mathcal S \in \mathcal S_i^\beta,\, \mathcal{S} \neq \varnothing}\alpha_{i, \mathcal S} \prod_{j \in \mathcal S} 1
    = \frac{1}{n}\sumn\sum_{\mathcal S \in \mathcal S_i^\beta, \,\mathcal{S} \neq \varnothing} \alpha_{i, \mathcal S}.
\end{align}

To estimate $\TTE$, \cite{CortezRodriguezEichhornYu+2023} propose an estimator, the Structured Neighborhood Interference Polynomial Estimator (SNIPE), defined as follows.
\begin{estimator}[Unadjusted SNIPE estimator] The Structured Neighborhood Interference Polynomial Estimator (SNIPE) for $\TTE$ is given by
\begin{equation}
\label{eqn:SNIPE_def}
   \SNIPE  = \frac{1}{n}\sum_{i=1}^n Y_i\sum_{\mathcal S\in \mathcal S_i^\beta}g(\mathcal S)\prod_{j\in\mathcal S}\frac{Z_j-p_j}{p_j(1-p_j)},
\end{equation}
where $g(\mathcal S) = \prod_{j \in\mathcal S }(1-p_j) - \prod_{j \in\mathcal S }(-p_j)$.     
\end{estimator}

\citet{CortezRodriguezEichhornYu+2023} show that $\SNIPE$ is unbiased for $\TTE$. They further establish that the SNIPE estimator satisfies a variance bound that scales inversely with the sample size $n$, polynomially with the network degree $d$, and exponentially with the interaction order $\beta$, and that it is asymptotically normal under suitable graph sparsity conditions.

We now build intuition for the estimator and its unbiasedness. Note from \eqref{equ: defTTE} that the estimand $\TTE$ is a linear function of the parameters $\alpha_{i,\mathcal S}$. Therefore, it suffices to construct unbiased estimators for each $\alpha_{i,\mathcal S}$.
We begin with the case $\beta = 1$. From \eqref{equ: Ydeg1}, we have
$Y_i = \alpha_{i,\varnothing} + \sum_{j \in \mathcal{N}_i} \alpha_{i,\{j\}} Z_j$.
Suppose we are interested in estimating $\alpha_{i,\{j\}}$. Multiplying both sides by $(Z_j - p_j)$ yields
\[
Y_i (Z_j - p_j)
= \alpha_{i,\varnothing}(Z_j - p_j)
+ \sum_{k \in \mathcal{N}_i} \alpha_{i,\{k\}} Z_k (Z_j - p_j).
\]
Taking expectations, all terms on the right-hand side have mean zero except the term corresponding to $k=j$, whose expectation is $\alpha_{i,\{j\}}\, p_j(1-p_j)$.
It follows that
$Y_i (Z_j - p_j)/(p_j(1-p_j))$
is an unbiased estimator for $\alpha_{i,\{j\}}$.
A similar argument applies when $\beta = 2$. From \eqref{equ: Ydeg2}, to estimate $\alpha_{i,\{j,k\}}$, we multiply both sides by $(Z_j - p_j)(Z_k - p_k)$. By independence and centering, all terms have mean zero except the one corresponding to $\{j,k\}$, which isolates $\alpha_{i,\{j,k\}}$.

More generally, for any $\beta$ and any subset $\mathcal S \in \mathcal S_i^\beta$, generalizing the above calculation yields the unbiased estimator
\begin{equation}
\label{eqn:alpha_hat_defi}
\hat \alpha_{i, \mathcal S}^{\text{unadj}}
= Y_i \prod_{j \in \mathcal S} \frac{-1}{p_j}
\sum_{\substack{\mathcal U \in \mathcal S_i^\beta \\ \mathcal U \supseteq \mathcal S}}
\prod_{l \in \mathcal U} \frac{p_l - Z_l}{1-p_l}.
\end{equation}
Aggregating these estimators gives
\[
\TTEhat_{\text{unadj}}
= \frac{1}{n}\sum_{i=1}^n \sum_{\substack{\mathcal S \in \mathcal S_i^\beta \\ \mathcal S \neq \varnothing}}
\hat{\alpha}_{i, \mathcal S}^{\text{unadj}},
\]
which coincides with the SNIPE estimator in \eqref{eqn:SNIPE_def}.

\begin{lemma}[Unbiasedness of $\hat{\alpha}_{i, \mathcal S}^{\text{unadj}}$]\label{lem:unbias}
Under Assumptions \ref{as:interference}--\ref{as:model}, for each unit $i$ and set $\mathcal{S} \in \mathcal{S}_i^\beta$, $\E(\hat{\alpha}_{i, \mathcal S}^{\text{unadj}}) = \alpha_{i, \mathcal S}.$   
\end{lemma}
We provide a detailed derivation of this result in Appendix~\ref{appendix:proof_lemma_unbiased}. This decomposition is especially useful for constructing covariate-adjusted versions of $\SNIPE$.


\subsection{A general covariate-adjusted SNIPE estimator}
\label{sec:general_cova_SNIPE}
Looking closely at the definition of $\SNIPE$ in \eqref{eqn:SNIPE_def}, we observe that it can be expressed as a weighted average of the outcomes $Y_i$. Specifically, 
$$\TTEhat_{\text{unadj}}=\frac{1}{n}\sum_{i=1}^n \omega_iY_i, \qquad
\omega_i=\sum_{S\in\mathcal S_i^\beta} g(S)\prod_{j\in S}\frac{Z_j-p_j}{p_j(1-p_j)}.$$
 Observe that the expectation of each weight is zero:
\begin{align*}
\E( \omega_i ) = 
\E\Big( \sum_{\mathcal S\in \mathcal S_i^\beta}g(\mathcal S)\prod_{j\in\mathcal S}\frac{Z_j-p_j}{p_j(1-p_j)} \Big) = 0.
\end{align*}

A natural way to incorporate covariate information is to subtract a function of the covariates from the outcome. Specifically, we define a covariate-adjusted estimator based on \cite{CortezRodriguezEichhornYu+2023}'s estimator $\SNIPE$ for $\TTE$ as
\begin{align}\label{equ: ttehat}
\TTEhat(\bs \theta)=\frac{1}{n}\sum_{i=1}^n \omega_i\left(Y_i - \bs \theta^\top \bX_i \right ). 
\end{align}
Since each $\omega_i$ is mean-zero, the added term is also mean-zero for any fixed $\bs \theta$. As a result, because the original unadjusted estimator $\SNIPE$ is unbiased for $\TTE$, the adjusted estimator $\TTEhat(\bs \theta)$ remains unbiased for any fixed choice of $\bs \theta$. 

This adjustment resembles the classical control variates technique, where auxiliary variables with known or mean-zero expectation are used to reduce variance without introducing bias \citep{glasserman2004monte, lemieux2014control, botev2017variance}. In this context, the added term $\frac{1}{n} \sum_{i=1}^n \omega_i \bs{\theta}^\top \bX_i$ serves as a control variate: it does not affect the expectation of the estimator but can potentially reduce its variance. While any fixed choice of $\bs{\theta}$ yields an unbiased estimator, choosing $\bs{\theta}$ carefully can lead to substantial variance reduction. In the following sections, we present our proposed choices for $\bs{\theta}$.

The covariate-adjusted estimator $\TTEhat(\bs \theta)$ also has a close connection to the Augmented Inverse Probability Weighting (AIPW) estimator, a canonical method in the doubly robust estimation literature (see, for example, \citet{ding2024first} for an introduction). In particular, in the no-interference setting\footnote{Throughout this paper, the “no interference” or SUTVA setting means both that the potential outcomes satisfy SUTVA, where each unit’s treatment does not affect other units’ outcomes, and that the interference network used by $\SNIPE$ contains only self-loops and no other edges.}, the unadjusted estimator $\hat\tau_{\text{unadj}}$ simplifies to
\begin{align*}
 \TTEhat_{\text{unadj}} = \frac{1}{n}\sum_{i=1}^n\frac{(Z_i-p_i)Y_i}{p_i(1-p_i)},
\end{align*}
which is the classical Inverse Probability Weighting (IPW) estimator. The covariate-adjusted estimator $\TTEhat(\bs \theta)$ then takes the form
\begin{align*}
    \TTEhat(\bs \theta) &= \frac{1}{n}\sum_{i=1}^n\frac{(Z_i-p_i)Y_i}{p_i(1-p_i)} - \frac{1}{n}\sum_{i=1}^n\frac{(Z_i-p_i)\bs \theta^\top \bs X_i}{p_i(1-p_i)}\\
    &=\frac{1}{n}\sumn \left[\frac{Z_i(Y_i - \bs \theta^\top \bs X_i) }{p_i} - \frac{(1-Z_i)(Y_i - \bs \theta^\top \bs X_i) }{1-p_i}\right].
\end{align*}
If $\bs \theta^\top \bs X_i$ is used as an estimate of the conditional mean outcome in the AIPW construction, this expression coincides with the AIPW estimator.

We also note that the unadjusted estimator $\SNIPE$ corresponds to $\TTEhat(\bs \theta)$ with $\bs \theta = \bs 0$, that is, $\TTEhat_{\text{unadj}} = \TTEhat(\bs 0)$. From this point forward, we use the notations $\TTEhat_{\text{unadj}}$ and $\TTEhat(\bs 0)$ interchangeably.

Finally, in Appendix~\ref{appendix:alternative_perspective}, we reinterpret $\TTEhat(\bs \theta)$ from a regression perspective.

\subsection{Regression-based covariate adjustment}\label{sec:Reg-SNIPE}

Choosing an effective $\bs{\theta}$ requires a good understanding of the variance of $\TTEhat(\bs{\theta})$. However, due to cross-unit interference, characterizing or accurately estimating this variance is highly nontrivial. As a first step, we approximate the variance of $\TTEhat(\bs{\theta})$ by its variance under the simplifying assumption of no interference. We then aim to select a value of $\bs{\theta}$ that minimizes this approximate variance. 

When there is no interference, the variance of $\TTEhat(\bs \theta)$ is written as
\begin{align*}
\Var\left(\TTEhat(\bs \theta)\right) &= \frac{1}{n^2}\sum_{i=1}^n \Var\sqb{\omega_i\left(Y_i - \bs \theta^\top \bX_i \right )}
\\
&=\frac{1}{n^2} \sumn\E\left[\omega^2_i\left(Y_i - \bs \theta^\top \bX_i \right )^2\right]  -\frac{1}{n^2}\sumn \left\{\E\left[\omega_i\left(Y_i - \bs \theta^\top \bX_i \right )\right]\right\}^2\\
&=\E\left[\frac{1}{n^2} \sumn\omega^2_i\left(Y_i - \bs \theta^\top \bX_i \right )^2\right]  -\frac{1}{n^2}\sumn \left[\E\left(\omega_iY_i\right)\right]^2.
\end{align*}
Since the second term in the expression above does not contain $\bs{\theta}$, minimizing the approximate variance reduces to minimizing the first term. This yields the regression-based covariate-adjusted estimator.
\begin{estimator}[Regression-based covariate adjustment] We define the regression-based covariate-adjusted estimator as follows:
\begin{align*}
\TTEhat(\rbhat)=\frac{1}{n}\sum_{i=1}^n \omega_i\left(Y_i - \bs \rbhat^\top \bX_i \right ),
\end{align*} where 
\begin{align*}
   \rbhat &= \argmin_{\bs \theta} \frac{1}{n} \sumn\omega^2_i\left(Y_i - \bs \theta^\top \bX_i \right )^2 = \Big(\frac{1}{n} \sumn \omega_i ^2\bX_i\bX_i^\top\Big)^{-1} \Big(\frac{1}{n} \sumn \omega_i^2\bX_iY_i\Big).
\end{align*}    
Recall that $\omega_i = \sum_{\mathcal S\in \mathcal S_i^\beta}g(\mathcal S)\prod_{j\in\mathcal S}\frac{Z_j-p_j}{p_j(1-p_j)}$.
\end{estimator}

We refer to this estimator as a regression-based estimator because $\rbhat$ corresponds to the weighted least squares estimator of the regression coefficients for $\bX_i$ in the linear model $Y_i \sim \bX_i$, with weights $\omega_i^2$ for each unit $i$.

In the absence of interference, under standard assumptions, we can show that the regression-based adjustment reduces variance relative to the unadjusted estimator asymptotically. Furthermore, the estimator is closely related to Lin's estimator \citep{Lin2013}, which is known to improve precision through covariate adjustment. In particular, Lin’s estimator can be rewritten in a control variate form: it is the difference-in-means estimator plus a control variate term with coefficient $\hat{\bs \theta}_{\text{Lin}}$. We can show that, asymptotically, the coefficient $\hat{\bs \theta}_{\text{Lin}}$ coincides with $\rbhat$. See Appendix \ref{appendix:reg_estimator} for details. 

However, in the presence of interference, this conclusion may no longer hold. The variance of $\TTEhat(\bs\theta)$ generally includes both variance and covariance components across units:
\begin{align*}
\Var\left(\TTEhat(\bs \theta)\right)
&= \frac{1}{n^2}\sum_{i=1}^n \Var\sqb{\omega_i\left(Y_i - \bs \theta^\top \bX_i \right )} \\
&\quad + \frac{1}{n^2}\sum_{i=1}^n \sum_{j=1}^n \Cov\sqb{\omega_i\left(Y_i - \bs \theta^\top \bX_i \right ), \omega_j\left(Y_j - \bs \theta^\top \bX_j \right )}.
\end{align*}
While the regression-based choice $\rbhat$ minimizes the marginal variance terms, it does not account for the covariance terms induced by interference. As a result, the overall variance can increase relative to the unadjusted estimator if these covariance contributions are sufficiently large.

We now provide a toy example to illustrate the possibility of such an overall variance increase.
\begin{example}[Toy example]\label{exp:increasevar_small}
Consider an undirected graph with $n=3$ units where $1$ is connected to $2$ and $3$ is isolated
(Figure~\ref{fig:single_three}). Let $\beta=1$ and assign treatments independently with $p=0.5$.
Potential outcomes are
\[
Y_1(\bs z)=z_1+z_2,\qquad
Y_2(\bs z)=-2+z_1+z_2,\qquad
Y_3(\bs z)=-0.5+z_3,
\]
and covariates are $\bX_1=0.5$, $\bX_2=0$, $\bX_3=-0.5$.

\begin{figure}[t]
\centering
\begin{subfigure}{0.48\textwidth}
\centering
\begin{tikzpicture}[>=Stealth, node distance=2.4cm, thick]
  \definecolor{control}{RGB}{204,204,255}
  \node[circle, draw, fill=control, minimum size=0.9cm] (1) {1};
  \node[circle, draw, fill=control, minimum size=0.9cm, right of=1] (2) {2};
  \node[circle, draw, fill=control, minimum size=0.9cm, right of=2] (3) {3};
  \draw[->] (1) -- (2);
  \draw[->] (2) -- (1);
\end{tikzpicture}
\caption{One group.}
\label{fig:single_three}
\end{subfigure}\hfill
\begin{subfigure}{0.48\textwidth}
\centering
\begin{tikzpicture}[>=Stealth, thick]
  \definecolor{lightred}{RGB}{255, 200, 200}
  \node[circle, draw, fill=lightred, minimum size=0.75cm] (1a) at (0,0) {1};
  \node[circle, draw, fill=lightred, minimum size=0.75cm] (2a) at (1.7,0) {2};
  \node[circle, draw, fill=lightred, minimum size=0.75cm] (3a) at (3.4,0) {3};
  \draw[->] (1a) -- (2a);
  \draw[->] (2a) -- (1a);
  \node at (1.7,-0.9) {$\vdots$};
  \node[circle, draw, fill=lightred, minimum size=0.75cm] (1c) at (0,-1.8) {1};
  \node[circle, draw, fill=lightred, minimum size=0.75cm] (2c) at (1.7,-1.8) {2};
  \node[circle, draw, fill=lightred, minimum size=0.75cm] (3c) at (3.4,-1.8) {3};
  \draw[->] (1c) -- (2c);
  \draw[->] (2c) -- (1c);
\end{tikzpicture}
\caption{Many independent groups.}
\label{fig:many_three}
\end{subfigure}
\caption{Toy network and repeated i.i.d.\ copies}
\label{fig:toy_many}
\end{figure}

It is straightforward to verify that $\TTE=5/3$. Direct calculation (Appendix~\ref{app:toy_example_details})
gives the closed forms of $\TTEhat_{\text{unadj}}=\TTEhat(\bs 0)$ and $\TTEhat(\bs\theta)$ and yields
\begin{equation}\label{eq:toy_var_identity}
\Var\{\TTEhat(\bs\theta)\}=\frac{16}{9}+\frac{1}{3}\bs\theta^2,
\end{equation}
so for any $\bs\theta\neq \bs 0$ the covariate-adjusted estimator has strictly larger variance than the
unadjusted estimator.

Consider i.i.d.\ copies of the three-unit group (Figure~\ref{fig:many_three}). In this setting, we can show that $\rbhat \stackrel{p}{\to} \rb = \frac{4}{3}$ as the number of copies grows (see Appendix~\ref{appendix:small_thetaReg}). Consequently, the regression-based covariate-adjusted estimator yields strictly larger variance than the unadjusted estimator:
$\Var(\TTEhat(\rb)) > \Var(\TTEhat(\bs 0)).$
\end{example}

\subsection{Variance-improvement–maximized covariate adjustment}\label{sec:VIM-SNIPE}

As discussed in the previous section, in the presence of interference, the regression-based covariate-adjusted estimator does not guarantee variance reduction compared to $\TTEhat_{\text{unadj}}$. 
Our goal now is to identify an alternative choice of $\bs \theta$ that guarantees the variance will be no greater than that of $\TTEhat_{\text{unadj}}$.

A natural first idea is to construct a consistent estimator of the variance and choose $\bs \theta$ to minimize it. This would ideally yield a value of $\bs \theta$ close to the optimizer of the true variance $\Var\!\left(\TTEhat(\bs \theta)\right)$. However, obtaining a consistent variance estimator is challenging in our setting. In particular, we can write
\begin{align}\label{equ: vartte}
&\Var\left(\TTEhat(\bs \theta)\right) =  \Var\left(\frac{1}{n}\sumn \omega_i\left(Y_i  - \bs \theta^\top \bX_i \right )\right) \nonumber \\
&=\frac{1}{n^2}\sumn\sum_{i'=1}^n \sum_{\mathcal S \in \mathcal S_i^\beta}\sum_{\mathcal S' \in \mathcal S_{i'}^\beta} \alpha_{i, \mathcal{S}}\alpha_{i', \mathcal{S}'}\E\left[\omega_i \omega_{i'} \prod_{k \in \mathcal{S}} \prod_{k' \in \mathcal{S}'}Z_{k}Z_{k'}\right] - \left(\frac{1}{n}\sumn\sum_{\mathcal S \in \mathcal S_i^\beta, \mathcal{S} \neq \varnothing} \alpha_{i, \mathcal S}\right)^2\\
&\quad+\E\left[\left(\frac{1}{n}\sumn  \omega_i \bs \theta^\top \bX_i \right)^2\right] - \frac{2}{n^2} \sumn \sum_{i':  \mathcal N_i \cap \mathcal N_{i'} \neq \varnothing}\sum_{\mathcal S\in\mathcal S_i^\beta} \alpha_{i, \mathcal{S}}\E\left[\omega_i \omega_{i'} \bs \theta^\top \bX_{i'} \prod_{k \in \mathcal{S}}Z_{k}\right]. \label{equ: vartte2}
\end{align}
The expression above decomposes the variance into two parts. 
The first part, in \eqref{equ: vartte}, consists of second-order terms in $\alpha_{i,\mathcal S}$, including $\alpha_{i,\mathcal S}^2$ and $\alpha_{i,\mathcal S}\alpha_{i',\mathcal S'}$. 
The second part, in \eqref{equ: vartte2}, consists of first-order and zeroth-order terms in $\alpha_{i,\mathcal S}$.
As discussed in Section~\ref{sec:low-order_SNIPE}, we can construct unbiased estimators for $\alpha_{i,\mathcal S}$. However, it is generally difficult to estimate all second-order terms without bias. We will return to this issue in Section~\ref{sec:var_estimator}, where we discuss strategies for constructing conservative estimators for these terms and hence conservative variance estimators.

To circumvent this difficulty, we instead consider the variance difference between $\TTEhat(\bs 0)$ and $\TTEhat(\bs \theta)$. In this difference, all second-order terms cancel, since \eqref{equ: vartte} does not depend on $\bs \theta$. The remaining terms correspond to the difference between two instances of \eqref{equ: vartte2}, involving only first-order and zeroth-order terms. 
This simplification is useful because these lower-order terms admit unbiased estimation. In particular, by replacing each $\alpha_{i,\mathcal S}$ with its unbiased estimator $\hat{\alpha}_{i,\mathcal S}^{\text{unadj}}$, we obtain an unbiased estimator of the variance difference. In Section~\ref{section:theory}, we further show that this estimator is consistent under standard assumptions.
An alternative approach is to directly minimize a conservative variance estimator; 
however, this approach is generally less effective than using a consistent estimator of the variance difference.

Formally, the variance difference between $\TTEhat(\bs 0)$ and $\TTEhat(\bs \theta)$ can be written as
\begin{equation}
\begin{split}
\label{eqn:variance_diff_defi}
\Delta(\bs \theta) &= \Var\left(\TTEhat(\bs 0)\right) - \Var\left(\TTEhat(\bs \theta)\right)\\
& = -\E\left[\left(\frac{1}{n}\sumn  \omega_i \bs \theta^\top \bX_i \right)^2\right] + \frac{2}{n^2} \sumn \sum_{i':  \mathcal N_i \cap \mathcal N_{i'} \neq \varnothing}\sum_{\mathcal S\in\mathcal S_i^\beta} \alpha_{i, \mathcal{S}}\E\left[\omega_i \omega_{i'} \bs \theta^\top \bX_{i'} \prod_{k \in \mathcal{S}}Z_{k}\right].
\end{split}
\end{equation}

We then substitute $\hat{\alpha}_{i,\mathcal S}^{\text{unadj}}$ for $\alpha_{i,\mathcal S}$ and define the variance-improvement-maximized adjustment coefficient $\hat{\bs\theta}_{\mathrm{VIM}}$ as the maximizer of the resulting empirical objective. Solving this optimization problem explicitly leads to the following formal definition of the adjusted estimator.

\begin{estimator}[Variance-improvement–maximized (VIM) covariate adjustment] We define the variance-improvement-maximized (VIM) covariate-adjusted estimator as follows:
\begin{align*}
\TTEhat(\estchat)=\frac{1}{n}\sum_{i=1}^n \omega_i\left(Y_i - \bs \estchat^\top \bX_i \right ),
\end{align*} where
\begin{align*}
    \estchat 
    &=\E\Big(\sumn\sum_{i':  \mathcal N_i \cap \mathcal N_{i'} \neq \varnothing} \omega_i\omega_{i'} \bX_i\bX_{i'}^\top\Big)^{-1}\sumn \sum_{\mathcal S\in\mathcal S_i^\beta} \hat \alpha_{i, \mathcal S}^{\text{unadj}}\E\Big(\sum_{i':  \mathcal N_i \cap \mathcal N_{i'} \neq \varnothing} \omega_{i} \omega_{i'} \bX_{i'} \prod_{k \in \mathcal{S}} Z_{k}\Big).
\end{align*}   
Recall that $\omega_i = \sum_{\mathcal S\in \mathcal S_i^\beta}g(\mathcal S)\prod_{j\in\mathcal S}\frac{Z_j-p_j}{p_j(1-p_j)}$ and $\hat \alpha_{i, \mathcal S}^{\text{unadj}} = Y_i\prod_{j \in \mathcal{S}} \frac{-1}{p_j}\sum_{\mathcal{U} \in \mathcal S_i^{\beta}, \mathcal S \subseteq \mathcal{U}}\prod_{l \in \mathcal{U}}\frac{p_l - Z_l}{1-p_l}$ for every unit $i$ and set $\mathcal{S} \in \mathcal{S}_i^\beta$. Note that the expectations in the definition of $\estchat$ are computable under the known design.
\end{estimator}

The adjustment coefficient $\estchat$ is more network-aware than $\rbhat$, in the sense that it explicitly incorporates estimates of $\alpha_{i,\mathcal S}$ together with cross-unit interaction terms that reflect interference and network structure.

It is useful to relate $\VIM$ to the regression-based adjustment discussed in the previous section. In the absence of interference, under standard assumptions, $\Reg$, $\VIM$, and Lin's estimator are closely related. In particular, when Lin's estimator is written in a control variate form, its coefficient $\hat{\bs\theta}_{\mathrm{Lin}}$ is asymptotically equivalent to both $\rbhat$ and $\estchat$. Consequently, in this setting, all three estimators achieve asymptotic variance reduction relative to the unadjusted estimator. See Appendix~\ref{appendix:VIM_discussion} for details.

This equivalence does not extend to settings with interference. In general, the three estimators target different directions, and their performance can differ substantially. 
In Section~\ref{section:theory}, we show that $\VIM$ satisfies a no-harm guarantee: its variance is no greater than that of $\SNIPE$. Such a guarantee is not available for $\Reg$. In Section~\ref{sec:simulation}, we compare their empirical performance and show that, depending on the interference pattern, $\VIM$ can outperform $\Reg$.

To build intuition, we revisit the toy example in Section~\ref{sec:Reg-SNIPE}, where $\Reg$ performs worse than $\SNIPE$. In the same setting, $\VIM$ does not perform worse than $\SNIPE$, illustrating how targeting variance improvement protects against the variance inflation that may arise for $\Reg$ under interference.
\begin{example}[Toy example continued]
We continue from Example \ref{exp:increasevar_small} introduced in Section \ref{sec:Reg-SNIPE}, where the variance of $\TTEhat(\bs \theta)$ for any $\bs \theta \in \mathbb{R}$ is
\[
\Var\sqb{\TTEhat(\bs \theta)}
= \frac{16}{9} + \frac{1}{3} {\bs \theta}^2. 
\]
 By definition, we compute the weight for each unit as 
 \begin{align*}
     \omega_1 = 4(Z_1 - 0.5) + 4(Z_2 - 0.5), \quad
     \omega_2 = 4(Z_1 - 0.5) + 4(Z_2 - 0.5), \quad
     \omega_3 = 4(Z_3 - 0.5).
\end{align*}
Proposition~\ref{prop:consistent} shows that $\estchat \stackrel{p}{\to} \estc$ (see Section~\ref{section:theory}), where
\begin{align*}
     \estc
=
\E\Big(\frac{1}{n^2}\sum_{i,i':\,\mathcal N_i\cap\mathcal N_{i'}\neq\varnothing}
\omega_i\omega_{i'}\,\bX_i\bX_{i'}^\top\Big)^{-1}
\E\Big(\frac{1}{n^2}\sum_{i,i':\,\mathcal N_i\cap\mathcal N_{i'}\neq\varnothing}
\omega_i\omega_{i'}\,\bX_i Y_{i'}\Big)
=0,
 \end{align*}   
 This implies that the asymptotic variance of $\VIM$ is the same as that of the unadjusted estimator $\SNIPE$. In contrast, as demonstrated in Example \ref{exp:increasevar_small}, the variance of $\Reg$ is asymptotically strictly greater than that of $\SNIPE$. $\VIM$, by explicitly incorporating the interference structure, avoids this issue and guarantees a variance no greater than that of the unadjusted estimator. 
\end{example}

\section{Conservative Variance Estimation}\label{sec:var_estimator}

\subsection{Variance estimator for the covariate-adjusted estimator}

Having constructed improved estimators for the total treatment effect, we now turn to variance estimation for inference. Our goal is to obtain a conservative (but not overly conservative) variance estimator for the general covariate-adjusted estimator $\TTEhat(\bs \theta)$, for arbitrary $\bs \theta$. Plugging in specific choices of $\bs \theta$ recovers variance estimators for the regression-based and VIM-based covariate-adjusted estimators.

Recall from \eqref{eqn:variance_diff_defi} that
\begin{align*}
    \Var(\TTEhat(\bs \theta)) = \Var(\SNIPE) - \Delta(\bs \theta),
\end{align*}
and that an unbiased estimator for $\Delta(\bs \theta)$ was developed in Section~\ref{sec:VIM-SNIPE}. In particular,
\begin{equation}
\label{eqn:Delta_hat_defi}
\begin{split} 
\hat \Delta(\bs \theta) 
&= -\frac{1}{n^2}{\bs \theta}^\top \E\Bigg(\frac{1}{n^2}\sumn \sum_{i': \mathcal N_i \cap \mathcal N_{i'} \neq \varnothing} \omega_i \omega_{i'} \bX_i \bX_{i'}^\top \Bigg) {\bs \theta} \\
&\qquad \qquad + \frac{2}{n^2} {\bs \theta}^\top \sumn \sum_{i': \mathcal N_i \cap \mathcal N_{i'} \neq \varnothing}
\sum_{\mathcal S \in \mathcal S_i^\beta} \hat\alpha^{\text{unadj}}_{i, \mathcal{S}}
\E\Bigg[\omega_i \omega_{i'} \bX_{i'} \prod_{k \in \mathcal{S}} Z_k \Bigg].
\end{split}
\end{equation}
Therefore, given any estimator $\widehat{\Var}(\SNIPE)$, we can construct a variance estimator for $\TTEhat(\bs \theta)$ via
\begin{align*}
    \widehat{\Var}(\TTEhat(\bs \theta)) = \widehat{\Var}(\SNIPE) - \hat{\Delta}(\bs \theta).
\end{align*}

\subsection{Variance estimator for the SNIPE estimator}

We now focus on constructing a conservative variance estimator for $\SNIPE$. This remains challenging: under interference, cross-unit dependence induced by the network complicates variance estimation.
\citet{CortezRodriguezEichhornYu+2023} propose a theoretically valid conservative estimator for $\Var(\SNIPE)$, building on worst-case bounding arguments from \citet{aronow2013conservative, aronow2017estimating}. While this estimator guarantees validity, it can be highly conservative in practice, often leading to confidence intervals that are much wider than necessary.
Our goal is to construct an alternative estimator that retains conservativeness while reducing over-coverage by leveraging the low-order interactions structure.

We begin by providing some intuition for the main challenge and how we address it. 
Recall from Section~\ref{sec:VIM-SNIPE} that $\Var(\SNIPE)$ involves both first- and second-order terms in ${\alpha}_{i, \mathcal S}$. 
While unbiased estimators for ${\alpha}_{i, \mathcal S}$ are readily available, estimating the products 
${\alpha}_{i, \mathcal S} {\alpha}_{i', \mathcal S'}$ is more subtle.

A key observation is that many such products can, in fact, be estimated unbiasedly. To build intuition, consider the case $\beta = 1$, where
$Y_i = \alpha_{i,\varnothing} + \sum_{j \in \mathcal{N}_i} \alpha_{i,\{j\}} Z_j$ for unit $i$ and $Y_{i'} = \alpha_{i',\varnothing} + \sum_{k \in \mathcal{N}_{i'}} \alpha_{i',\{k\}} Z_k$ for another unit $i'$.
We first note that, interestingly, the product $Y_i Y_{i'}$ follows a second-order interaction model:
\[
Y_i Y_{i'} 
= \alpha_{i,\varnothing}\alpha_{i',\varnothing} 
+ \alpha_{i',\varnothing}\sum_{j \in \mathcal{N}_i} \alpha_{i,\{j\}} Z_j
+ \alpha_{i,\varnothing}\sum_{k \in \mathcal{N}_{i'}} \alpha_{i',\{k\}} Z_k
+ \sum_{j \in \mathcal{N}_i} \sum_{k \in \mathcal{N}_{i'}} \alpha_{i,\{j\}} \alpha_{i',\{k\}} Z_j Z_k.
\]
For $j \neq k$, to estimate $\alpha_{i,{j}} \alpha_{i',{k}}$, we can apply the same idea used in constructing the SNIPE estimator. Multiply both sides of the above equation by $(Z_j - p_j)(Z_k - p_k)$. Then, all terms on the right-hand side have mean zero except for $\alpha_{i,\{j\}} \alpha_{i',\{k\}} Z_j Z_k (Z_j - p_j)(Z_k - p_k)$, which implies that
$Y_i Y_{i'} (Z_j - p_j)(Z_k - p_k) /\p{p_j(1-p_j)p_k(1-p_k)}$
is an unbiased estimator of $\alpha_{i,\{j\}} \alpha_{i',\{k\}}$.

Things become more subtle when one of $\mathcal S$ and $\mathcal S'$ is empty, in which case the orthogonalization argument breaks down. In these cases, we resort to conservative bounds based on Cauchy--Schwarz. A key point is that applying Cauchy--Schwarz locally leads to substantial conservativeness; instead, we apply it at a more aggregated level, as described below.

We now present a detailed construction of the variance estimator. Recall that $\SNIPE$ can be expressed as the difference between the full low-order expansion in the $\hat{\alpha}$'s and the baseline component, which yields
\begin{equation}
\label{eqn:key_cauchy_schwarz}
\begin{split}
\Var(\SNIPE)
&= \Var\Big(
\frac{1}{n}\sum_{i=1}^n \sum_{\mathcal S \in \mathcal S_i^\beta} \hat{\alpha}_{i, \mathcal S}^{\text{unadj}}
-
\frac{1}{n}\sum_{i=1}^n \hat{\alpha}_{i, \varnothing}^{\text{unadj}}
\Big) \\
&\leq 2\Bigg[
\Var\Big(
\frac{1}{n}\sum_{i=1}^n \sum_{\mathcal S \in \mathcal S_i^\beta} \hat{\alpha}_{i, \mathcal S}^{\text{unadj}}
\Big)
+
\Var\Big(
\frac{1}{n}\sum_{i=1}^n \hat{\alpha}_{i, \varnothing}^{\text{unadj}}
\Big)
\Bigg].
\end{split}
\end{equation}
We estimate the two variance components in \eqref{eqn:key_cauchy_schwarz} separately.

For variance of the interaction component $\Var\Big(
\frac{1}{n}\sum_{i=1}^n \sum_{\mathcal S \in \mathcal S_i^\beta} \hat{\alpha}_{i, \mathcal S}^{\text{unadj}}
\Big)$, recall that $\hat{\alpha}_{i, \mathcal S}^{\text{unadj}}$ depends only on $Y_i$ and $\{Z_j: j \in \mathcal S\}$ with $\mathcal S \subseteq \mathcal N_i$. 
Hence,
$\Cov\big(
\hat{\alpha}_{i, \mathcal S}^{\text{unadj}},
\hat{\alpha}_{i', \mathcal S'}^{\text{unadj}}
\big) = 0$ if $\mathcal N_i \cap \mathcal N_{i'} = \varnothing$. 
Therefore,
\begin{align*}
\Var\Big(
\frac{1}{n}\sum_{i=1}^n \sum_{\mathcal S \in \mathcal S_i^\beta} \hat{\alpha}_{i, \mathcal S}^{\text{unadj}}
\Big)
&= \frac{1}{n^2}
\sum_{i=1}^n
\sum_{i':\, \mathcal N_i \cap \mathcal N_{i'} \neq \varnothing}
\sum_{\mathcal S \in \mathcal S_i^\beta}
\sum_{\mathcal S' \in \mathcal S_{i'}^\beta}
\Cov\big(
\hat{\alpha}_{i, \mathcal S}^{\text{unadj}},
\hat{\alpha}_{i', \mathcal S'}^{\text{unadj}}
\big) \\
&= \frac{1}{n^2}
\sum_{i=1}^n
\sum_{i':\, \mathcal N_i \cap \mathcal N_{i'} \neq \varnothing}
\sum_{\mathcal S \in \mathcal S_i^\beta}
\sum_{\mathcal S' \in \mathcal S_{i'}^\beta}
\Big[
\E\big(
\hat{\alpha}_{i, \mathcal S}^{\text{unadj}}
\hat{\alpha}_{i', \mathcal S'}^{\text{unadj}}
\big)
-
{\alpha}_{i, \mathcal S}
{\alpha}_{i', \mathcal S'}
\Big].
\end{align*}

We estimate the sum of $\E\big(
\hat{\alpha}_{i, \mathcal S}^{\text{unadj}}
\hat{\alpha}_{i', \mathcal S'}^{\text{unadj}}
\big)$ terms using the plug-in second-moment estimator:
\[
\frac{1}{n^2}
\sum_{i=1}^n
\sum_{i':\, \mathcal N_i \cap \mathcal N_{i'} \neq \varnothing}
\sum_{\mathcal S \in \mathcal S_i^\beta}
\sum_{\mathcal S' \in \mathcal S_{i'}^\beta}
\hat{\alpha}_{i, \mathcal S}^{\text{unadj}}
\hat{\alpha}_{i', \mathcal S'}^{\text{unadj}}.
\]

For the terms involving the ${\alpha}_{i, \mathcal S}
{\alpha}_{i', \mathcal S'}$, we will use the unbiased-product construction described above. 
Specifically, define the pseudo-outcome $\tilde Y_{ii'}:=Y_i Y_{i'}$ for pair of units $(i,i')$. Since
$Y_i=\sum_{\mathcal S\in\mathcal S_i^\beta}\alpha_{i,\mathcal S}\prod_{j\in\mathcal S} Z_j$,
it follows that
\[
\tilde Y_{ii'}
=\sum_{\mathcal S\in\mathcal S_i^\beta}\sum_{\mathcal S'\in\mathcal S_{i'}^\beta}
\alpha_{i,\mathcal S}\alpha_{i',\mathcal S'}\prod_{j\in\mathcal S\cup\mathcal S'} Z_j
=\sum_{\mathcal T\in\mathcal T_{ii'}^{\beta}}\gamma_{ii',\mathcal T}\prod_{j\in\mathcal T} Z_j,
\]
where $\mathcal T_{ii'}^{\beta}:=\{\mathcal S\cup\mathcal S':\mathcal S\in\mathcal S_i^\beta,\ \mathcal S'\in\mathcal S_{i'}^\beta\}$ and
$\gamma_{ii',\mathcal T}:=\sum_{\mathcal S \in\mathcal S_i^\beta,\mathcal S'\in\mathcal S_{i'}^\beta:\,\mathcal S\cup\mathcal S'=\mathcal T}\alpha_{i,\mathcal S}\alpha_{i',\mathcal S'}$.
Therefore, $\sum_{\mathcal S\in\mathcal S_i^\beta}\sum_{\mathcal S'\in\mathcal S_{i'}^\beta}\alpha_{i,\mathcal S}\alpha_{i',\mathcal S'}
=\sum_{\mathcal T\in\mathcal T_{ii'}^{\beta}}\gamma_{ii',\mathcal T}$. 
We estimate these terms by applying the same unadjusted estimator to $\tilde Y_{ii'}$, 
yielding $\{\hat\gamma^{\mathrm{unadj}}_{ii',\mathcal T}\}$, and summing 
$\sum_{\mathcal T}\hat\gamma^{\mathrm{unadj}}_{ii',\mathcal T}$ over pairs $(i,i')$ with 
$\mathcal N_i \cap \mathcal N_{i'} \neq \varnothing$. This leads to the estimator:
\begin{equation}
\begin{split}
\widehat{\Var}\Big(
\frac{1}{n}\sum_{i=1}^n \sum_{\mathcal S \in \mathcal S_i^\beta} \hat{\alpha}_{i, \mathcal S}^{\text{unadj}}
\Big)
= \frac{1}{n^2}
\sum_{i=1}^n
\sum_{i':\, \mathcal N_i \cap \mathcal N_{i'} \neq \varnothing}
\sum_{\mathcal S \in \mathcal S_i^\beta}
\sum_{\mathcal S' \in \mathcal S_{i'}^\beta}
\hat{\alpha}_{i, \mathcal S}^{\text{unadj}}
\hat{\alpha}_{i', \mathcal S'}^{\text{unadj}} 
- \frac{2}{n^2}\sumn\sum_{i': \mathcal N_i \cap \mathcal N_{i'}} \sum_{\mathcal T \in \mathcal T_{ii'}^{\beta}} \hat\gamma_{ii', \mathcal T}^{\text{unadj}},
  \end{split}  
\end{equation}
where 
\begin{equation}
\hat\gamma_{ii', \mathcal T}^{\text{unadj}} = 
 Y_i Y_{i'} \prod_{j \in \mathcal T} \frac{-1}{p_j}
\sum_{\substack{\mathcal U \in \mathcal T_{i i'}^\beta \\ \mathcal U \supseteq \mathcal T}}
\prod_{l \in \mathcal U} \frac{p_l - Z_l}{1-p_l}.
\end{equation}

The variance of the baseline component,
$\Var\Big(
\frac{1}{n}\sum_{i=1}^n \hat{\alpha}_{i, \varnothing}^{\mathrm{unadj}}
\Big)$,
is treated analogously.

Combining the two components yields the variance estimator stated below.
\begin{varestimator}[Variance Estimator for SNIPE]
\begin{align*}
\widehat{\Var}\left(\TTEhat(\bs 0)\right) &= \frac{2}{n^2}\sumn\sum_{i': \mathcal N_i \cap \mathcal N_{i'}} \sum_{\mathcal S \in \mathcal S_i^\beta} \sum_{\mathcal S' \in \mathcal S_{i'}^\beta} \hat\alpha_{i, \mathcal S}^{\text{unadj}}\hat\alpha_{i', \mathcal S'}^{\text{unadj}}
- \frac{2}{n^2}\sumn\sum_{i': \mathcal N_i \cap \mathcal N_{i'}} \sum_{\mathcal T \in \mathcal T_{ii'}^{\beta}} \hat\gamma_{ii', \mathcal T}^{\text{unadj}}\\
 &\qquad  \qquad 
 + \frac{2}{n^2}\sumn\sum_{i': \mathcal N_i \cap \mathcal N_{i'}} \hat\alpha_{i, \varnothing}^{\text{unadj}}\hat\alpha_{i', \varnothing}^{\text{unadj}}
 - \frac{2}{n^2}\sumn\sum_{i': \mathcal N_i \cap \mathcal N_{i'}} \hat\gamma_{ii', \varnothing }^{\text{unadj}},
\end{align*}
where 
\end{varestimator}
$\mathcal T_{ii'}^{\beta}:=\{\mathcal S\cup\mathcal S':\mathcal S\in\mathcal S_i^\beta,\ \mathcal S'\in\mathcal S_{i'}^\beta\}$ and 
$\hat\gamma_{ii', \mathcal T}^{\text{unadj}} = 
 Y_i Y_{i'} \prod_{j \in \mathcal T} \frac{-1}{p_j}
\sum_{\mathcal U \in \mathcal T_{i i'}^\beta, \mathcal U \supseteq \mathcal T}
\prod_{l \in \mathcal U} \frac{p_l - Z_l}{1-p_l}$. 

Finally, for general $\bs \theta$, we define the corresponding variance estimator for the covariate-adjusted estimator as follows:
\begin{varestimator}[Variance estimator for the covariate-adjusted estimator]
\label{varestimator_covariate}
\[
   \widehat{\Var}\left(\TTEhat(\bs \theta)\right) = \widehat{\Var}\left(\TTEhat(\bs 0)\right) - \hat{\Delta}(\bs \theta), 
\]
where $\hat{\Delta}(\bs \theta)$ is defined in \eqref{eqn:Delta_hat_defi}.
\end{varestimator}

\section{Large Sample Properties}\label{section:theory}
In this section, we study the large-sample properties of the proposed estimators. After introducing the assumptions, we first establish consistency of the estimated adjustment coefficients $\rbhat$ and $\estchat$, and hence consistency of the corresponding estimators $\TTEhat(\rbhat)$ and $\TTEhat(\estchat)$. We then show that the VIM-based estimator has an asymptotic no-harm property. Next, for a general class of covariate-adjusted estimators, we derive a variance upper bound and establish asymptotic normality under suitable conditions. Finally, we show that the proposed variance estimator is asymptotically conservative, thereby enabling valid large-sample inference.

\subsection{Assumptions}
\begin{assumption}[Boundedness]\label{as:bounded}
Let $X_{\max} = \max_{i \in [n]} \|\bX_i\|_1$ and $Y_{\max} = \max_{i \in [n]} \sum_{\mathcal S \in \mathcal S_{i}^\beta}|\alpha_{i, \mathcal S}|$. There exists a constant $C > 0$ such that $X_{\max} \leq C$ and $Y_{\max} \leq C$. The parameter $\beta$ is a fixed integer that does not vary with $n$. Moreover, there exists a constant $p \in (0, 0.5]$ such that the individual treatment probabilities satisfy $p \le p_i \le 1 - p$ for all $i \in [n]$.
\end{assumption}

Assumption \ref{as:bounded} imposes standard regularity conditions that avoid instability in estimation and ensure sufficient variation in treatment assignments.

\begin{assumption}[Sparsity]\label{as:sparsity}
The maximum of in- and out-degrees of the interference network satisfies
$d = O(1)$.
\end{assumption}
We impose a sparsity assumption on the interference network in Assumption~\ref{as:sparsity}. This assumption is reasonable in many empirical settings. For example, in the well-known study of \citet{cai2015social}, the interference network has maximum degree five. Moreover, when this assumption is mildly violated, we do not observe substantial empirical degradation in the estimator’s behavior. We therefore impose sparsity primarily to keep the theoretical analysis tractable.

\begin{assumption}[Invertibility]\label{as:invertible} Define $\bs M$ element-wise by 
\begin{equation}\label{equ: M}
M_{ii'} = \sum_{\mathcal S \in\mathcal S_i^\beta \cap S_{i'}^\beta } g(\mathcal S)^2\prod_{j \in \mathcal{S}} \frac{1}{p_j(1-p_j)}, 
\end{equation} 
and let $\bX = \left[\bX_1, \ldots, \bX_n\right]^\top$. 
Then there exists a positive constant $c_{\lambda_{\min}}$ such that the smallest absolute eigenvalues of 
$
\frac{1}{n}\bX^\top\bX$ and $\frac{1}{n} \bX^\top \bs M \bX
$
are bounded below by $c_{\lambda_{\min}}$. 

\end{assumption}
Assumption~\ref{as:invertible} imposes regularity conditions that are standard in the literature on causal inference under network interference. This assumption, which partly relies on Assumption~\ref{as:sparsity}, requires a lower bound on the smallest absolute eigenvalue of the average outer product of covariates. Such a condition rules out degeneracy in the covariate structure induced by the network topology. As with Assumption~\ref{as:sparsity}, this assumption is introduced mainly for analytical convenience; in practice, mild violations do not appear to substantially affect performance.

\begin{assumption}[Non-degeneracy]\label{as:variance} As $n \to \infty$, the following asymptotic convergence holds:
\begin{enumerate}[label=(\roman*)]
    \item \label{as:variance(i)}$$
\frac{1}{n}\sumn \sum_{\mathcal S \in \mathcal S_i^\beta} g^2(\mathcal S) \prod_{j \in \mathcal S}\frac{1}{p_j(1-p_j)}\bX_i\bX_i^\top \to \tilde V_{\bX}, \quad \frac{1}{n}\sumn\E\left( \omega_i^2 \bX_i Y_i\right) \to \tilde{V}_{\bX Y},
$$
for some finite $\tilde V_{\bX}$, and $\tilde V_{\bX Y}$;
    \item \label{as:variance(ii)}$$
\frac{1}{n} \bs  \bX^\top \bs M \bs X \to V_{\bX},\quad
\frac{1}{n} \sumn \sum_{i':  \mathcal N_i \cap \mathcal N_{i'} \neq \varnothing} \E\left(\omega_i \omega_{i'}\bX_{i'}Y_i\right) \to V_{\bX Y},
$$
for some finite $V_{\bX}$, and $V_{\bX Y}$, where $\bs M$ is defined in \eqref{equ: M}.
\end{enumerate}
\end{assumption}
The boundedness of each term is already implied by Assumptions \ref{as:interference}-\ref{as:invertible}. Moreover, combined with Assumption \ref{as:invertible}, it implies that $\|V_{\bX}\| >0$. This assumption rules out degeneracy of the limiting design matrices and guarantees that the required terms converge to finite limits. Under SUTVA, \hyperref[as:variance(i)]{6(i)} and \hyperref[as:variance(i)]{6(ii)} are equivalent.

\subsection{Consistency of \texorpdfstring{$\TTEhat(\rbhat)$ and $\TTEhat(\estchat)$}{tauhat(rb) and tauhat(VIM)}}
\begin{proposition}[Consistency of $\rbhat$]\label{prop:consistrb}
Define
\begin{align*}
  \rb =    \E\left(\frac{1}{n}\sumn \omega_i^2\bX_i\bX_i^\top\right)^{-1}\E\left(\frac{1}{n}\sumn \omega_i^2 \bX_i Y_i\right),
    \end{align*} 
where $\omega_i = \sum_{\mathcal S\in \mathcal S_i^\beta}g(\mathcal S)\prod_{j\in\mathcal S}\frac{Z_j-p_j}{p_j(1-p_j)}$. Under Assumptions \ref{as:interference}--\ref{as:variance}, then 
$\rbhat - \rb \overset{p}{\to} 0$. Moreover, $\rb$ converges to a finite limit denoted by $\rb^\ast$, and hence $\rbhat \overset{p}{\to} \rb^\ast$.
\end{proposition}

\begin{proposition}[Consistency of $\estchat$]\label{prop:consistent}
Let  
\begin{align*}
    \estc = \E\Big(\frac{1}{n^2}\sumn\sum_{i':  \mathcal N_i \cap \mathcal N_{i'} \neq \varnothing} \omega_i\omega_{i'} \bX_i\bX_{i'}^\top\Big)^{-1}\E\Big(\frac{1}{n^2}\sumn\sum_{i':  \mathcal N_i \cap \mathcal N_{i'} \neq \varnothing} \omega_i\omega_{i'} \bX_i Y_{i'}\Big),
\end{align*}
where $\omega_i = \sum_{\mathcal S\in \mathcal S_i^\beta}g(\mathcal S)\prod_{j\in\mathcal S}\frac{Z_j-p_j}{p_j(1-p_j)}$. Under Assumptions \ref{as:interference}--\ref{as:variance},
$$
\estchat - \estc \overset{p}{\to} 0.
$$
Moreover,  $\estc$ has a finite limit, denoted by $\estc^\ast$, and hence $\estchat \overset{p}{\to} \estc^\ast$ in probability.
\end{proposition}

Propositions~\ref{prop:consistrb} and~\ref{prop:consistent} establish the consistency of $\rbhat$ and $\estchat$, respectively. As a direct corollary, the corresponding estimators $\TTEhat(\rbhat)$ and $\TTEhat(\estchat)$ are asymptotically unbiased and consistent. In particular,
\[
\TTEhat(\rbhat) \xrightarrow{p} \TTE
\qquad\text{and}\qquad
\TTEhat(\estchat) \xrightarrow{p} \TTE.
\]

\subsection{Asymptotic no-harm property of \texorpdfstring{$\VIM$}{tauhat(VIM)}}\label{sec:vim_asymptotic}
In what follows, we show that $\VIM$ has asymptotic variance no greater than that of $\SNIPE$, a property generally not enjoyed by $\Reg$. This mirrors a key property of \cite{Lin2013}’s estimator in the no-interference setting. 

\begin{theorem}[No worse variance]\label{thm:vimvar}
Under Assumptions \ref{as:interference}--\ref{as:variance}, the variance of $\VIM$ is asymptotically no worse than $\SNIPE$. Specifically,
$$n\left[\Var\left(\TTEhat_{\text{unadj}}\right) - \Var\left(\TTEhat(\estchat)\right)\right] \to {\estc^\ast}^\top V_{\bX} \estc^\ast \geq 0,$$ 
where $V_{\bX}$ is the positive semidefinite matrix defined in Assumption \ref{as:variance}, and ${\estc^\ast}$ is defined in Proposition \ref{prop:consistent}. Moreover, the variance of $\VIM$ is asymptotically no worse than any $\bs \theta$-adjusted estimator:
$n\left[ \Var\left(\TTEhat(\bs \theta)\right)-  \Var\left(\TTEhat(\estchat)\right)\right] \to \left(\bs \theta^\ast - \estc^\ast\right) ^\top V_{\bX}\left(\bs \theta^\ast - \estc^\ast\right)  \geq 0,$ 
for any ${\bs\theta}$ that converges to a finite limit $\bs \theta^\ast$. 
\end{theorem}

Theorem~\ref{thm:vimvar} shows that $\VIM$ attains asymptotic variance no larger than that of $\SNIPE$. The theorem also shows that $\VIM$ is optimal within our class of general covariate-adjusted estimators parameterized by $\bs \theta$. This result provides a theoretical guarantee for using the maximized-improvement framework: while naive regression adjustments may inflate variance, $\VIM$ ensures that covariates can only help, never hurt, asymptotically.

\subsection{General covariate-adjusted estimator}\label{sec:general_adj-SNIPE}
In this subsection, we focus on a general covariate-adjusted estimator.

\begin{theorem}[Variance upper bound]\label{thm:ubound}
 Under Assumptions \ref{as:interference}-\ref{as:bounded}, for any fixed $\theta$, the estimator $\hat{\tau}(\theta)$ is unbiased, and
 \[
 \Var\p{\TTEhat(\bs\theta)} \leq \frac{4d_\text{in}d_\text{out}}{n}\Big(\max_{i \in [n]} \big(|\alpha_{i, \mathcal \varnothing} - \bs \theta^\top \bX_i| + \sum_{\substack{\mathcal S \in \mathcal S_{i}^\beta\\
\mathcal S \neq \varnothing}}|\alpha_{i, \mathcal S}| \big) \Big)^2 \left(
\frac{ed_\text{in}}{\beta}\max\left\{4\beta^2,\frac{1}{p(1-p)}\right\}\right)^\beta.\]
\end{theorem}
Theorem~\ref{thm:ubound} provides an upper bound on the variance of the general covariate-adjusted estimator for any fixed $\bs \theta$. It is closely related to the variance upper bound established by \citet{CortezRodriguezEichhornYu+2023}. In particular, it illustrates that consistency of the covariate-adjusted estimator does not require the maximum degree of the interference network to remain bounded by a constant. Instead, the variance bound only requires that the degrees grow at a controlled polynomial rate in $n$. This highlights that our estimator remains consistent under more general network structures than those implied by Assumption \ref{as:sparsity}. We view the bound as sufficient for our theoretical development, but we do not claim it is sharp; tighter bounds may be achievable under additional structural restrictions. The proof strategy of Theorem~\ref{thm:ubound} largely follows that of Theorem 1 in \citet{CortezRodriguezEichhornYu+2023}.

Next, we study the large-sample properties of $\TTEhat(\hat{\bs\theta})$, where $\hat{\bs\theta}$ may be data-dependent. We first introduce two additional assumptions.
\begin{assumption}[No outcome degeneracy]\label{as:no_degeneracy} As $n \to \infty$,
$
\frac{1}{n}\Var\left(\sumn\omega_iY_i\right) \to V_Y
$
for some finite $V_{Y}$.
\end{assumption}

Assumption \ref{as:no_degeneracy} extends a standard regularity condition commonly imposed in the literature (e.g., Assumption 3 in \cite{CortezRodriguezEichhornYu+2023}) to our setting. 
In contrast to Assumption 3 in \cite{CortezRodriguezEichhornYu+2023}, which requires the variance of the unadjusted treatment effect estimator to converge to a strictly positive constant, here the variance of the weighted sum of outcomes is allowed to vanish in the limit. Instead, we will later impose the similar requirement that the asymptotic variance of the corresponding covariate-adjusted estimator is strictly positive. 

\begin{assumption}[Convergence of $\hat{\bs \theta}$]\label{as:thetahat}
There exists a finite $\bs \theta^\ast \in \mathbb R^{d_{\bs X}}$ such that
$\hat{\bs \theta} \overset{p}{\to} \bs \theta^\ast$. 
\end{assumption}
This assumption states that the estimator $\hat{\boldsymbol{\theta}}$ converges in probability to a well-defined population limit $\boldsymbol{\theta}^\ast$. 
Specifically, both $\rbhat$ and $\estchat$ satisfy this assumption under regularity conditions; see Propositions \ref{prop:consistrb} and \ref{prop:consistent} in Appendix \ref{appendix:additional_discussion}.

For some fixed finite $\bs \theta^\ast$ independent of $n$, we define
\begin{equation}
\label{eqn:V_formula}
V(\bs \theta^\ast) = V_{Y} +  {\bs \theta^\ast}^\top V_{\bs X} \bs \theta^\ast - 2{\bs \theta^\ast}^\top V_{\bs XY}, 
\end{equation}
where $V_Y, V_{\bX}, V_{\bX Y}$ are defined in Assumptions \ref{as:variance} and \ref{as:no_degeneracy}.

\begin{theorem}[CLT]\label{thm:clt}
Let $\TTEhat(\hat{\bs \theta})$ denote the covariate-adjusted estimator based on an estimated parameter $\hat{\bs \theta}$ (see \eqref{equ: ttehat} for the definition). 
Under Assumptions~\ref{as:interference}--\ref{as:sparsity} and \ref{as:variance}--\ref{as:no_degeneracy}, suppose that $\hat{\bs \theta}$ satisfies Assumption~\ref{as:thetahat}, and that $V(\bs \theta^\ast)$ defined in \eqref{eqn:V_formula} is strictly positive. 
Then
\[
  \sqrt{n} \p{\TTEhat(\hat{\bs\theta}) - \TTE}
  \xrightarrow{d} \mathcal{N}\!\left(0,\, V(\bs \theta^\ast)\right).
\]
\end{theorem}

Theorem~\ref{thm:clt} can be viewed as the covariate-adjusted analogue of Theorem~3 in \citet{CortezRodriguezEichhornYu+2023}. It establishes that, under the stated conditions, the general covariate-adjusted estimator is asymptotically normal. The proof adapts techniques from \citet{CortezRodriguezEichhornYu+2023}.

\begin{corollary}[CLT for $\TTEhat(\rbhat)$ and $\TTEhat(\estchat)$]\label{thm:vimclt}
Under Assumptions \ref{as:interference}--\ref{as:no_degeneracy}, 
\begin{itemize}
    \item [(a)] If $V(\rb^\ast) > 0$, 
    $\sqrt{n}(\TTEhat(\rbhat) - \TTE) \stackrel{d}{\to} \mathcal{N}(0,V(\rb^\ast))$.
    \item [(b)] If $V(\estc^\ast) > 0$, $\sqrt{n}(\TTEhat(\estchat) - \TTE) \stackrel{d}{\to} \mathcal{N}(0,V(\estc^\ast))$. 
\end{itemize}
\end{corollary}
Corollary~\ref{thm:vimclt} follows directly from Theorem~\ref{thm:clt} applied to $\TTEhat(\rbhat)$ and $\TTEhat(\estchat)$. Under Assumptions \ref{as:interference}--\ref{as:variance}, Assumption~\ref{as:thetahat} is automatically satisfied; see Propositions~\ref{prop:consistrb} and \ref{prop:consistent}.

\subsection{Conservative variance estimator}
\label{sec:theory_variance}
We now study the large-sample properties of the variance estimator introduced in Section~\ref{sec:var_estimator}. 
Theorem~\ref{thm:var_est} shows that this estimator is asymptotically conservative. 
In particular, the result applies to both the $\Reg$ and $\VIM$ adjustment schemes considered in this paper.

\begin{theorem}[Conservative variance estimator]\label{thm:var_est}
Let $\TTEhat(\hat{\bs \theta})$ denote the covariate-adjusted estimator based on an estimated parameter $\hat{\bs \theta}$ (see \eqref{equ: ttehat} for the definition). Define $\hat V(\hat{\bs \theta}) = n\widehat{\Var}\left(\TTEhat(\hat{\bs \theta})\right)$, where $\widehat{\Var}\left(\TTEhat(\hat{\bs \theta})\right)$ is given in Section~\ref{sec:var_estimator}. Under Assumptions~\ref{as:interference}--\ref{as:sparsity} and \ref{as:variance}--\ref{as:no_degeneracy}, suppose that $\hat{\bs \theta}$ satisfies Assumption~\ref{as:thetahat}. Then
\[
\hat V(\hat{\bs\theta})\xrightarrow{p}\tilde V(\bs\theta^\star)\ge V(\bs\theta^\star),
\]
where $\tilde V(\bs\theta^\star) = 2\left[\Var\left(\frac{1}{n}\sumn\sum_{\mathcal S \in \mathcal S_i^\beta} \hat{\alpha}_{i, \mathcal S}^{\text{unadj}}\right) + \Var\left(\frac{1}{n}\sumn \hat{\alpha}_{i, \varnothing}^{\text{unadj}}\right)\right] - \Delta(\bs\theta^\star)$.
\end{theorem}

\section{Simulation Study}\label{sec:simulation}

In this section, we run simulation studies\footnote{Code is available at \url{https://github.com/Cynlia/Covariate-Adjustment-Based-on-SNIPE}.} to evaluate the finite-sample performance of $\Reg$ and $\VIM$ under four experimental factors: sample size ($n$), treatment probability ($p$), the indirect-to-direct effect ratio ($r$), and the fraction of observed covariates ($\rho$). For each factor, we consider two network models and two interaction orders ($\beta\in\{1,2\}$). We compare $\Reg$ and $\VIM$ with $\SNIPE$, the estimator of \citet{Lin2013} (see Estimator~\ref{estimator:lin} in Appendix~\ref{appendix:reg_estimator} for details), and the naive difference-in-means (DM). Each setting is repeated independently $500$ times.

Both the estimator of \citet{Lin2013} and the difference-in-means estimator rely on SUTVA, and are therefore expected to be biased in the presence of interference.

\paragraph{Covariates.}
For each replicate, generate independent $3$-dimensional covariates
$\tilde{\bX}_i^{\text{obs}},\tilde{\bX}_i^{\text{unobs}}\stackrel{\text{i.i.d.}}{\sim}\mathcal N(\bm 0,\bm I_3)$ for $i=1,\ldots,n$.
Only $\tilde{\bX}_i^{\text{obs}}$ is observed.
Let $\bX_i^{\text{obs}}$ and $\bX_i^{\text{unobs}}$ denote their centered versions, and define $\bX_i^{\text{true}}
= \rho\,\bX_i^{\text{obs}}
+ \sqrt{1-\rho^2}\,\bX_i^{\text{unobs}}$.

\paragraph{Treatment.}
Each node is independently assigned to treatment with probability $p$.

\paragraph{Interference network.}
We consider a directed Erd\H{o}s--R\'enyi graph \citep{Erdos1959} and a directed soft random geometric graph \citep{Penrose2003}.
The Erd\H{o}s--R\'enyi graph is generated independently of covariates, whereas the soft random geometric graph induces covariate-dependent link formation.
For the Erd\H{o}s--R\'enyi graph, each ordered pair $(i,j)$ forms an edge independently with probability $p^{\text{ER}}=10/n$.
For the soft random geometric graph, let $d_{ij}$ be the pairwise Euclidean distance between $\bX_i^{\text{true}}$ and $\bX_j^{\text{true}}$, normalized by $\max_{i,j} d_{ij}$, and sample edges independently with $p^{\text{SRGG}}_{ij}=\exp\!\left(-\frac{d_{ij}}{\sigma}\right)$, where $\sigma>0$ controls the decay rate.
For $\beta=1$, we fix the connectivity parameter at $0.02$ for all $n$ to study the regime where neighborhoods grow with $n$.
For $\beta=2$, we tune $\sigma$ to keep the average number of neighbors approximately stable across $n$, using $\{0.02,0.018,0.016,0.016,0.014,0.014\}$ for $n\in\{5000,6000,7000,8000,9000,10000\}$.

\paragraph{Outcome.}
We construct the potential outcomes model for degree $\beta = 1, 2$ as:
\begin{equation}\label{equ: outcome}
Y_i(\mathbf{z}) = \alpha_{i,\varnothing} + \sum_{j \in \mathcal{N}_i} \alpha^{\text{linear}}_{ij} z_j  +\I(\beta=2)Q_i(\mathbf z)+ \bs \theta^\top \bX^\text{true}_i,
\end{equation}
where
\begin{align*}
    Q_i(\mathbf z)&= \left( \frac{\sum_{j \in \mathcal{N}_i} \alpha^{\text{quad}}_{ij} z_j}{\sum_{j \in \mathcal{N}_i} \alpha^{\text{quad}}_{ij}} \right)^2 - \frac{\sum_{j \in \mathcal{N}_i} \left(\alpha^{\text{quad}}_{ij} z_j\right)^2}{\left(\sum_{j \in \mathcal{N}_i} \alpha^{\text{quad}}_{ij}\right)^2} 
\end{align*}
captures the second-order interactions on the outcome. The coefficients $\alpha_{ij}^\text{linear}$ are determined from the following process. First, we generate $\alpha_{i,\varnothing}$ from $ \mathcal{U}[0,1]$. Next, based on the adjacency matrix $\bm A$ of the graph, we compute $\tilde{\bs A}=\bm D_\text{in}(\bm A-\bm I)$, where $\bm D_\text{in}$ is the diagonal matrix with each entry being the in-degree of each node. Further, we introduce a transformation matrix $\bm \Psi$, and we decide $\alpha_{ij}^\text{linear}$ from the entries of the matrix $\mathrm{Rescale}_1(\tilde{\bs A})+\bm A\odot\mathrm{Rescale}_2(\bX^\text{true}\bm \Psi)$, where $\mathrm{Rescale}_1$ and $\mathrm{Rescale}_2$ are operators that rescale diagonal and off-diagonal entries with different strengths governed by a hyperparameter $r$ and $\odot$ denotes elementwise multiplication; see details in Algorithm~\ref{alg:alpha-linear}. Finally, we generate $\alpha_{ij}^{\text{quad}} = \mathrm{Rescale}_1(\tilde{\bs A})_{ij}$ if $i \neq j$ and $\alpha_{ii}^{\text{quad}} = u_i(\bm 1^\top \bX^\text{true}_i+\texttt{diag})$, where $u_i \sim \mathcal{U}[0,1]$ and $\texttt{diag}$ is a constant offset; see details in Algorithm~\ref{alg:alpha-quad}.

We now present the simulation results for the four settings. We use relative bias and mean squared error (MSE) to evaluate each method. Specifically, we define the relative bias to be $(\E(\TTEhat)-\TTE)/\TTE$. In the simulations, the expectation $\E(\TTEhat)$ is approximated by the average estimate across repetitions. Moreover, the relative MSE is defined as the average squared error of each repetition normalized by the magnitude of the true $\TTE$.

\subsection{Erd\H{o}s--R\'enyi graph with first-order interactions}
\label{sec:4.1}
In this setting, we generate directed Erd\H{o}s--R\'enyi graphs and outcomes from (\ref{equ: outcome}) with $\beta=1$. Here, the graphs are generated independently of the covariate information. The following Figure \ref{fig:setting1} summarizes the results of this setting.

\FloatBarrier
\begin{figure}[htbp]
  \centering
  \begin{subfigure}[b]{0.24\textwidth}
    \includegraphics[width=\textwidth]{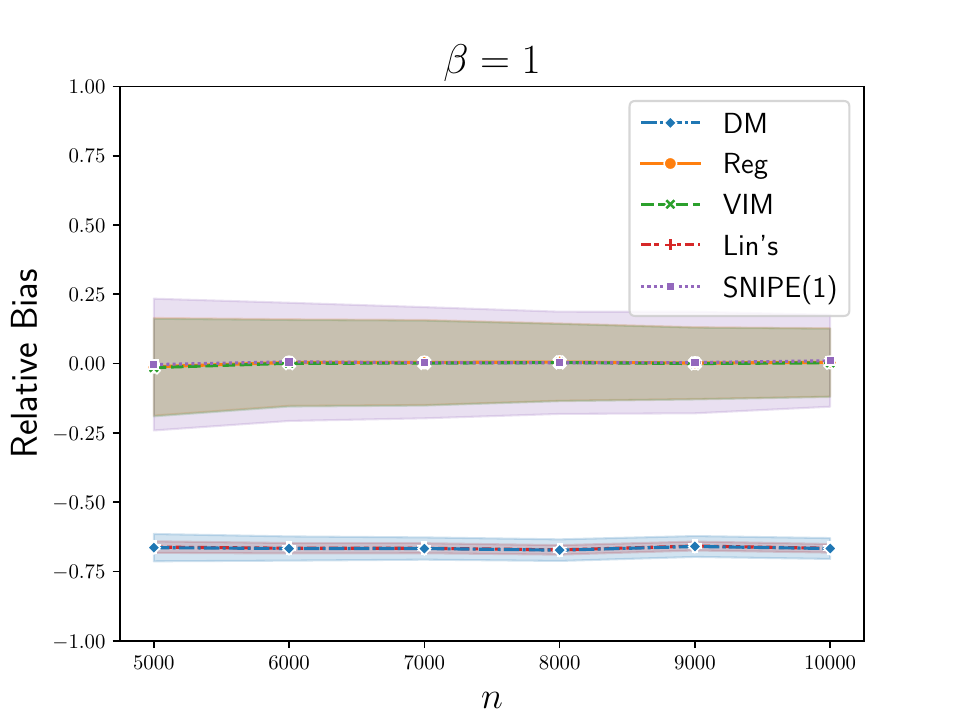}
  \end{subfigure}
  \hfill
  \begin{subfigure}[b]{0.24\textwidth}
    \includegraphics[width=\textwidth]{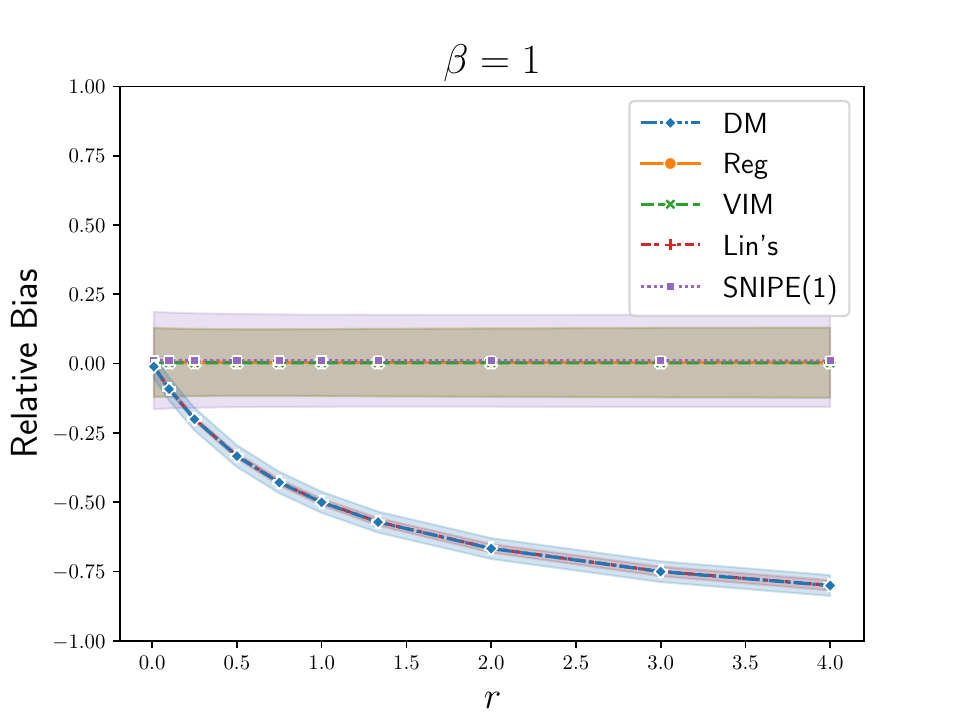}
  \end{subfigure}
  \hfill
  \begin{subfigure}[b]{0.24\textwidth}
    \includegraphics[width=\textwidth]{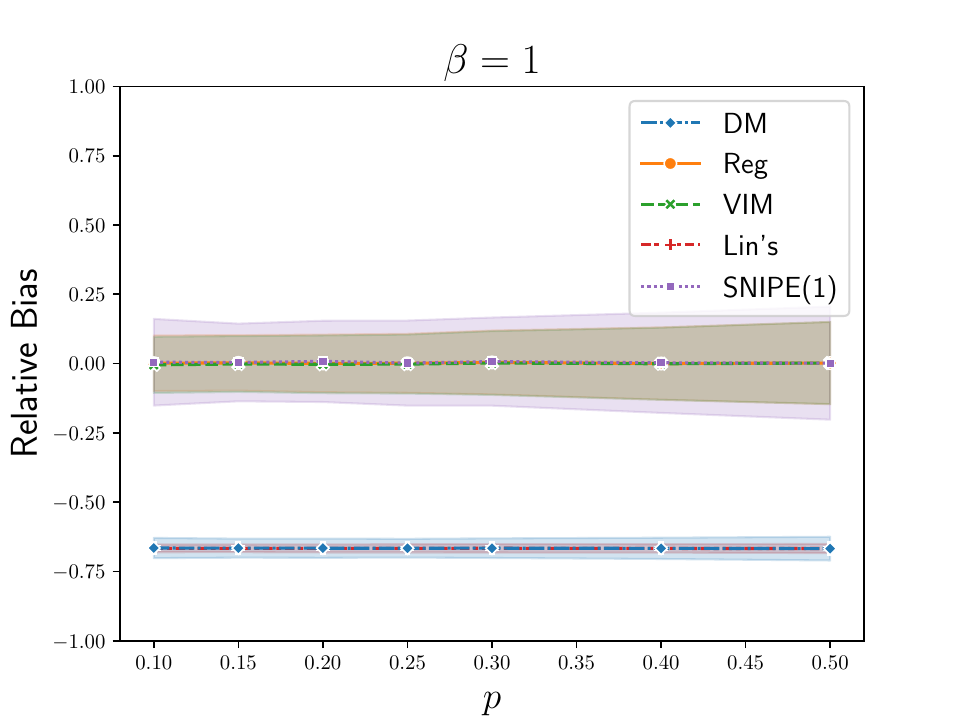}
  \end{subfigure}
    \hfill
  \begin{subfigure}[b]{0.24\textwidth}
    \includegraphics[width=\textwidth]{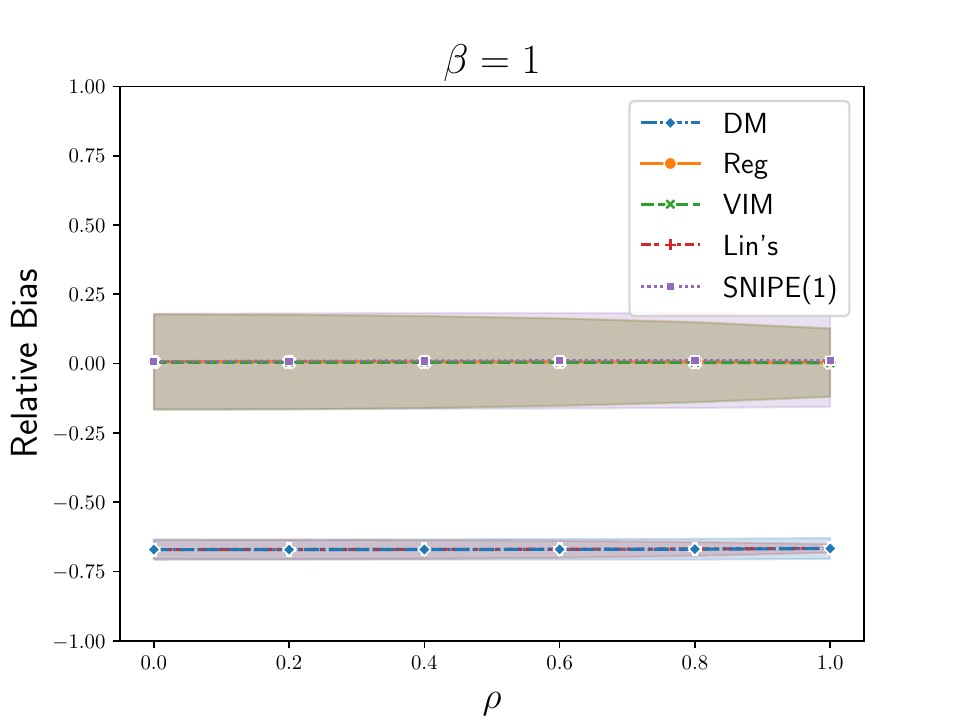}
  \end{subfigure}
    \vspace{0.2cm}

   \begin{subfigure}[b]{0.24\textwidth}
    \includegraphics[width=\textwidth]{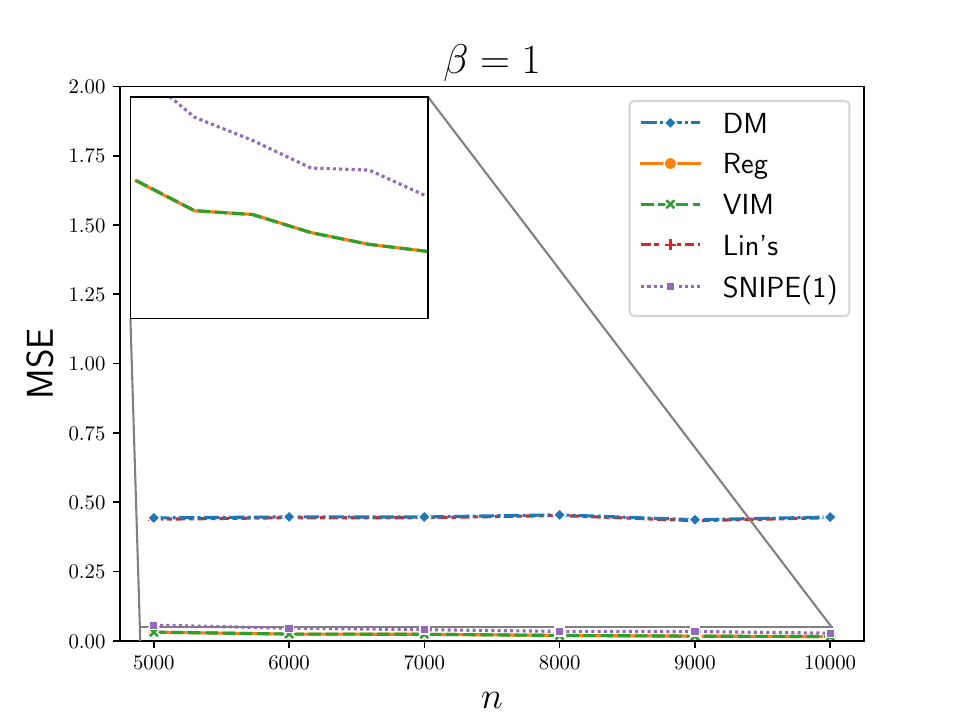}
  \end{subfigure}
  \hfill
  \begin{subfigure}[b]{0.24\textwidth}
    \includegraphics[width=\textwidth]{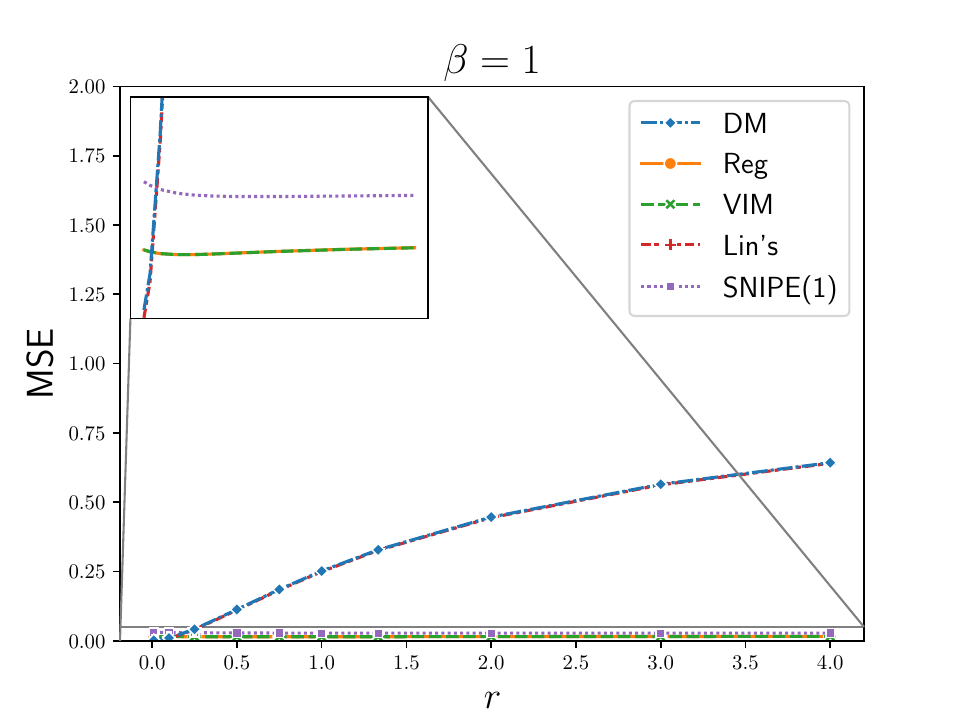}
  \end{subfigure}
  \hfill
  \begin{subfigure}[b]{0.24\textwidth}
    \includegraphics[width=\textwidth]{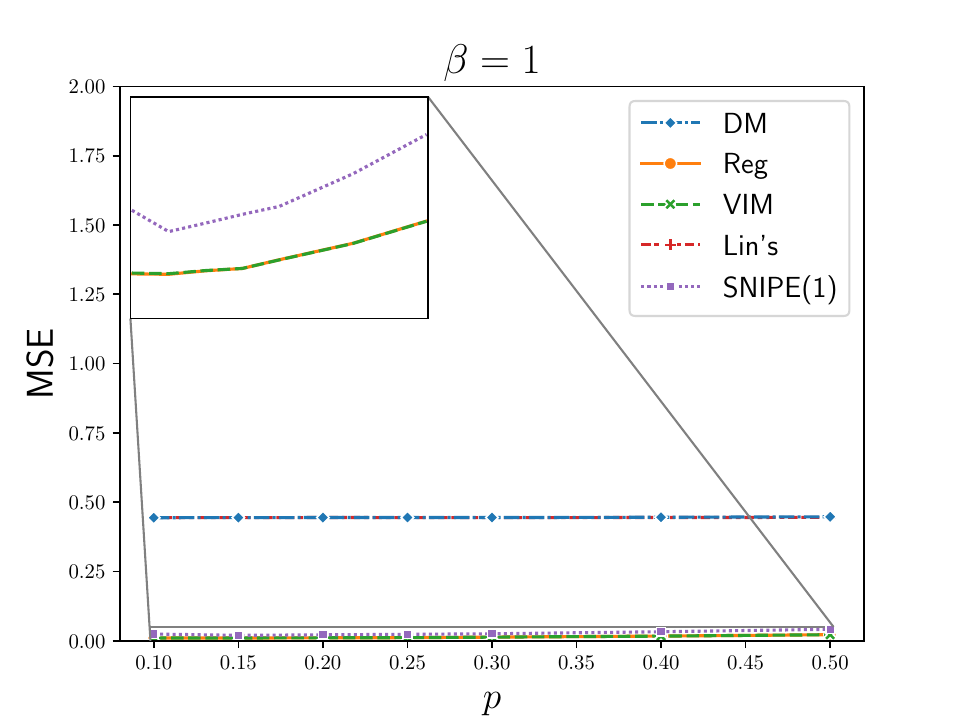}
  \end{subfigure}
  \hfill
  \begin{subfigure}[b]{0.24\textwidth}
    \includegraphics[width=\textwidth]{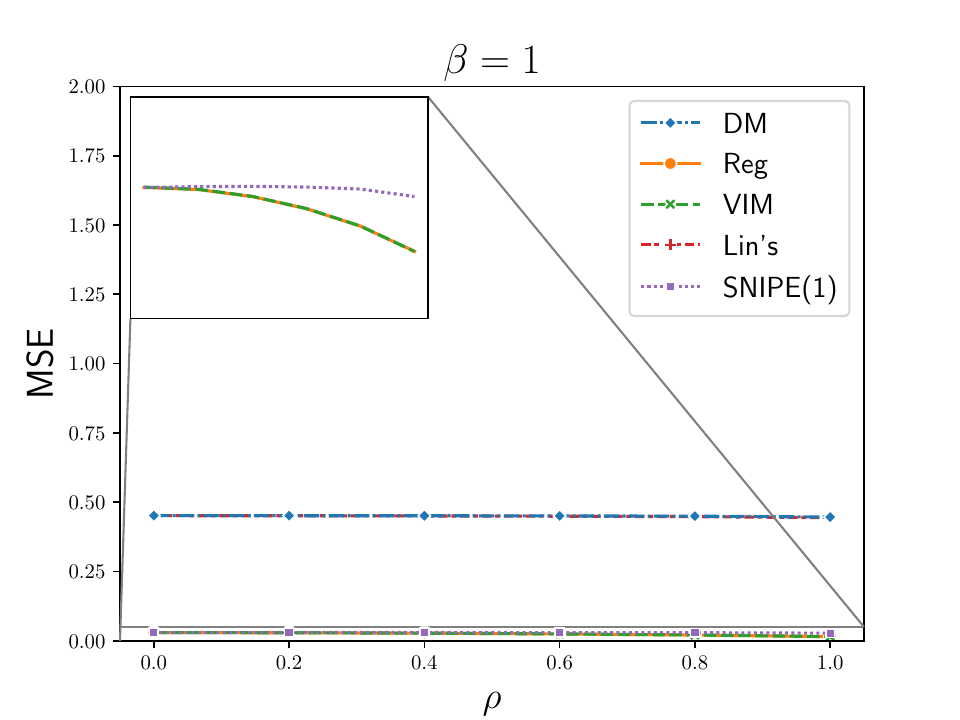}
    \end{subfigure}
  \caption{Relative bias (top row) and mean squared error (MSE; bottom row) of DM, $\Reg$ (Reg), $\VIM$ (VIM), \cite{Lin2013}'s estimator, and $\SNIPE$ under Erd\H{o}s--R\'enyi with $\beta = 1$ (SNIPE(1)). 
  }
  \label{fig:setting1}
\end{figure}
\FloatBarrier

Figure \ref{fig:setting1} shows that $\SNIPE$, $\Reg$, and $\VIM$ are unbiased across all settings. However, DM and \cite{Lin2013}'s estimator are biased under all settings. The bias tends to increase as indirect effects become stronger as expected. $\Reg$ and $\VIM$ outperform $\SNIPE$ in terms of relative MSE, particularly when there is a greater proportion of observed covariates. The relative MSEs of DM and \cite{Lin2013}'s estimator are dominated by bias and are much larger than that of $\SNIPE$, $\Reg$, and $\VIM$. 

\subsection{Erd\H{o}s--R\'enyi graph with second-order interactions}\label{sec:4.2}
This setting generates outcomes from (\ref{equ: outcome}) with $\beta=2$ while keeping the Erd\H{o}s--R\'enyi setting for generation of directed graphs.

\FloatBarrier
\begin{figure}[htbp]
  \centering
  \begin{subfigure}[b]{0.24\textwidth}
    \includegraphics[width=\textwidth]{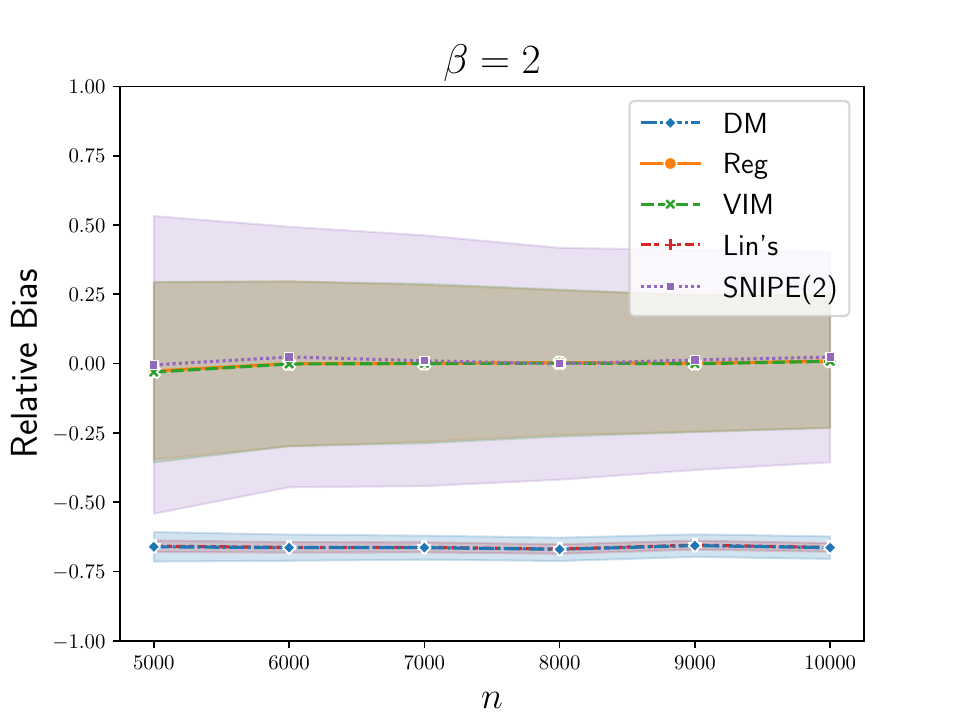}
  \end{subfigure}
  \hfill
  \begin{subfigure}[b]{0.24\textwidth}
    \includegraphics[width=\textwidth]{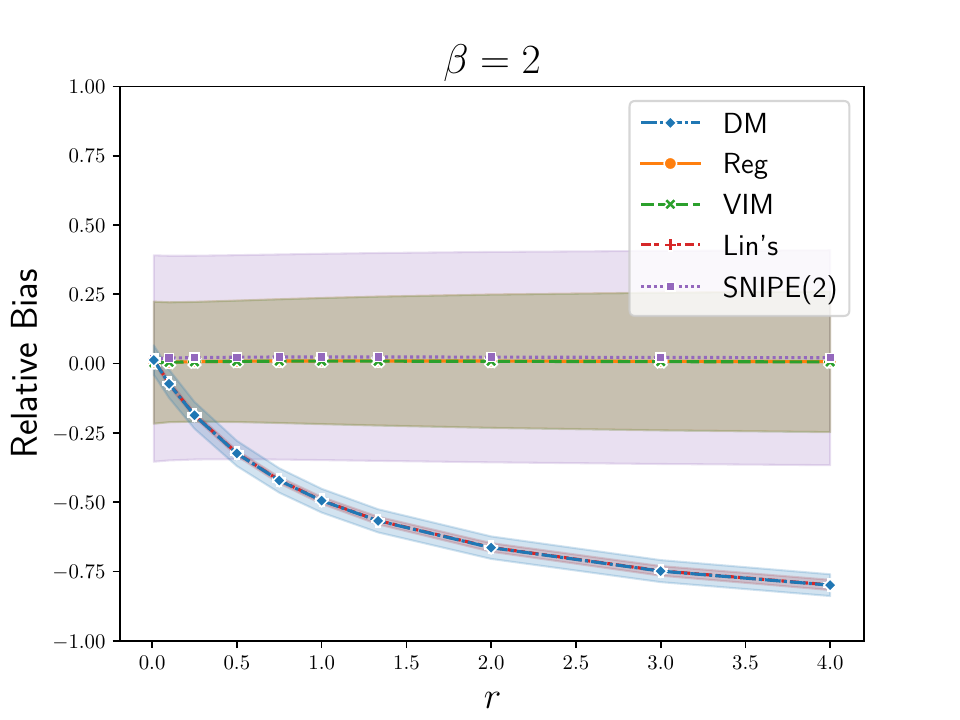}
  \end{subfigure}
  \hfill
  \begin{subfigure}[b]{0.24\textwidth}
    \includegraphics[width=\textwidth]{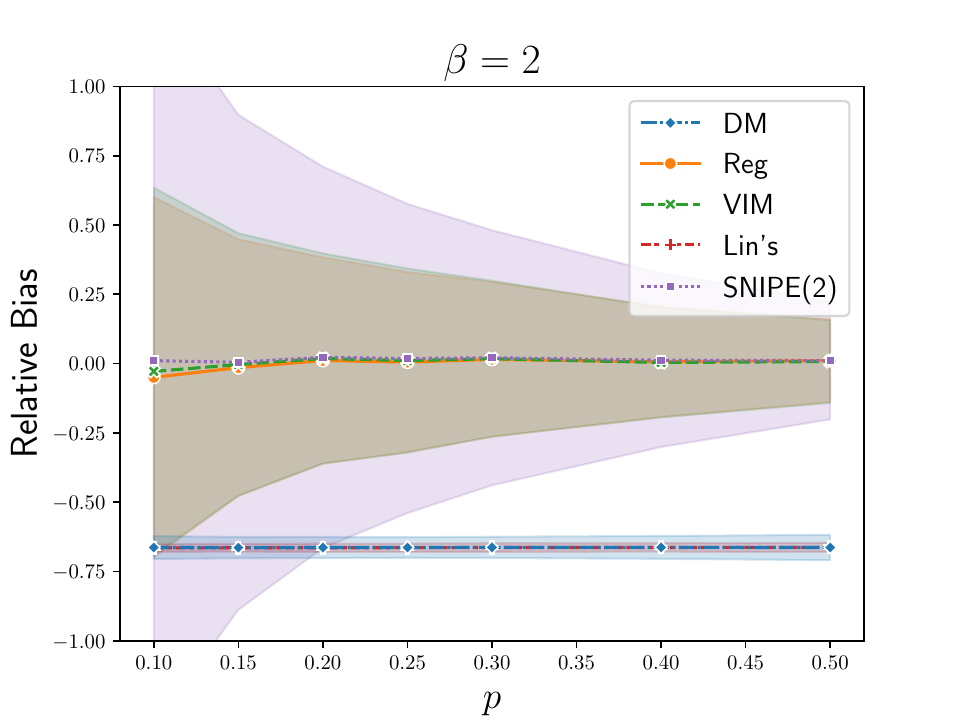}
  \end{subfigure}
    \hfill
  \begin{subfigure}[b]{0.24\textwidth}
    \includegraphics[width=\textwidth]{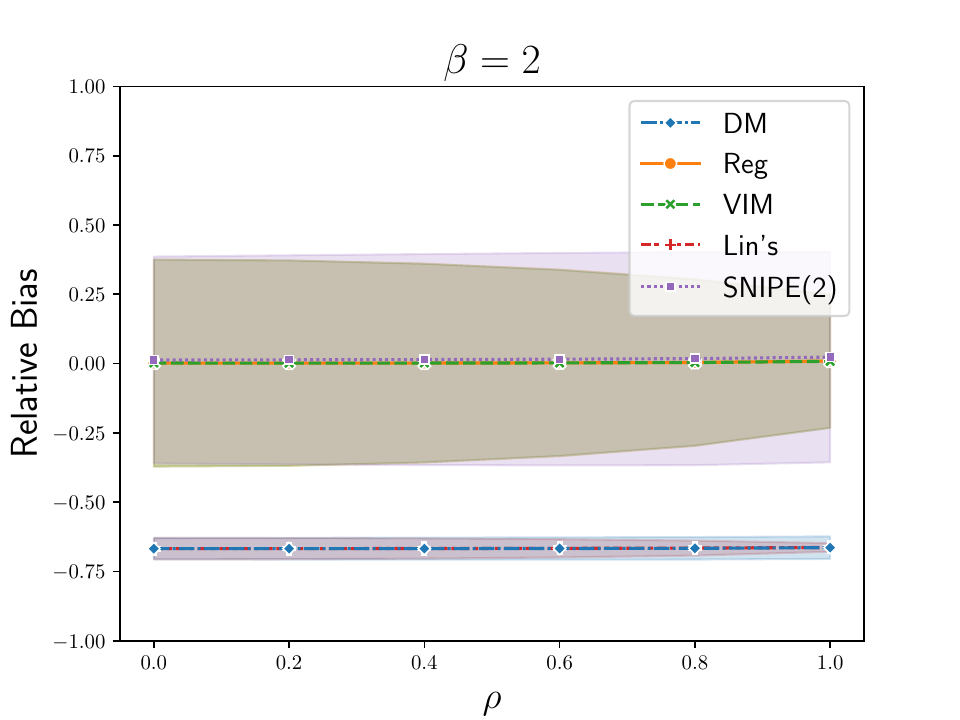}
  \end{subfigure}
    \vspace{0.2cm}

   \begin{subfigure}[b]{0.24\textwidth}
    \includegraphics[width=\textwidth]{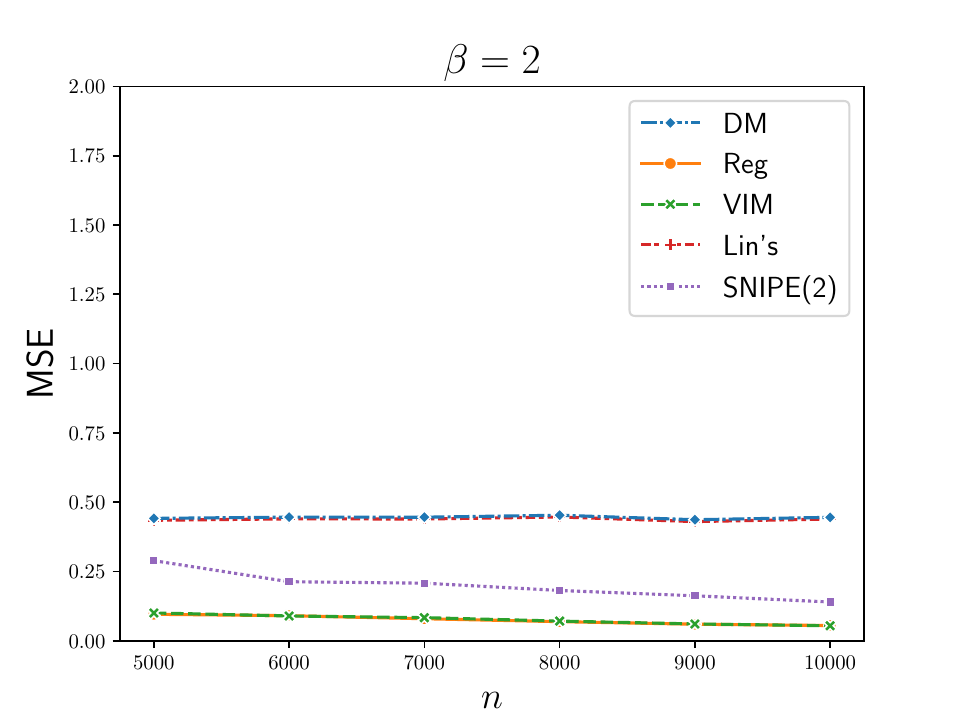}
  \end{subfigure}
  \hfill
  \begin{subfigure}[b]{0.24\textwidth}
    \includegraphics[width=\textwidth]{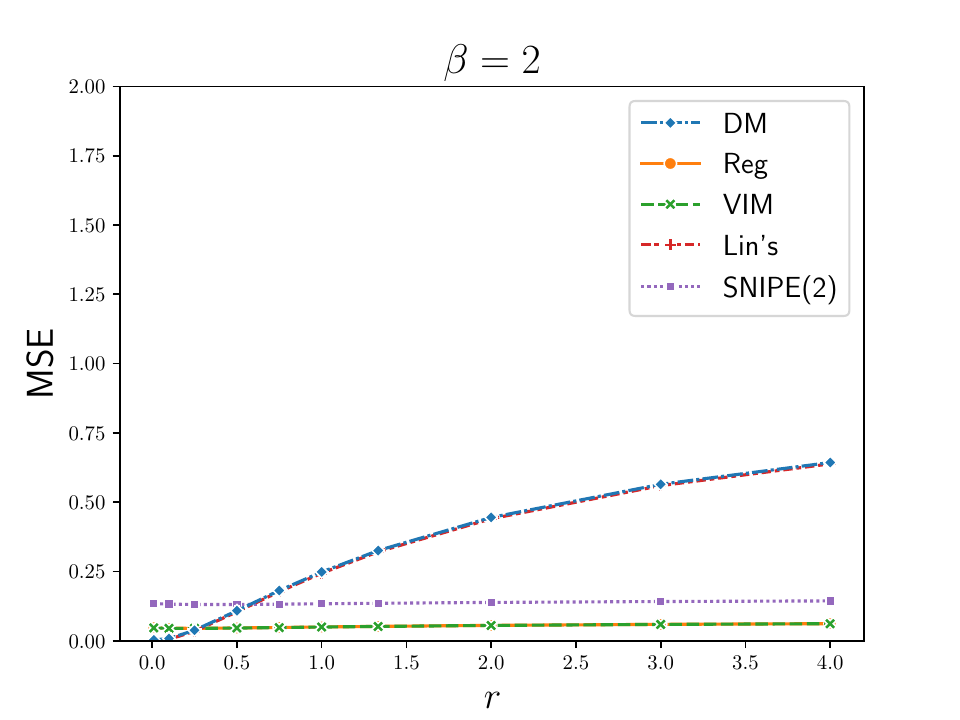}
  \end{subfigure}
  \hfill
  \begin{subfigure}[b]{0.24\textwidth}
    \includegraphics[width=\textwidth]{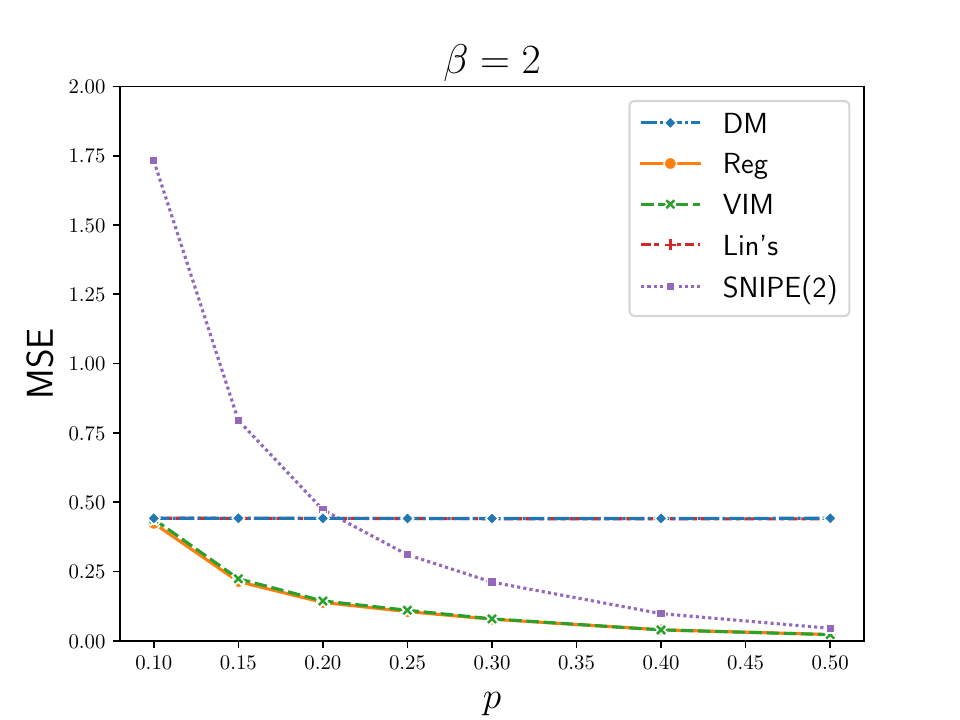}
  \end{subfigure}
  \hfill
  \begin{subfigure}[b]{0.24\textwidth}
    \includegraphics[width=\textwidth]{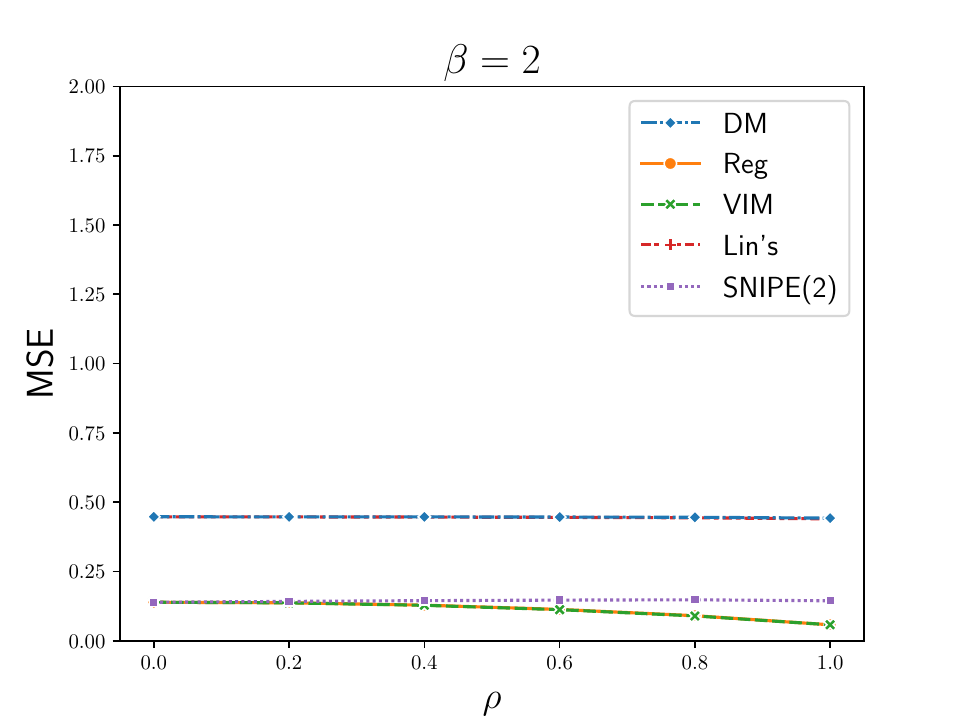}
    \end{subfigure}
    
  \caption{Relative bias (top row) and mean squared error (MSE; bottom row) of DM, $\Reg$ (Reg), $\VIM$ (VIM), \cite{Lin2013}'s estimator, and $\SNIPE$ under Erd\H{o}s--R\'enyi with $\beta = 2$ (SNIPE(2)). 
  }
  \label{fig:setting2}
\end{figure}
\FloatBarrier
As shown in Figure \ref{fig:setting2}, the overall patterns of relative bias and relative MSE closely resemble those observed in the previous setting (Section \ref{sec:4.1}). However, the performance gap between estimators is more explicit: $\Reg$ and $\VIM$ exhibit clear improvements over $\SNIPE$ in terms of relative MSE. Notably, when the treatment probability is relatively low, the MSE of $\SNIPE$ can even exceed that of the two asymptotically biased estimators -- DM and \cite{Lin2013}'s estimator.
\subsection{Soft RGG with first-order interactions}
In Setting 3, we adopt a soft RGG to generate the underlying network structure and use (\ref{equ: outcome}) with $\beta=1$. As described previously, the network structure is correlated with the covariate information. Specifically, units who are more alike in terms of covariates, such as having similar ages, shared interests, or common daily routines, tend to have a higher chance of being connected.   
\FloatBarrier
\begin{figure}[htbp]
  \centering
  \begin{subfigure}[b]{0.24\textwidth}
    \includegraphics[width=\textwidth]{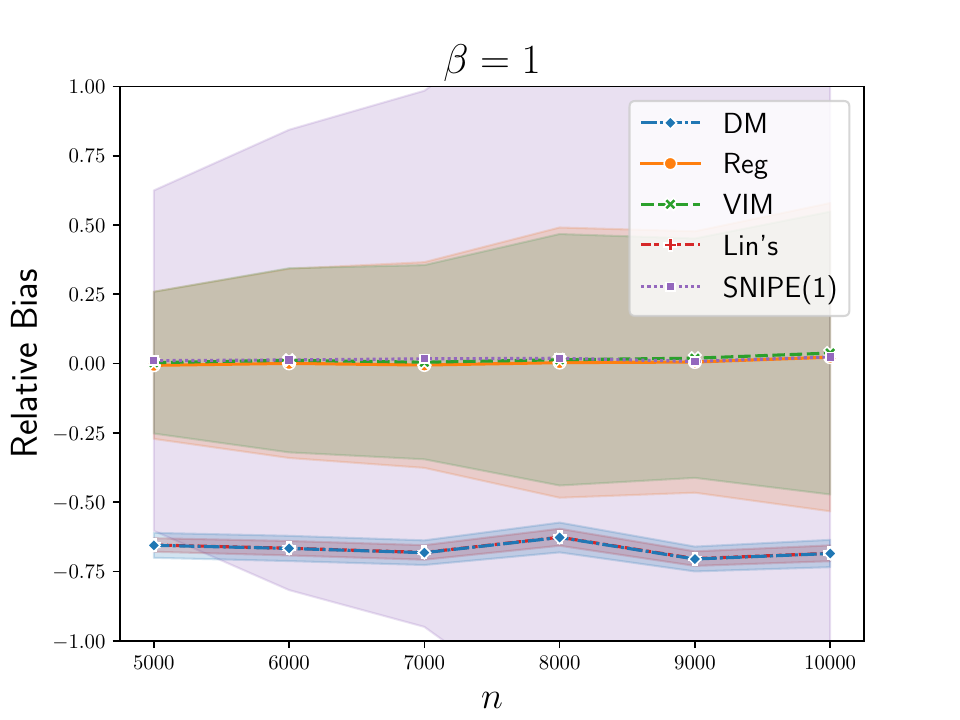}
  \end{subfigure}
    \hfill
  \begin{subfigure}[b]{0.24\textwidth}
    \includegraphics[width=\textwidth]{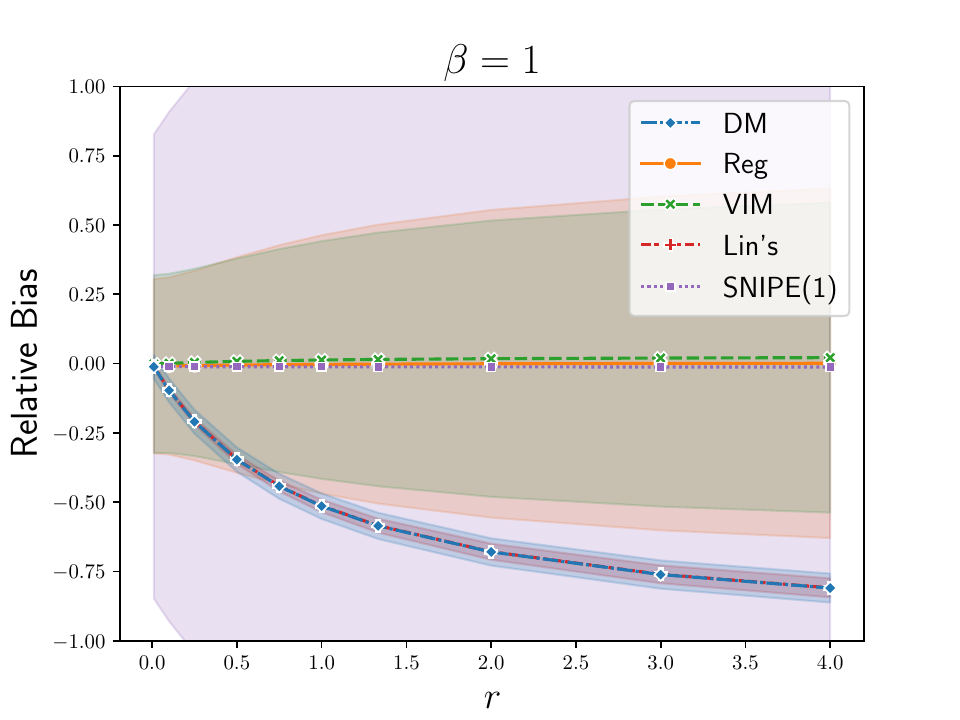}
  \end{subfigure}
  \hfill
  \begin{subfigure}[b]{0.24\textwidth}
    \includegraphics[width=\textwidth]{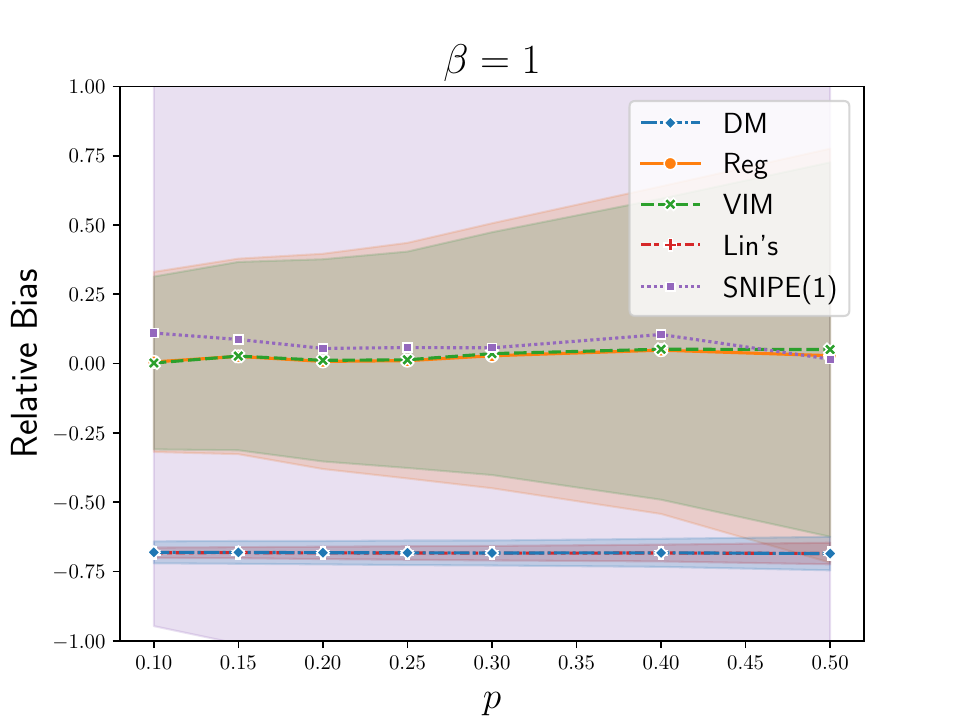}
  \end{subfigure}
  \hfill
  \begin{subfigure}[b]{0.24\textwidth}
    \includegraphics[width=\textwidth]{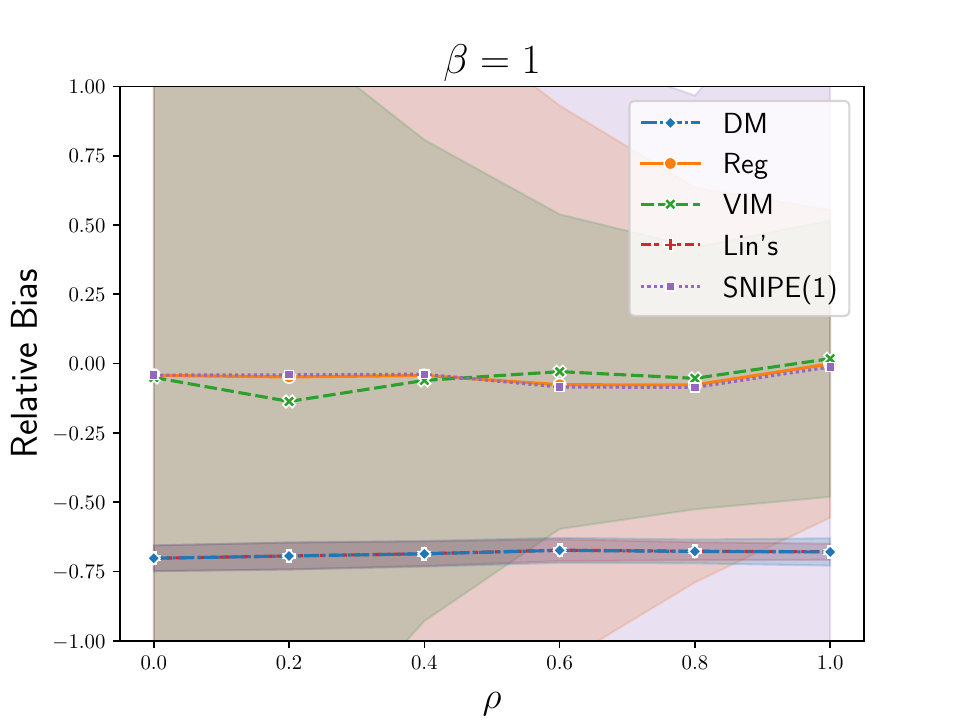}
  \end{subfigure}
    \vspace{0.2cm}

   \begin{subfigure}[b]{0.24\textwidth}
    \includegraphics[width=\textwidth]{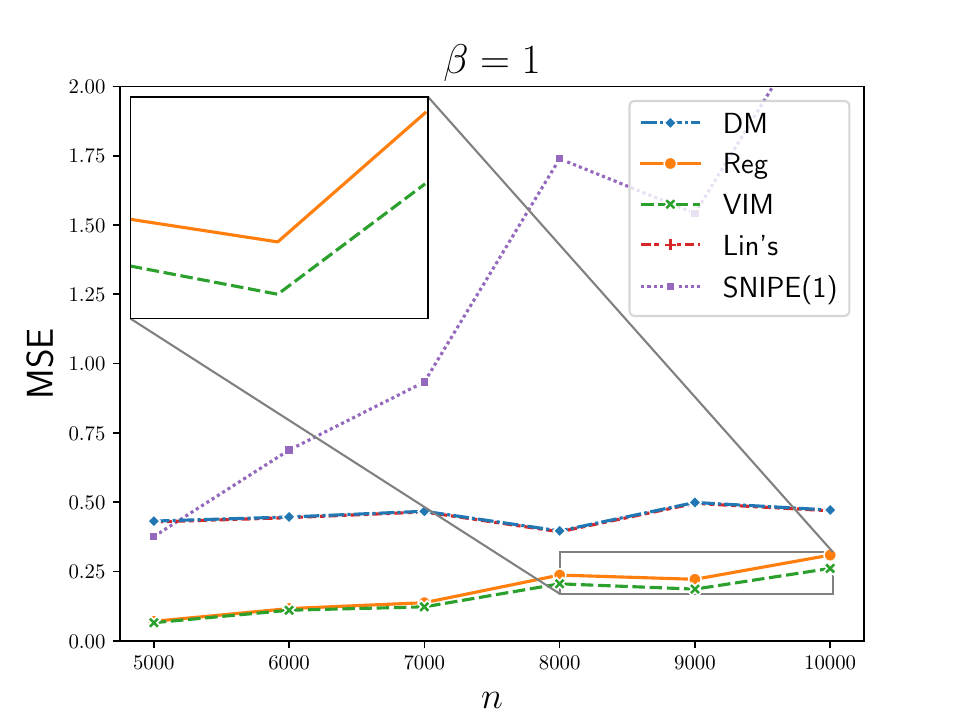}
  \end{subfigure}
  \hfill
  \begin{subfigure}[b]{0.24\textwidth}
    \includegraphics[width=\textwidth]{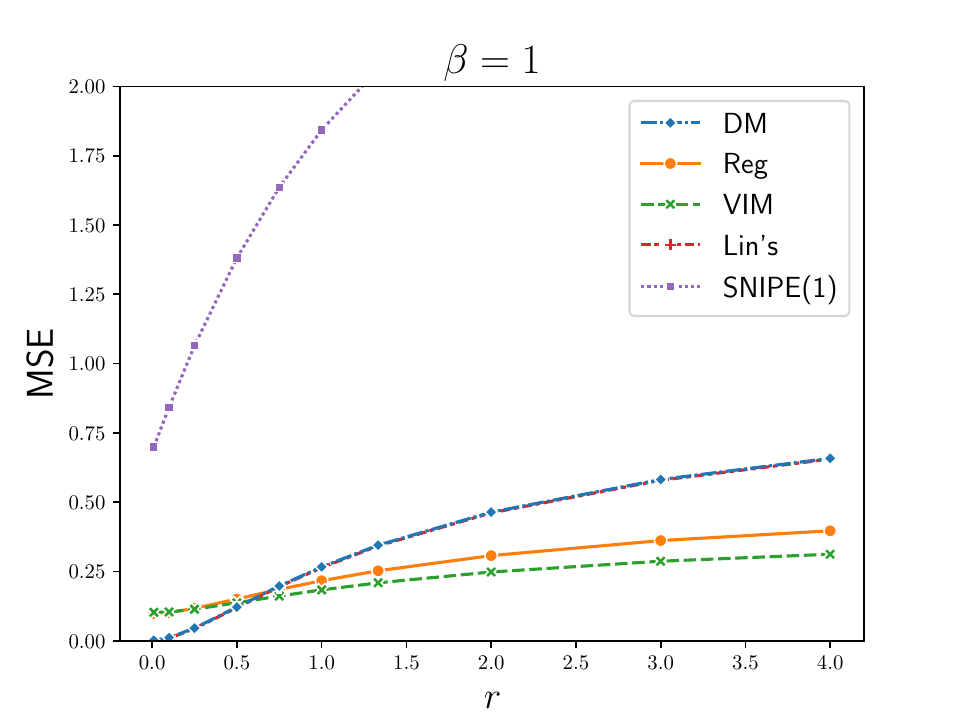}
  \end{subfigure}
  \hfill
  \begin{subfigure}[b]{0.24\textwidth}
    \includegraphics[width=\textwidth]{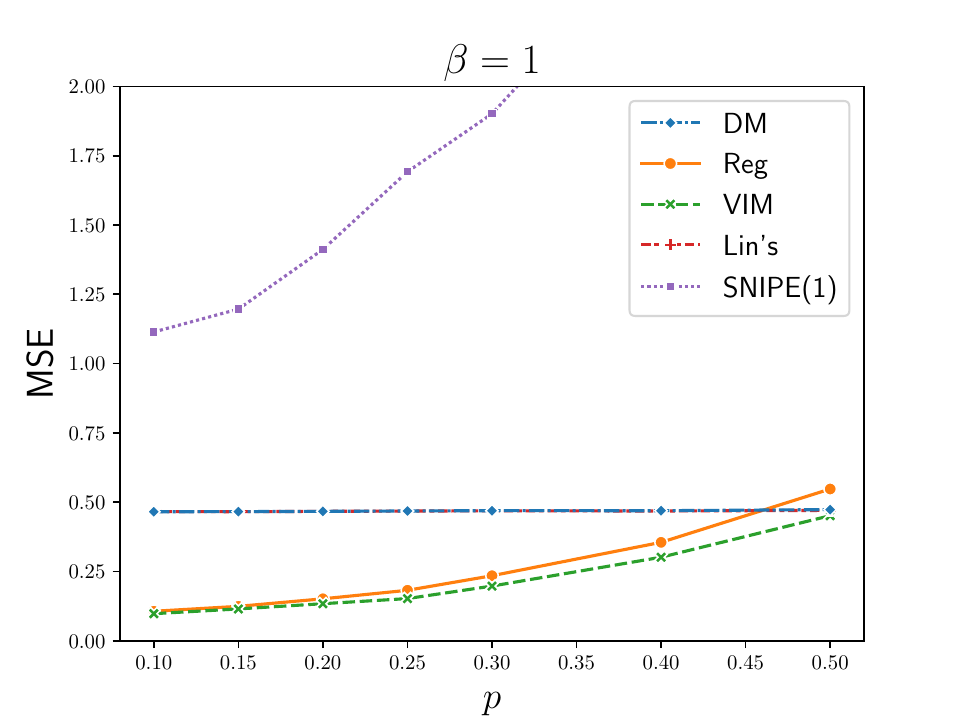}
  \end{subfigure}
  \hfill
  \begin{subfigure}[b]{0.24\textwidth}
    \includegraphics[width=\textwidth]{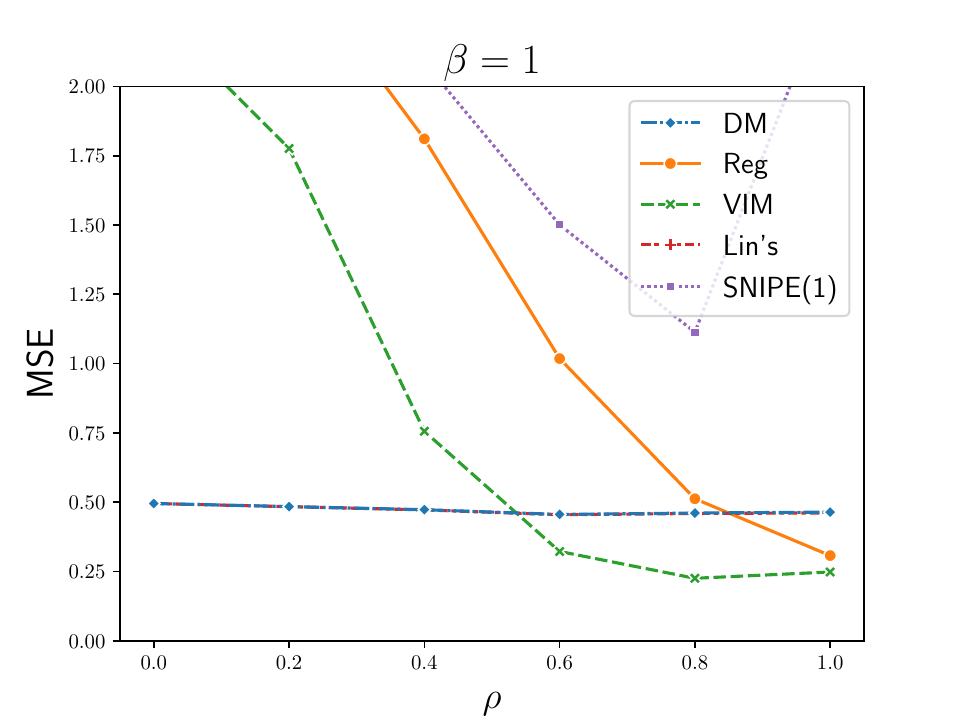}
  \end{subfigure}

  \caption{Relative bias (top row) and mean squared error (MSE; bottom row) of DM, $\Reg$ (Reg), $\VIM$ (VIM), \cite{Lin2013}'s estimator, and $\SNIPE$ under soft RGG with $\beta = 1$ (SNIPE(1)). 
  }
  \label{fig:setting3}
\end{figure}
\FloatBarrier
The relative bias patterns in Setting 3 are similar to those observed in the previous settings. Both DM and \cite{Lin2013}'s estimator remain biased, with their MSEs largely driven by this bias. However, the relative MSEs of the other estimators show more substantial differences. As shown in Figure \ref{fig:setting3}, $\VIM$ achieves a lower MSE than $\Reg$, which is consistent with their large sample properties. As shown in the plot across different network sizes, when the decay parameter $\sigma$ is held constant, increasing the network size leads to a higher average number of neighbors. In this regime, $\VIM$ increasingly outperforms $\Reg$ in terms of MSE. Moreover, $\SNIPE$ performs worse than all other estimators, even the DM estimator.

\subsection{Soft RGG with second-order interactions}
Finally, Setting 4 combines soft RGG and second-order interactions. 
\FloatBarrier
\begin{figure}[htbp]
  \centering
  \begin{subfigure}[b]{0.24\textwidth}
    \includegraphics[width=\textwidth]{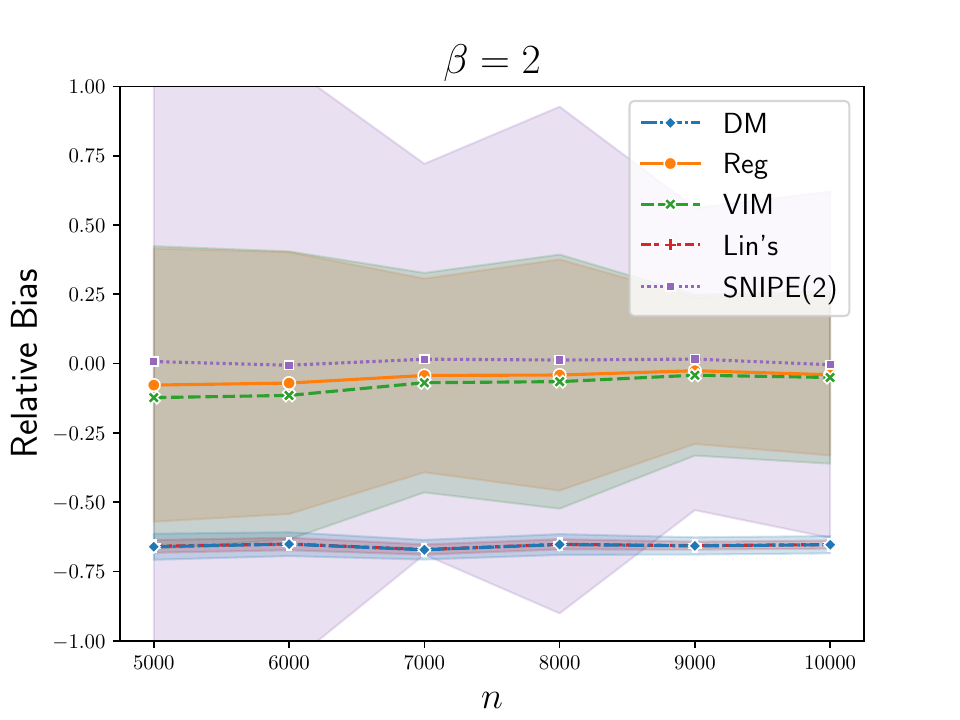}
  \end{subfigure}
  \hfill
  \begin{subfigure}[b]{0.24\textwidth}
    \includegraphics[width=\textwidth]{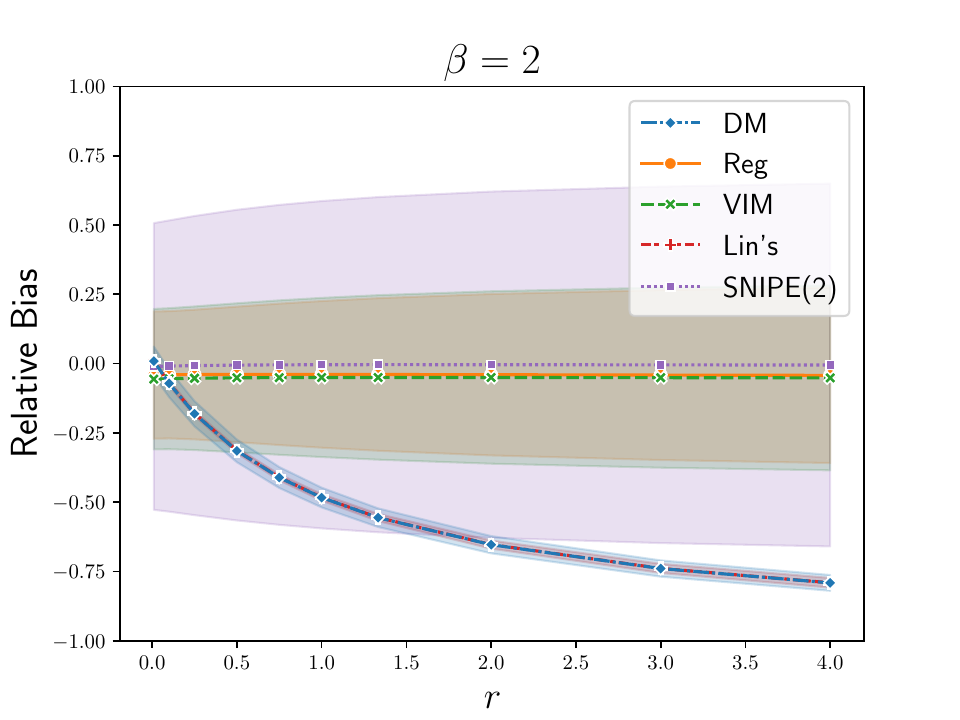}
  \end{subfigure}
  \hfill
  \begin{subfigure}[b]{0.24\textwidth}
    \includegraphics[width=\textwidth]{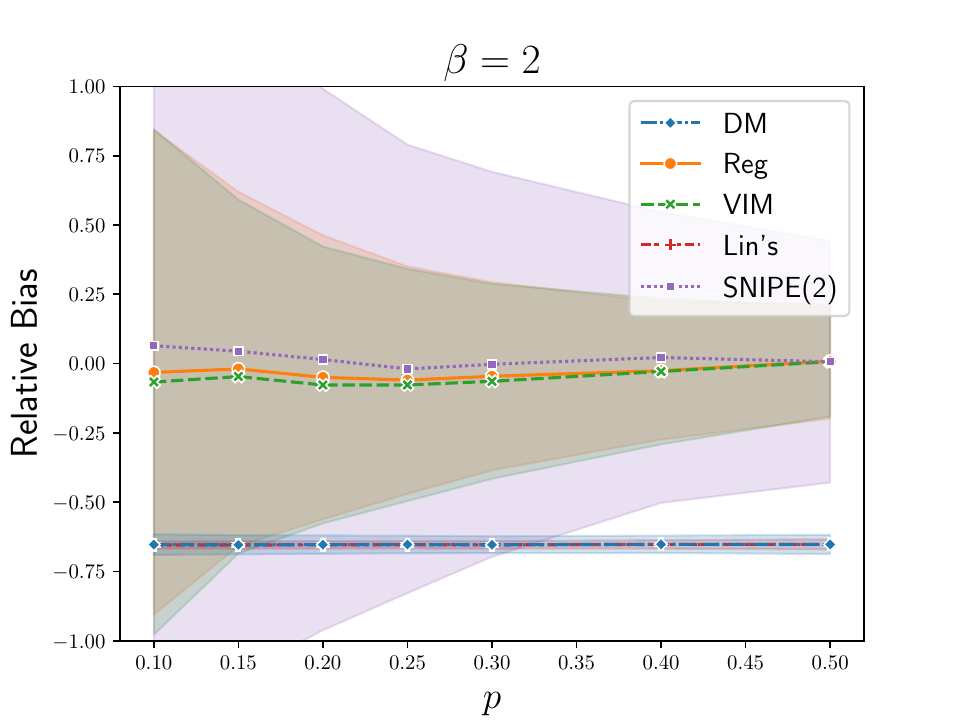}
  \end{subfigure}
  \hfill
  \begin{subfigure}[b]{0.24\textwidth}
    \includegraphics[width=\textwidth]{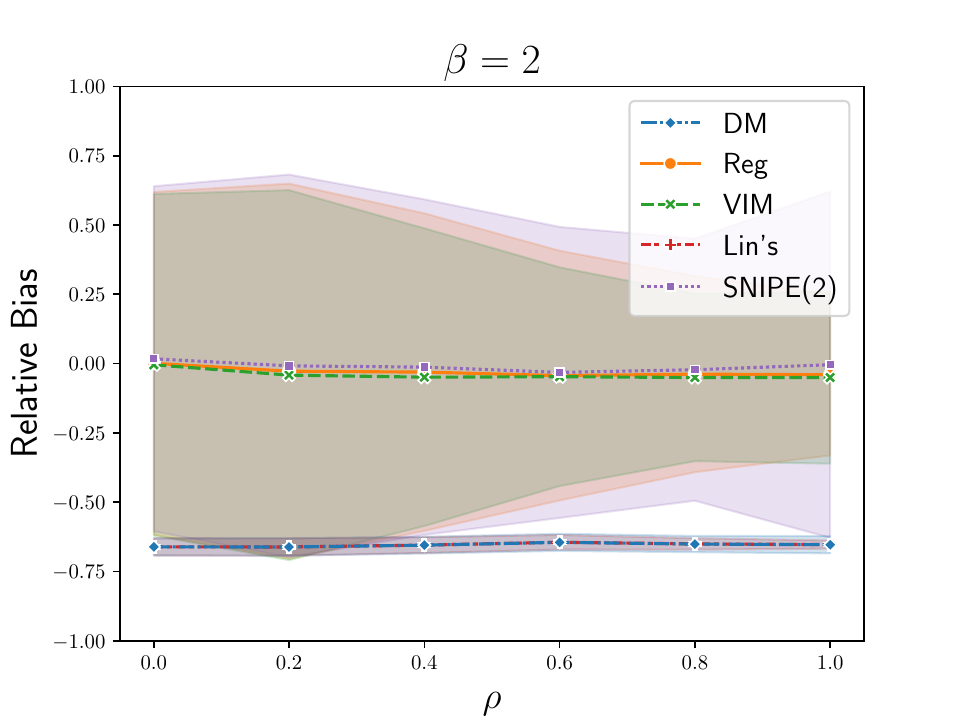}
  \end{subfigure}
  
    \vspace{0.2cm}

   \begin{subfigure}[b]{0.24\textwidth}
    \includegraphics[width=\textwidth]{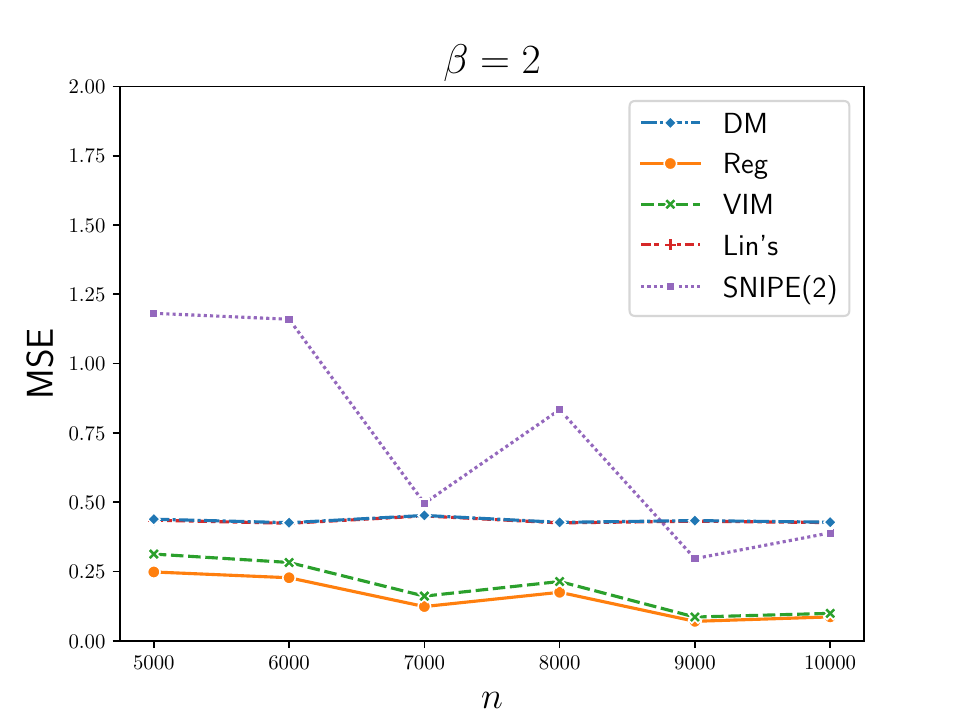}
  \end{subfigure}
  \hfill
  \begin{subfigure}[b]{0.24\textwidth}
    \includegraphics[width=\textwidth]{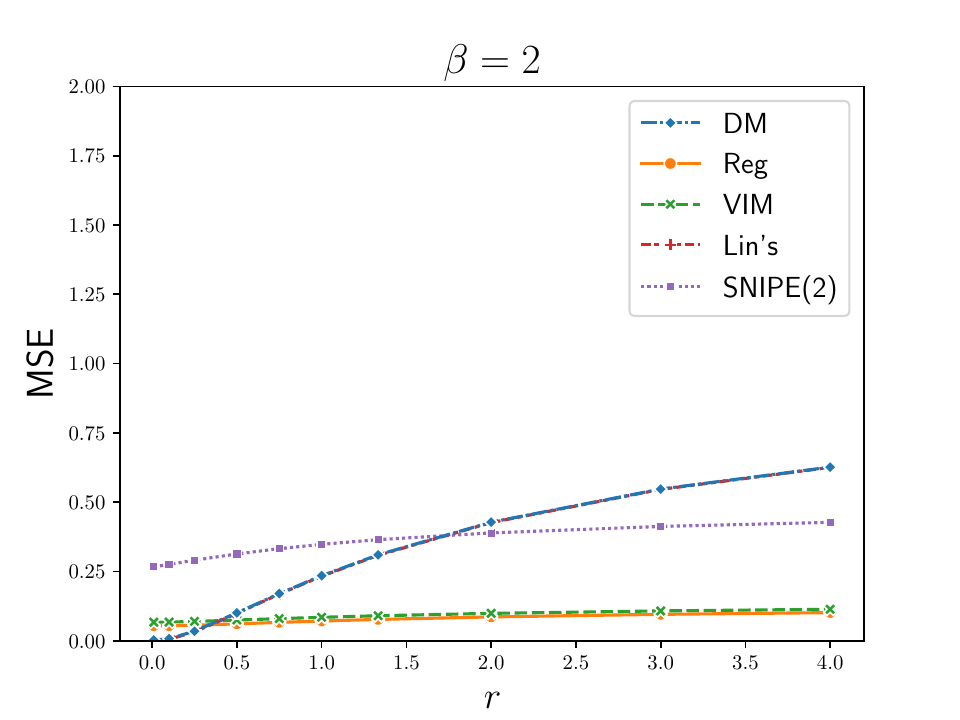}
  \end{subfigure}
  \hfill
  \begin{subfigure}[b]{0.24\textwidth}
    \includegraphics[width=\textwidth]{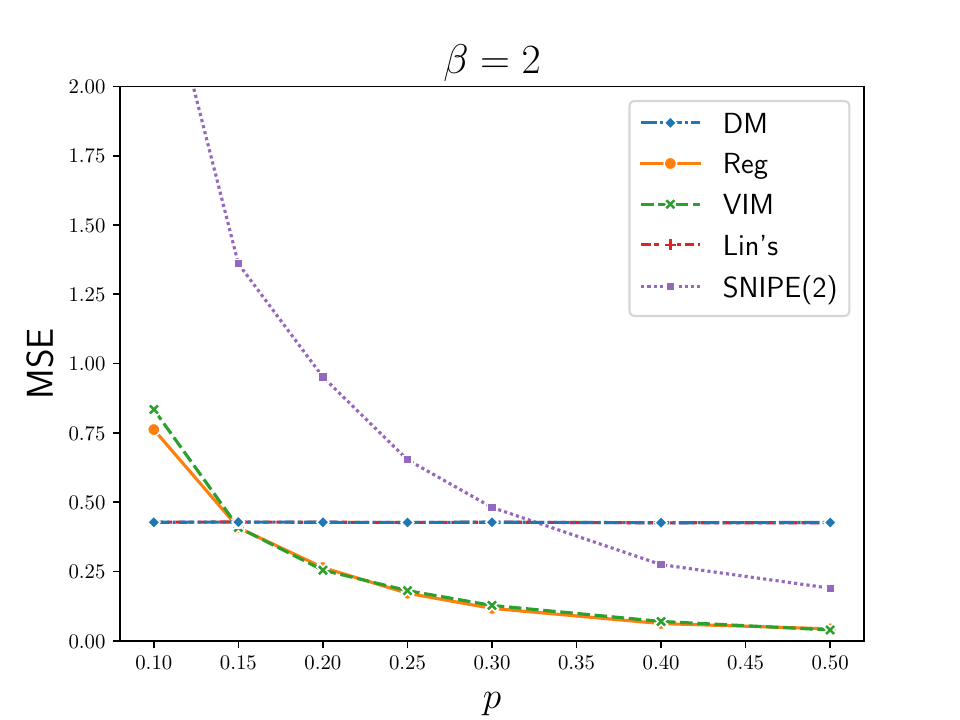}
  \end{subfigure}
  \hfill
  \begin{subfigure}[b]{0.24\textwidth}
    \includegraphics[width=\textwidth]{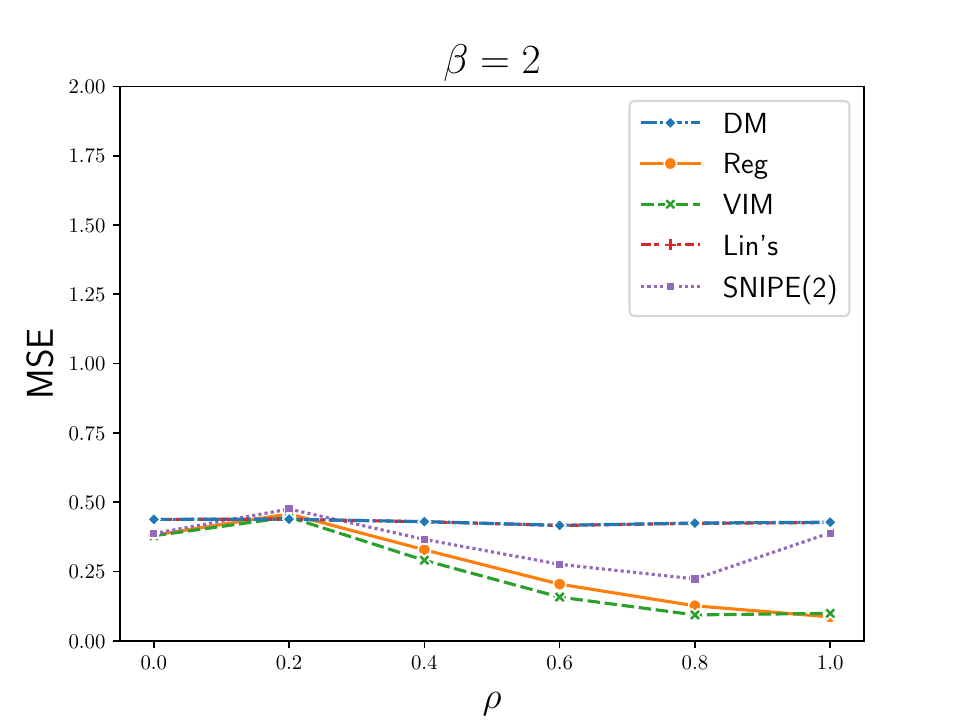}
  \end{subfigure}

  \caption{Relative bias (top row) and mean squared error (MSE; bottom row) of DM, $\Reg$ (Reg), $\VIM$ (VIM), \cite{Lin2013}'s estimator, and $\SNIPE$ under soft RGG with $\beta = 2$ (SNIPE(2)). 
  }
  \label{fig:setting4}
\end{figure}
\FloatBarrier
Figure \ref{fig:setting4} shows that the bias patterns remain similar to those in previous settings. Both DM and \cite{Lin2013}'s estimator are biased with their MSEs controlled by this bias. In this setting, we vary the decay parameter $\sigma$ to maintain a roughly constant average number of neighbors across different network sizes. As a result, the network size plot shows that $\VIM$ converges slightly more slowly than $\Reg$ as expected. The plot varying the proportion of observed covariates indicates that $\VIM$ is more robust to partial covariate observability. Overall, these two methods consistently yield the best performance. In contrast, $\SNIPE$ exhibits high variance in many configurations, resulting in MSEs that are worse than those of the biased estimators, DM and \cite{Lin2013}'s estimator.

\subsection{Comparison of variance estimators}
In this subsection, we compare the conservative variance estimator proposed in this paper with the Monte Carlo variance and the conservative variance estimator of \citet{CortezRodriguezEichhornYu+2023}. 
For each simulation setting, we construct Wald-type confidence intervals using each variance estimator. 
To facilitate comparison, we report the logarithm of the ratio of confidence interval lengths,
\[
\log\big(\mathrm{CI}_{\mathrm{new}} / \mathrm{CI}_{\mathrm{MC}}\big)
\quad \text{and} \quad
\log\big(\mathrm{CI}_{\mathrm{old}} / \mathrm{CI}_{\mathrm{MC}}\big),
\]
as well as the corresponding variance estimates, across simulation settings with $\beta = 1$. 
Here, “new” refers to the variance estimator proposed in this paper, and “old” refers to that of \citet{CortezRodriguezEichhornYu+2023}.

\begin{figure}[t]
  \centering

  \begin{subfigure}[t]{0.48\textwidth}
    \centering
    \includegraphics[width=\textwidth]{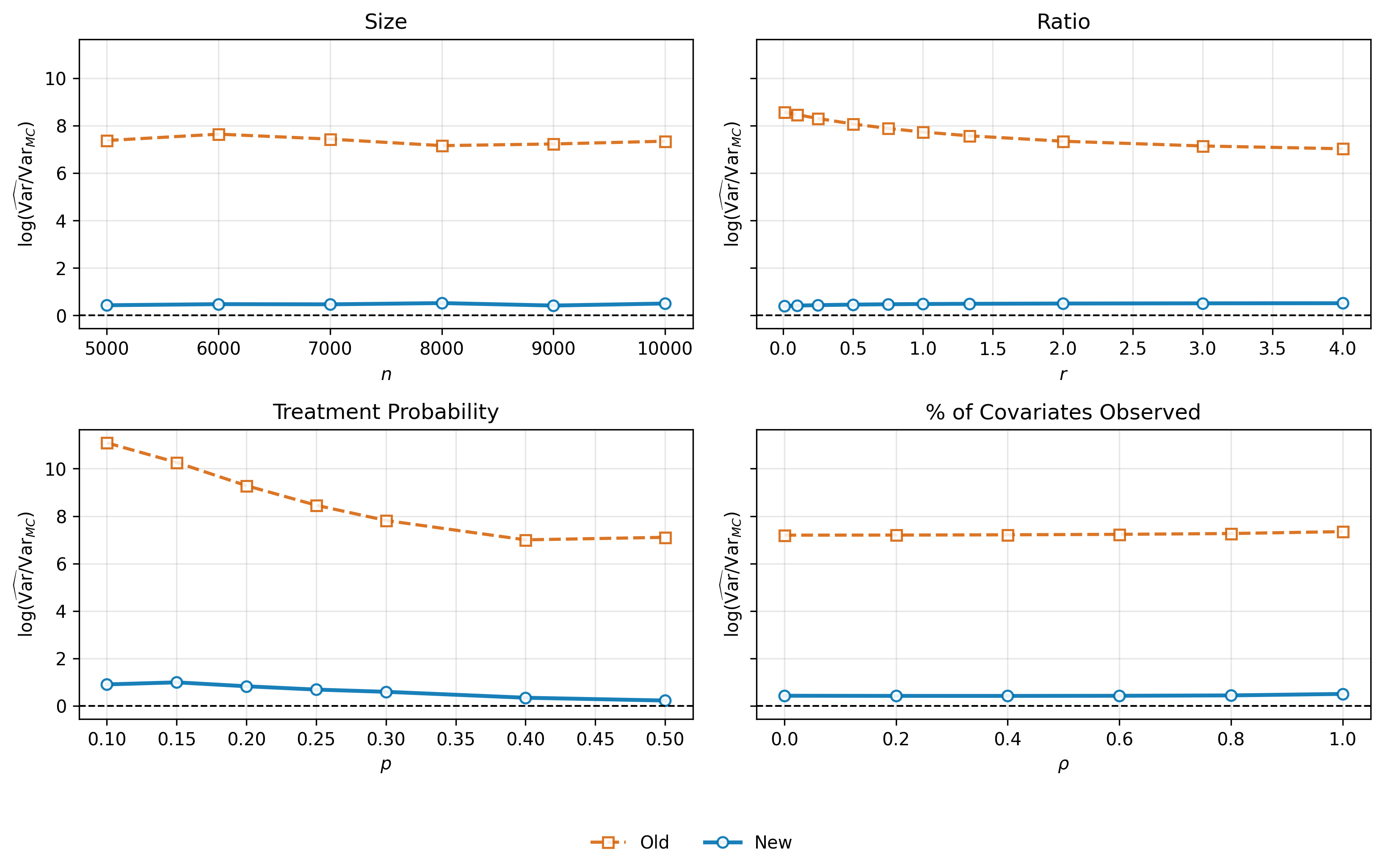}
    \caption{Erd\H{o}s--R\'enyi graph, $\beta = 1$.}
    \label{fig:ci_er_b1}
  \end{subfigure}
  \hfill
  \begin{subfigure}[t]{0.48\textwidth}
    \centering
    \includegraphics[width=\textwidth]{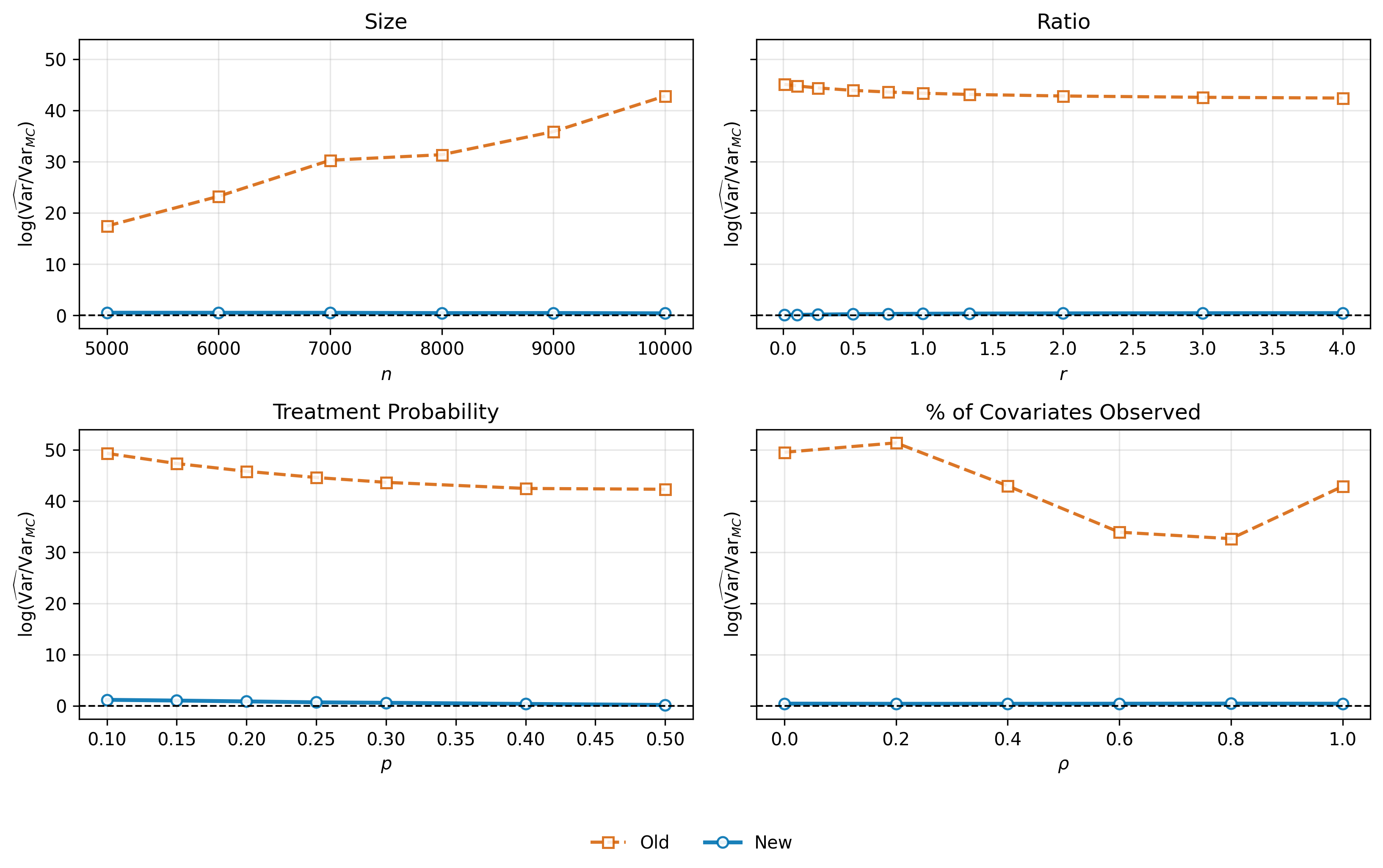}
    \caption{Soft RGG, $\beta = 1$.}
    \label{fig:ci_srgg_b1}
  \end{subfigure}

  \caption{Log ratio of confidence interval lengths,
    $\log(\mathrm{CI}_{\text{old}}/\mathrm{CI}_{\text{MC}})$ and $\log(\mathrm{CI}_{\text{new}}/\mathrm{CI}_{\text{MC}})$ for $\SNIPE$.}
  \label{fig:ci}
\end{figure}

\begin{figure}[t]
  \centering

  \begin{subfigure}[t]{0.48\textwidth}
    \centering
    \includegraphics[width=\textwidth]{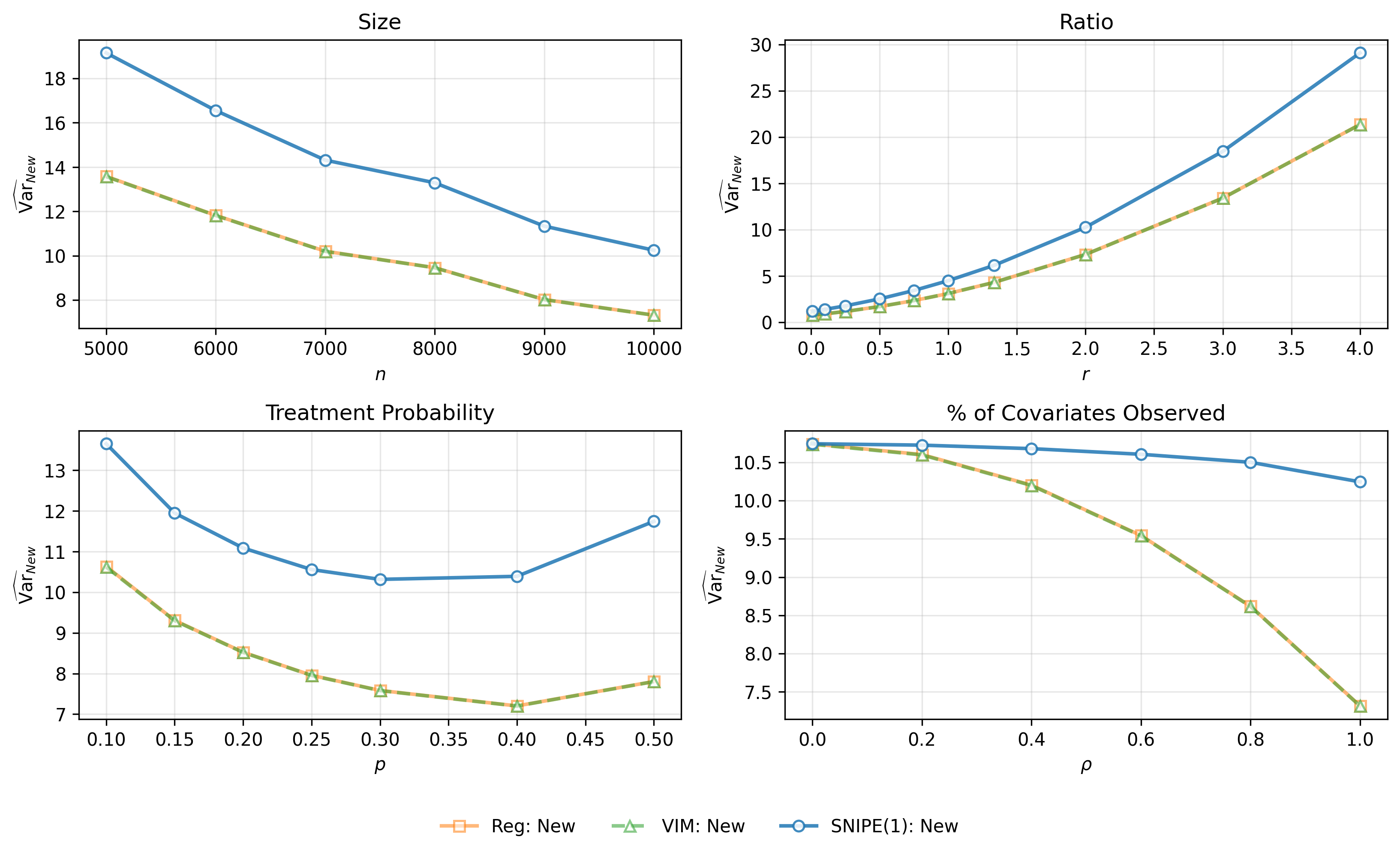}
    \caption{Erd\H{o}s--R\'enyi graph, $\beta = 1$.}
    \label{fig:var_er_b1}
  \end{subfigure}
  \hfill
  \begin{subfigure}[t]{0.48\textwidth}
    \centering
    \includegraphics[width=\textwidth]{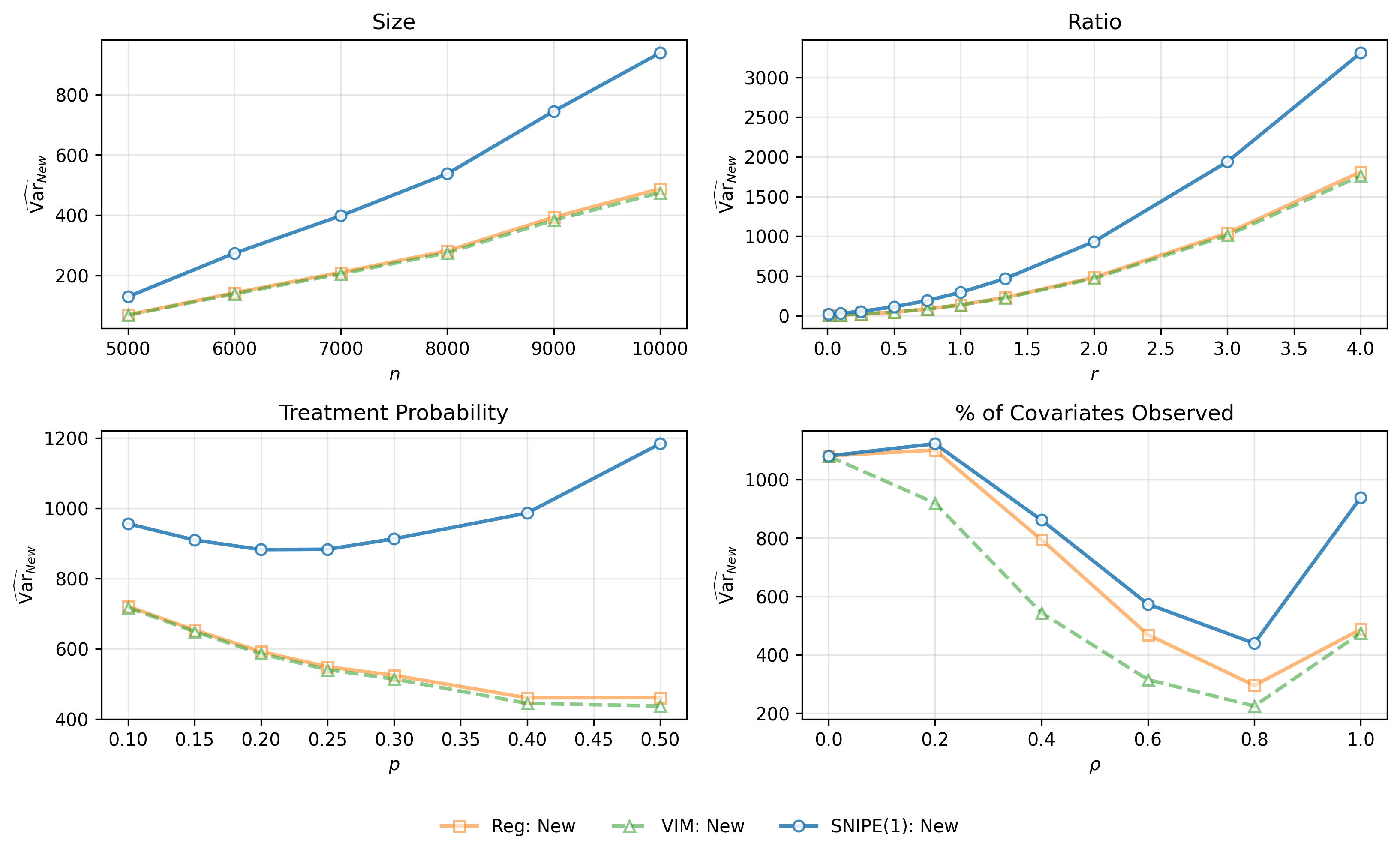}
    \caption{Soft RGG, $\beta = 1$.}
    \label{fig:var_srgg_b1}
  \end{subfigure}

  \caption{Proposed variance estimators for $\SNIPE$ (SNIPE(1)), $\Reg$ (Reg), and $\VIM$ (VIM).}
  \label{fig:var}
\end{figure}

Figure~\ref{fig:ci} provides empirical evidence supporting the theoretical discussion in Section \ref{sec:theory_variance}. Across all designs, both the variance estimator proposed in this paper and the conservative variance estimator of \citet{CortezRodriguezEichhornYu+2023} are conservative relative to the Monte Carlo variance. Moreover, confidence intervals constructed using the conservative variance estimator of \citet{CortezRodriguezEichhornYu+2023} are substantially wider, often by orders of magnitude. These findings are consistent with Table~2 of \citet{CortezRodriguezEichhornYu+2023}, where the conservative variance estimators exceed the empirical variance by several orders of magnitude (e.g., $3.34$ vs.\ $2270.81$ when $n=5000$). Figure~\ref{fig:ci} reports only the log ratios for $\SNIPE$; the corresponding results for $\TTEhat(\rbhat)$ and $\TTEhat(\estchat)$ are in Appendix~\ref{appendix:additional_simulation}. Figure~\ref{fig:var} presents the conservative variance estimators across simulation settings. 
The variance estimator for the VIM-based covariate-adjusted estimator is uniformly the smallest, and the variance estimators for the covariate-adjusted estimators are smaller than that of the unadjusted estimator.

Several features of the results are noteworthy. First, the difference in confidence interval length persists as the sample size increases, indicating that the conservativeness of the existing variance estimator is not a finite-sample artifact but rather a structural consequence of its worst-case bounding construction. Second, the effect is particularly significant for the Soft RGG design with $\beta =1$, where the average number of neighbors increases as sample size increases. The conservative variance estimator of \citet{CortezRodriguezEichhornYu+2023} exhibits high-order polynomial dependence on neighborhood size, which leads to increasing instability as the graph becomes denser. In contrast, the variance estimator proposed in this paper remains relatively stable.

\section{Discussion}\label{section:discussion}

Covariate adjustment is one of the most effective ways to improve precision in randomized experiments. This paper shows that similar gains remain available under interference, provided the adjustment is constructed in a way that respects the dependence induced by the network. Building on the estimator of \citet{CortezRodriguezEichhornYu+2023}, we proposed a general covariate-adjusted estimator $\TTEhat(\bs\theta)$ together with two data-driven choices of the adjustment coefficient, $\Reg$ and $\VIM$. Under the low-order interaction outcome model and suitable sparsity and regularity conditions, both estimators are asymptotically unbiased and asymptotically normal. Moreover, $\VIM$ enjoys a no-harm guarantee: its asymptotic variance is no larger than that of the unadjusted estimator, and it is asymptotically optimal within the class indexed by $\bs\theta$ in terms of mean squared error. In addition, we developed a variance estimator for $\TTEhat(\bs\theta)$ that is asymptotically conservative and empirically much less conservative than the benchmark variance estimator of \citet{CortezRodriguezEichhornYu+2023}, leading to substantially shorter confidence intervals in our simulations.

An important practical issue is how to construct the covariates $\bX_i$. Our theory imposes relatively mild requirements: the covariates may be dependent across units, may depend on the observed network, and need not be identically distributed; the key requirement is that they be independent of the treatment assignment vector. This flexibility leaves room for many useful constructions. As discussed in Appendix~\ref{appendix:construction_covariates}, one may use raw pre-treatment covariates directly, apply nonlinear transformations such as polynomial terms, splines, interactions, kernels, or ReLU-style features, or construct network-based covariates such as degrees and spectral embeddings. One may also combine raw covariates and network structure through procedures such as graph neural network embeddings, or use pre-experiment outcomes, which often have especially strong predictive power. We expect the best construction to depend heavily on the scientific application. A natural direction for future work is to develop principled guidance for this choice, both theoretically and empirically.
Relatedly, our analysis keeps the covariate dimension fixed. This is a natural starting point, but modern applications often generate large collections of candidate covariates or features. It would therefore be valuable to understand high-dimensional adjustment under interference: when can the dimension of $\bX_i$ grow with $n$; what forms of regularization preserve the no-harm property; and how should one select covariates in finite samples? 

While the paper focuses on Bernoulli experiments and the low-order interaction model, the underlying idea is not restricted to this setting. Our results build on a baseline estimator that is tailored to low-order interactions, but the adjustment principle is more general. Whenever one has a primitive estimator that is unbiased or asymptotically unbiased for a target estimand, together with a mean-zero adjustment term constructed from covariates, one can ask how to choose the adjustment coefficient to maximize variance reduction. In this sense, we hope the paper provides a template that can be combined with other baseline estimators and other experimental designs. For example, \citet{eichhorn2024low} extend the model of \citet{CortezRodriguezEichhornYu+2023} to more general experimental designs and show that carefully designed clustered experiments can themselves reduce variance. Our adjustment framework can, in principle, be combined with such designs to obtain further gains.

Another natural extension is to move beyond the total treatment effect. Under the low-order interaction model, the primitive building blocks are the coefficients $\alpha_{i,\mathcal S}$, and many causal estimands can be written as linear combinations of these quantities. This makes the extension of our methodology conceptually straightforward. For example, for any exposure level $q \in [0,1]$, let
$\mu(q)=\frac{1}{n}\sum_{i=1}^n \E_{Z_i \stackrel{\operatorname{i.i.d.}}{\sim} \operatorname{Bernoulli}(q)}\left[Y_i(\bs Z)\right]$. Under the low-order interaction model,
$\mu(q)=\frac{1}{n}\sum_{i=1}^n \sum_{\mathcal S \in \mathcal S_i^\beta}\alpha_{i,\mathcal S} q^{|\mathcal S|}$.
Hence the contrast between two exposure levels $q_1$ and $q_0$ can be written as
\[
\tau(q_1,q_0)=\mu(q_1)-\mu(q_0)
=\frac{1}{n}\sum_{i=1}^n \sum_{\mathcal S \in \mathcal S_i^\beta:\mathcal S\neq\varnothing}
\alpha_{i,\mathcal S}\bigl(q_1^{|\mathcal S|}-q_0^{|\mathcal S|}\bigr).
\]
Since this estimand is again a linear functional of the $\alpha_{i,\mathcal S}$, one obtains a primitive estimator by replacing $\alpha_{i,\mathcal S}$ with their corresponding estimators, and the same mean-zero covariate adjustment can then be added to improve efficiency. The same logic applies to other linear contrasts, including average direct effects, average indirect effects, and other policy-relevant exposure contrasts. We therefore expect the adjustment framework developed here to be useful well beyond the TTE.

More broadly, our framework is conceptually related to the literature on efficient covariate adjustment in randomized experiments; see, for example, \citet{roth2023efficient} for a relevant discussion. At a high level, consider estimators of the form
\[
\hat\tau(\bs\theta)=\hat\tau_0-\bs{\Gamma}^\top \bs\theta,
\]
where $\hat\tau_0$ is a primitive unbiased estimator of the target estimand, $\bs{\Gamma}$ is a mean-zero adjustment term, and $\bs\theta$ is the adjustment coefficient. Within this class, the variance-minimizing choice is
\[
\bs\theta^*=\operatorname{Var}(\bs{\Gamma})^{-1}\operatorname{Cov}(\bs{\Gamma},\hat\tau_0).
\]
In our setting, $\bs{\Gamma} = \frac{1}{n}\sum_{i=1}^n \omega_i \bX_i$.
This perspective is quite general: it neither relies on interference nor depends on the low-order interaction model. Those ingredients enter instead through the construction and estimation of the relevant moments in our problem. One can obtain an asymptotically optimal adjusted estimator by consistently estimating $\bs\theta^*$ and then plugging it into $\hat\tau(\bs\theta)$. Doing so requires consistent estimators of $\operatorname{Var}(\bs{\Gamma})$ and $\operatorname{Cov}(\bs{\Gamma},\hat\tau_0)$, but notably not of the full variance $\operatorname{Var}(\hat\tau_0)$. This distinction is important in our setting. Estimating the full variance of the primitive estimator involves difficult second-order terms in the estimated $\alpha$'s, whereas estimating $\operatorname{Cov}(\bs{\Gamma},\hat\tau_0)$ only involves first-order terms and is therefore substantially more tractable. This viewpoint also clarifies the main conceptual focus of the paper. Rather than analyzing the variance of each adjusted estimator separately, we study the variance reduction induced by adjustment relative to the unadjusted estimator, treating $\bs\theta$ as the optimization variable.

\section*{Acknowledgement}
The authors thank Mayleen Cortez-Rodriguez, Peter Hull, Soonwoo Kwon, Xin Lu, Peng Ding, Jonathan Roth and Christina Yu for helpful discussions.

\newpage
\bibliographystyle{apalike} 
\bibliography{references}

\newpage
\begin{appendix}
\section{Details of the Simulation Design}

\begin{algorithm}[H]

\caption{Generation of Weighted Network Matrix $\alpha^{\text{linear}}$}
\begin{algorithmic}[1]
\REQUIRE Adjacency matrix $\bs A \in \{0,1\}^{n \times n}$, true covariates $\bX^{\text{true}} \in \mathbb{R}^{n \times p}$, matrix $\bs \Psi \in \mathbb{R}^{p \times p}$, constants \texttt{diag}, $\texttt{offdiag}=r\cdot\texttt{diag}$, vectors $\bs{v},\bs u\in\mathbb R^n$ (optional)
\STATE Generate $\bm v \sim \text{Unif}([0,1]^n)$ if not provided
\STATE $\bm c_{\text{offdiag}} \gets \texttt{offdiag} \cdot \bm v$
\STATE Set $\bs d$ by $d_i \gets \sum_{j=1}^n A_{ij}$ for $i = 1,\dots,n$
\STATE $\bm D_\text{in} \gets \text{diag}(\bs d)$
\STATE $\tilde{\bm A} \gets \bm D_\text{in}(\bm A - \bm I)$
\STATE Set $\bs s$ by $s_j \gets \sum_{i=1}^n \tilde{A}_{ij}$; if $s_j = 0$ then set $s_j \gets 1$
\STATE $\bm S \gets \text{diag}(\bm c_{\text{offdiag}} / \bm s)$
\STATE $\bm C \gets \tilde{\bm A}\bm S$
\STATE Generate $\bm u \sim \text{Unif}([0,1]^n)$ if not provided
\STATE Set $C_{ii} \gets \texttt{diag} \cdot u_i$ for $i = 1, \dots, n$
\STATE $\bm X_{\Psi}\gets \bX^{\text{true}}\bs \Psi$
\STATE $\bX_{\Psi}\gets \bX_{\Psi} / \sum_{i,j} |(X_{\Psi})_{ij}| \cdot n^2 / 5$
\STATE $\bX_{\text{temp}} \gets \texttt{offdiag} \cdot \bX_{\Psi}$
\STATE Set $(X_{\text{temp}})_{ii}\gets (X_{\Psi})_{ii} \cdot \texttt{diag}$ for $i = 1,\dots,n$
\STATE $\bX_{\text{mod}} \gets \bm A \odot\bX_{\text{temp}}$
\STATE \textbf{Output} $\alpha^{\text{linear}} \gets \bm C + \bX_{\text{mod}}$
\end{algorithmic}\label{alg:alpha-linear}
\end{algorithm}

\begin{algorithm}[H]
\caption{Generation of Weighted Network Matrix $\alpha^{\text{quad}}$ (Degree-Dependent)}
\begin{algorithmic}[1]
\REQUIRE Adjacency matrix $\bs A \in \{0,1\}^{n \times n}$, true covariates $\bX^{\text{true}} \in \mathbb{R}^{n \times p}$,  constants \texttt{diag}, $\texttt{offdiag}=r\cdot\texttt{diag}$, vectors $\bs{v},\bs u\in\mathbb R^n$ (optional)
\STATE Generate $\bs v \sim \text{Unif}([0,1]^n)$ if not provided
\STATE $\bm c_{\text{offdiag}} \gets \texttt{offdiag} \cdot \bm v$
\STATE Set $\bs d$ by $d_i \gets \sum_{j=1}^n A_{ij}$ for $i = 1,\dots,n$
\STATE $\bs D_\text{in} \gets \text{diag}(\bs d)$
\STATE $\tilde{\bs A} \gets \bs D_\text{in} (\bs A - \bs I)$
\STATE Set $\bs s$ by $s_j \gets \sum_{i=1}^n \tilde{A}_{ij}$; if $s_j = 0$ then set $s_j \gets 1$
\STATE $\bs S \gets \text{diag}(\bs c_{\text{offdiag}} / \bs s)$
\STATE $\alpha^{\text{quad}} \gets \tilde{\bs A}\bs S$
\STATE Generate $\bs u \sim \text{Unif}([0,1]^n)$ if not provided
\STATE Set $\alpha^{\text{quad}}_{ii} \gets \left( \sum_{j=1}^p X^{\text{true}}_{ij} + \texttt{diag} \right) \cdot u_i$ for $i = 1,\dots,n$
\STATE \textbf{Output} $\alpha^{\text{quad}}$
\end{algorithmic}\label{alg:alpha-quad}
\end{algorithm}

\section{Additional Discussions}
\label{appendix:additional_discussion}

\subsection{An alternative perspective on the covariate-adjusted estimator}
\label{appendix:alternative_perspective}
To take advantage of the covariates that we mentioned previously in the estimation of TTE, we introduce the following \textit{working} model:
\begin{align*}
     Y_i(\bs Z) = \sum_{\mathcal{S} \in \mathcal{S}_i^\beta, \mathcal{S} \neq \varnothing}\alpha_{i, \mathcal S} \prod_{j \in \mathcal S} Z_j + c_{i, \varnothing} + \bs 
        \theta^\top \bX_i. 
\end{align*}
\begin{remark}
Here, $\alpha_{i, \mathcal{S}}$ and $c_{i, \varnothing}$ may be correlated with $\bX_i$. This model is still equivalent with the original low-order interaction model, as it simply extracts the linear effect of $\bs X_i$ from $\alpha_{i, \varnothing}$. 
\end{remark}

Recall that our target is to estimate TTE:
\begin{align}\label{TTEadj}
    \tau:= \frac{1}{n}\sumn\sum_{\mathcal S \in \mathcal S_i^\beta, \mathcal{S} \neq \varnothing} \alpha_{i, \mathcal S}
\end{align}
For simplicity, let $\tilde {\bs Z}_i = [\prod_{j \in \mathcal S} Z_j]_{\mathcal S \in \mathcal S_i^\beta}$ denote the treatment interaction vectors. For example, when $\beta = 1$ and $\mathcal N_i = \{i, j\}$, $\tilde {\bs Z}_i = [1 \quad Z_i \quad Z_j]^\top$, when $\beta = 2$ and $\mathcal N_i = \{i, j\}$, $\tilde {\bs Z}_i = [1 \quad Z_i \quad Z_j \quad Z_iZ_j]^\top$. And our \textit{working} model can be expressed as
\begin{align*} 
Y_i(\bs Z) = \bs C_{i}^\top \tilde{\bs Z}_i + \bs \theta^\top \bX_i,     
\end{align*}
where $\bs C_{i} = [c_{i, \varnothing} \quad [\alpha_{i, \mathcal{S}}]^\top_{\mathcal S \in \mathcal S_i^\beta, \mathcal S \neq \varnothing}]^\top$.

Similar to \cite{CortezRodriguezEichhornYu+2023}, to motivate our adjusted estimator, consider a thought experiment in which we can conduct $M$ independent replications of the randomized experiment. That is, we can conduct independent randomized experiment for the same population in $M$ parallel worlds. In this setting, for each unit $i$, we observe $M$ independent treatment interactions vectors $\tilde {\bs Z_i}^{(1)}, \ldots, \tilde {\bs Z_i}^{(M)}$ and realizations of the potential outcome $Y_i^{(1)}, \ldots, Y_i^{(M)}$. With predetermined choices of $\bs \theta$, we adopt the least squares estimator as our estimates of $\bs C_i$, denoted as $\hat {\bs C}_i$, for $i = 1, \ldots n$:
    \begin{align}\label{equ: cihat}
     \hat {\bs C}_i &= \argmin_{{\bs C}_i} \sum_{m=1}^M(Y_i^{(m)}  - {\bs C}_i^\top \tilde {\bs Z}_i^{(m)} - \bs 
        \theta^\top \bX_i)^2,\nonumber\\
        &= \left(\frac{1}{M}\bs\Phi_i^\top\bs\Phi_i\right)^{-1}\cdot\frac{1}{M}\bs\Phi_i^\top \tilde {\bs Y}_i, 
    \end{align}
    where $\bs \Phi_i$ is the design matrix of unit $i$, and $\tilde {\bs Y}_i = [Y_i^{(1)} - \bs \theta^\top \bX_i, \ldots, Y_i^{(M)} - \bs \theta^\top \bX_i]^\top$. Inspired by \cite{CortezRodriguezEichhornYu+2023}, we replace $\left(\frac{1}{M}\bs\Phi_i^\top\bs\Phi_i\right)^{-1}$ by $\mathbb{E}\left(\tilde {\bs Z}_i \tilde {\bs Z}_i^\top\right)^{-1}$ and $\frac{1}{M}\bs\Phi_i\tilde {\bs Y}_i$ by $\tilde {\bs Z}_i(Y_i -\bs \theta^\top \bX_i)$ in (\ref{equ: cihat}). The first replacement is motivated by almost sure convergence, and the second replacement is motivated by the true realization. Therefore, we have
    \begin{align*}
    \hat{\bs C}_i &= \mathbb{E}\left(\tilde {\bs Z}_i \tilde {\bs Z}_i^\top\right)^{-1}\tilde {\bs Z}_i\left(Y_i -\bs \theta^\top \bX_i\right).
    \end{align*}
    Plug the above result into (\ref{TTEadj}), we have our form of general covariate-adjusted SNIPE estimator of TTE as
    \begin{align*}
    \TTEhat(\bs \theta)
    &=\frac{1}{n} \sum_{i = 1}^n \left<(\bs 1_i - \bs e_{1i}), \mathbb{E}\left(\tilde {\bs Z}_i \tilde {\bs Z}_i^\top\right)^{-1}\tilde {\bs Z}_i\left(Y_i -\bs \theta^\top X_i\right)\right>\\
    &= \frac{1}{n}\sum_{i=1}^nY_i\sum_{\mathcal S\subseteq \mathcal S_i^\beta}g(\mathcal S)\prod_{j\in\mathcal S}\frac{Z_j-p_j}{p_j(1-p_j)}-\frac{1}{n}\sum_{i=1}^n\bs 
        \theta^\top \bX_i\sum_{\mathcal S \subseteq\mathcal S_i^\beta}g(\mathcal S)\prod_{j\in\mathcal S}\frac{Z_j-p_j}{p_j(1-p_j)}.
    \end{align*}
Compared to the original SNIPE estimator, the first term in $\TTEhat(\bs \theta)$ remains unchanged. The second term is newly introduced to perform covariate adjustment.

\subsection{Additional properties of the regression-based covariate-adjusted estimator}
\label{appendix:reg_estimator}

To understand the properties of $\Reg$, we begin with the simple setting of no interference. Under SUTVA, the most widely used covariate-adjusted estimator in the literature is the one proposed by \citet{Lin2013}. We now examine connections between $\Reg$ and \citet{Lin2013}’s estimator.
\citet{Lin2013}’s estimator targets the average treatment effect (ATE). It builds on the difference-in-means (DM) estimator, incorporating covariates to improve precision. Specifically, it fits a linear regression of the observed outcome on the treatment indicator, the covariates, and all treatment–covariate interaction terms, and takes the fitted coefficient on the treatment indicator as the ATE estimate.
Formally, Lin’s estimator can be written as:
\begin{estimator}[\cite{Lin2013}’s estimator]
\label{estimator:lin}
\begin{align*}
\TTEhat_{\text{Lin}} &=  \frac{\sumn Z_i\left(Y_i - \hat{\bs \theta}_1^\top \bs X_i\right)}{\sumn Z_i} - \frac{\sumn (1-Z_i)\left(Y_i - \hat{\bs \theta}_0^\top \bs X_i\right)}{\sumn (1-Z_i)},
\end{align*}   
where $\hat{\bs \theta}_0$ and $\hat{\bs \theta}_1$ are the ordinary least squares coefficients on $\bX_i$
 from regressing $Y_i$ on $\bX_i$ in the treatment and control groups, respectively.
\end{estimator}

To facilitate comparison between $\TTEhat(\rbhat)$ and \cite{Lin2013}'s estimator, we rearrange terms and make use of the centered covariates $\bX_i$ to rewrite the latter as follows:
\begin{align*}
\TTEhat_{\text{Lin}}
&\equiv \frac{\sumn Z_iY_i }{\sumn Z_i} - \frac{\sumn (1-Z_i)Y_i }{\sumn (1-Z_i)} - \hat{\bs \theta}_{\text{Lin}}^\top \left[\frac{\sumn Z_i\bs X_i}{\sumn Z_i} - \frac{\sumn (1-Z_i)\bs X_i}{\sumn (1-Z_i)}\right], 
\end{align*}
where $\hat{\bs \theta}_{\text{Lin}} = \frac{\sumn (1-Z_i)}{n}\hat{\bs \theta}_1 + \frac{\sumn Z_i}{n}\hat{\bs \theta}_0$. 
This form combines the two group-specific adjustment coefficients into a single coefficient $\hat{\bs \theta}_{\text{Lin}}$ so the entire expression can be viewed as a DM estimator adjusted by $\hat{\bs \theta}_{\text{Lin}}$.
To put things in parallel, recall from Section~\ref{sec:general_cova_SNIPE} that if there is no interference, $\TTEhat(\rbhat)$ can be written as a covariate-adjusted IPW estimator:
\begin{align*}
    \TTEhat(\rbhat) &= \frac{1}{n}\sumn \left[\frac{Z_i\left(Y_i - \rbhat^\top \bs X_i\right)}{p_i} - \frac{(1-Z_i)\left(Y_i - \rbhat^\top \bs X_i\right)}{1-p_i}\right]\\
    &=\frac{1}{n}\sumn \left[\frac{Z_iY_i }{p_i} - \frac{(1-Z_i)Y_i }{1-p_i}\right] - \rbhat^\top\frac{1}{n}\sumn \left[\frac{Z_i\bs X_i}{p_i} - \frac{(1-Z_i)\bs X_i}{1-p_i}\right].
\end{align*}
Although IPW and DM have different asymptotic properties, this reformulation makes clear that both estimators subtract an adjustment term involving a single coefficient ($\hat{\bs \theta}_{\text{Lin}}$  for \citet{Lin2013}'s estimator and $\rbhat$ for $\Reg$). Interestingly, under regularity conditions and the additional assumption that treatment probabilities are identical across all units, Proposition~\ref{prop:regvimlin} shows that $\rbhat$ and $\hat{\bs \theta}_{\text{Lin}}$ are asymptotically equivalent; see Section~\ref{sec:VIM-SNIPE} for details.

A notable property of \citet{Lin2013}’s estimator is that, under SUTVA, its asymptotic variance is guaranteed to be no greater than that of the DM estimator, regardless of whether the true outcome model is linear in covariates. Analogously, under regularity conditions, we can show that the asymptotic variance of $\TTEhat(\rbhat)$ is no greater than that of $\TTEhat_{\text{unadj}}$ under SUTVA. As we shall see later, this result is a natural corollary of Proposition \ref{prop:regvimlin} and Theorem \ref{thm:vimvar}.

\subsection{Details of Example~\ref{exp:increasevar_small}}\label{app:toy_example_details}

Consider an undirected graph with $n = 3$ units. In this graph, Node 1 is connected to Node 2, while Node 3 is isolated and has no connections (See Figure~\ref{fig:single_three} for an illustration). We consider a low-order interaction outcome model with interaction order $\beta = 1$. 
The potential outcomes are given by
\begin{align*}
Y_1(\mathbf{z}) = z_1 + z_2, \qquad
Y_2(\mathbf{z}) = -2 + z_1 + z_2, \qquad
Y_3(\mathbf{z}) = -0.5 + z_3. 
\end{align*}
Each unit receives treatment independently with probability $p = 0.5$. The covariate values for the three units are:
\[
\bX_1 = 0.5, \qquad \bX_2 = 0, \qquad \bX_3 = -0.5. 
\]

\begin{figure}
\centering
\begin{tikzpicture}[>=Stealth, node distance=3cm, thick]

  \definecolor{control}{RGB}{204,204,255}

  \node[circle, draw, fill=control, minimum size=1cm] (1) {1};
  \node[circle, draw, fill=control, minimum size=1cm, right of=1] (2) {2};
  \node[circle, draw, fill=control, minimum size=1cm, right of=2] (3) {3};

  \draw[->] (1) -- (2);
  \draw[->] (2) -- (1);

\end{tikzpicture}
\caption{An interference network with three units.}

\end{figure}

\begin{figure}
\centering
\begin{tikzpicture}[>=Stealth, thick]

  \definecolor{lightred}{RGB}{255, 200, 200}

  \node[circle, draw, fill=lightred, minimum size=1cm] (1a) at (0,0) {1a};
  \node[circle, draw, fill=lightred, minimum size=1cm] (2a) at (2.5,0) {2a};
  \node[circle, draw, fill=lightred, minimum size=1cm] (3a) at (5,0) {3a};
  \draw[->] (1a) -- (2a);
  \draw[->] (2a) -- (1a);

  \node[circle, draw, fill=lightred, minimum size=1cm] (1b) at (0,-1.6) {1b};
  \node[circle, draw, fill=lightred, minimum size=1cm] (2b) at (2.5,-1.6) {2b};
  \node[circle, draw, fill=lightred, minimum size=1cm] (3b) at (5,-1.6) {3b};
  \draw[->] (1b) -- (2b);
  \draw[->] (2b) -- (1b);

  \node at (2.5,-2.6) {\Large $\vdots$};

  \node[circle, draw, fill=lightred, minimum size=1cm] (1c) at (0,-3.6) {1c};
  \node[circle, draw, fill=lightred, minimum size=1cm] (2c) at (2.5,-3.6) {2c};
  \node[circle, draw, fill=lightred, minimum size=1cm] (3c) at (5,-3.6) {3c};
  \draw[->] (1c) -- (2c);
  \draw[->] (2c) -- (1c);

\end{tikzpicture}
\caption{An interference network with many groups of three units.}
\end{figure}

In this setting, it is straightforward to verify that $\operatorname{TTE} = \frac{5}{3}$.
The unadjusted estimator $\SNIPE$ is given by
\[
\begin{split}
\TTEhat_{\text{unadj}}
& = \frac{4}{3} \p{ 2(Z_1 - 0.5 + Z_2-0.5)^2 + (Z_3 - 0.5)^2}.
\end{split}
\]
For any $\bs \theta \in \mathbb{R}$, the covariate-adjusted estimator defined in \eqref{equ: ttehat} is
\[
\begin{split}
\TTEhat(\bs \theta)
& = \TTEhat_{\text{unadj}} - \frac{2}{3} \bs \theta(Z_1 - 0.5 + Z_2-0.5 - Z_3 + 0.5).
\end{split}
\]

Straightforward calculations yield
$\Var\sqb{\TTEhat_{\text{unadj}}} = \frac{16}{9}.$
Since the term $(Z_1 - 0.5 + Z_2 - 0.5 - Z_3 + 0.5)$ is uncorrelated with $\TTEhat_{\text{unadj}} = \frac{4}{3} \big( 2(Z_1 - 0.5 +$ $Z_2-0.5)^2 + (Z_3 - 0.5)^2 \big)$, we have
\[
\Var\sqb{\TTEhat(\bs \theta)}
= \Var\sqb{\TTEhat_{\text{unadj}}} + \Var\sqb{\frac{2}{3} \bs \theta(Z_1 - 0.5 + Z_2-0.5 - Z_3 + 0.5)}
= \frac{16}{9} + \frac{1}{3} {\bs \theta}^2. 
\]
Therefore, unless $\bs \theta = \bs 0$, the variance of the covariate-adjusted estimator is strictly larger than that of the unadjusted estimator.

In particular, in this example, we can explicitly compute $\rb$ and find that $\rb = \frac{4}{3}$ (see the detailed calculation in Appendix~\ref{appendix:small_thetaReg}). Therefore, we have
\[
\Var\sqb{\TTEhat_{\text{unadj}}} < \Var\sqb{\TTEhat(\rb)}. 
\]
Of course, since this example involves only three units, asymptotic results do not apply, and Proposition~\ref{prop:consistrb}, which states that $\rbhat - \rb \overset{p}{\to} 0$, does not hold directly. However, consider a setting where we observe many independent groups of three units, each identical to the configuration above (see Figure~\ref{fig:many_three} for an illustration). In that case, we would still have
$\Var\sqb{\TTEhat_{\text{unadj}}} < \Var\sqb{\TTEhat(\rb)}$ and clearly have $\rbhat - \rb \overset{p}{\to} 0$. This suggests that, due to interference, the asymptotic variance of the regression-based covariate-adjusted estimator is strictly greater than that of the unadjusted estimator.

Finally, we provide some intuition for why this can happen. Why is the regression-based estimator not guaranteed to reduce variance as it does in the no-interference setting? What explains the difference between the no-interference and interference cases?
In the presence of interference, the terms $\omega_i\left(Y_i - \bs \theta^\top \bX_i \right)$ are generally not independent across units. As a result, the variance of $\TTEhat(\bs \theta)$ includes not only individual variance terms, but also non-negligible covariance terms. The choice of $\rb$ minimizes the sum of variance terms, but does not account for the impact on the covariance terms, which may increase and dominate.
In our toy example, the terms $\omega_1\left(Y_1 - \bs \theta^\top \bX_1 \right ) = 4(Y_1 - 0.5\bs \theta)(Z_1 - 0.5 + Z_2-0.5) $ and $\omega_2\left(Y_2 - \bs \theta^\top \bX_2 \right ) = 4Y_2(Z_1 - 0.5 + Z_2-0.5)$ are clearly correlated, since both depend on the same random variables $Z_1$ and $Z_2$. Although the choice $\rb = \frac{4}{3}$ does reduce the variance terms, it does not mitigate the resulting covariance contribution, which ultimately increases the overall variance.

In particular, in our toy example
\[
\begin{split}
\Var\sqb{\TTEhat(\bs \theta)}&= \frac{16}{9} \Big( \Var\sqb{(Y_1 - 0.5\bs \theta)(Z_1 - 0.5 + Z_2-0.5)} + \Var\sqb{Y_2(Z_1 - 0.5 + Z_2-0.5)}\\
&\qquad \qquad \qquad+ \Var\sqb{(Y_3 + 0.5 \bs \theta)(Z_3 - 0.5)} \Big)\\
&\qquad +  \frac{32}{9}\Cov\sqb{(Y_1 - 0.5\bs \theta)(Z_1 - 0.5 + Z_2-0.5), Y_2(Z_1 - 0.5 + Z_2-0.5)}\\
& = V_{\Var}(\bs \theta) + V_{\Cov}(\bs \theta),
\end{split}
\]
where $V_{\Var}(\bs \theta)$ denotes the sum of the variance terms and $V_{\Cov}(\bs \theta)$ the covariance term.
Moving from the unadjusted estimator ($\bs \theta = 0$) to the regression-based covariate-adjusted estimator ($\bs \theta = \rb$), the variance component decreases:
\[ V_{\Var}(\bs 0) = \frac{8}{3} = 2.67, \qquad \qquad V_{\Var}(\rb) =  \frac{56}{27} = 2.07,\]
but the increase in the covariance component outweighs the decrease in the variance component:
\[ V_{\Cov}(\bs 0) = -\frac{8}{9} = -0.89, \qquad \qquad V_{\Cov}(\rb) = \frac{8}{27} = 0.30. \]

More generally, beyond the toy example in Example~\ref{exp:increasevar_small}, the same phenomenon persists. Indeed, when the treatment probability is the same across all units (i.e., $p_1 = \cdots = p_n$) and $\beta = 1$, the difference between the variance of the unadjusted estimator and that of the regression-based covariate-adjusted estimator can be expressed as follows:
\[
\begin{split}
&\Var( \TTEhat(\bs 0)) - \Var( \TTEhat(\rb)) \\
&\qquad = \frac{1}{p_1(1-p_1)n^2}\sumn  \Bigg[ |\mathcal N_i|\rb^\top  \bX_i\bX_i^\top\rb \quad + \\
& \qquad \sum_{i': \mathcal N_i \cap \mathcal N_{i'} \neq \varnothing, i'\neq i}  \rb^\top \bX_{i'}\sum_{j \in \mathcal N_i \cap \mathcal N_{i'}} (2(1-2p_1)\alpha_{i, \{j\}} +2(\alpha_{i,\varnothing}+p_1\sum_{j' \in \mathcal N_i}\alpha_{i, \{j'\}}) - \bX_i^\top\rb)\Bigg],
\end{split}
\]
where the first term corresponds to the change in the sum of variances of $\omega_i\left(Y_i - \bs \theta^\top \bX_i \right)$, and the second term corresponds to the change in the sum of covariances between such terms (see derivation details in Appendix \ref{appendix:variance_difference}).
As in the toy example in Example~\ref{exp:increasevar_small}, the first term is always nonnegative: the choice of $\rb$ reduces the variance components. However, the second term can be either positive or negative, depending on the structure of the covariates. In particular, the second term may be negative when the covariates and potential outcomes of unit $i$ are related to those of other units.

\subsection{\texorpdfstring{Derivation of $\rb$ in Example \ref{exp:increasevar_small}}{Derivation of thetaReg in Example 1}}
\label{appendix:small_thetaReg}

We are given
$$
\rb = \mathbb{E} \left( \frac{1}{n} \sum_{i=1}^n \omega_i^2 \bX_i \bX_i^\top \right)^{-1} \mathbb{E} \left( \frac{1}{n} \sum_{i=1}^n \omega_i^2 Y_i \bX_i \right),
$$
with $n = 3$, covariates $\bX_1 = 0.5$, $\bX_2 = 0$, $\bX_3 = -0.5$, and outcomes
$$
Y_1 = Z_1 + Z_2, \quad Y_2 = -2 + Z_1 + Z_2, \quad Y_3 = -0.5 + Z_3.
$$
The weights are given by
$$
\omega_1 = \omega_2 = 4(Z_1 - 0.5 + Z_2 - 0.5) = 4(Z_1 + Z_2 - 1), \quad \omega_3 = 4(Z_3 - 0.5).
$$

For the denominator $\mathbb{E}\left[ \frac{1}{3} \sum_{i=1}^3 \omega_i^2 \bX_i^2 \right]$, note that
$\bX_1^2 = 0.25$, $\bX_2^2 = 0$, and $\bX_3^2 = 0.25$, and that
$\omega_1^2 = 16(Z_1 + Z_2 - 1)^2$.
We compute
$$
\mathbb{E}[(Z_1 + Z_2 - 1)^2] = 0.25(1)^2 + 0.5(0)^2 + 0.25(1)^2 = 0.5,
$$
so
$$
\mathbb{E}[\omega_1^2] = \mathbb{E}[\omega_2^2]= 16 \cdot 0.5 = 8.
$$
We also have
$$
\omega_3^2 = 16(Z_3 - 0.5)^2, \quad \mathbb{E}[\omega_3^2] = 16 \cdot 0.25 = 4.
$$
Therefore,
$$
\mathbb{E} \left[ \frac{1}{3} \sum_{i=1}^3 \omega_i^2 \bX_i^2 \right] = \frac{1}{3} \left( 8 \cdot 0.25 + 0 + 4 \cdot 0.25 \right) = \frac{1}{3}(2 + 1) = 1.
$$

For the numerator $\mathbb{E}\left[ \frac{1}{3} \sum_{i=1}^3 \omega_i^2 Y_i \bX_i \right]$, 
we compute each term individually: $\mathbb{E}[\omega_1^2 Y_1 \bX_1]$, $\mathbb{E}[\omega_2^2 Y_2 \bX_2]$, and $\mathbb{E}[\omega_3^2 Y_3 \bX_3]$.

Starting with $\mathbb{E}[\omega_1^2 Y_1 \bX_1]$:
$$
\omega_1^2 Y_1 \bX_1 = 16(Z_1 + Z_2 - 1)^2 (Z_1 + Z_2) \cdot 0.5 = 8(Z_1 + Z_2 - 1)^2 (Z_1 + Z_2).
$$
This is a discrete random variable with the following distribution:
\begin{align*}
Z_1 + Z_2 = 0 &\Rightarrow (0 - 1)^2 \cdot 0 = 0 \quad \text{(prob $0.25$)}, \\
Z_1 + Z_2 = 1 &\Rightarrow (1 - 1)^2 \cdot 1 = 0 \quad \text{(prob $0.5$)}, \\
Z_1 + Z_2 = 2 &\Rightarrow (2 - 1)^2 \cdot 2 = 2 \quad \text{(prob $0.25$)}.
\end{align*}
Therefore,
$$
\mathbb{E}[\omega_1^2 Y_1 \bX_1] = 8 \cdot (0 + 0 + 0.25 \cdot 2) = 8 \cdot 0.5 = 4.
$$

Next, since $\bX_2 = 0$, we have
$$
\mathbb{E}[\omega_2^2 Y_2 \bX_2] = 0.
$$

For $\mathbb{E}[\omega_3^2 Y_3 \bX_3]$, note that
$$
\omega_3^2 Y_3 \bX_3 = 16(Z_3 - 0.5)^2(-0.5 + Z_3)(-0.5),
$$
which again is a discrete random variable with distribution:
\begin{align*}
Z_3 = 0 &\Rightarrow 16(0.25)(-0.5)(-0.5) = 1 \quad \text{(prob $0.5$)}, \\
Z_3 = 1 &\Rightarrow 16(0.25)(0.5)(-0.5) = -1 \quad \text{(prob $0.5$)}.
\end{align*}
Thus,
$$
\mathbb{E}[\omega_3^2 Y_3 \bX_3] = 0.5 \cdot 1 + 0.5 \cdot (-1) = 0.
$$

Summing the three components:
$$
\mathbb{E} \left[ \frac{1}{3} \sum_{i=1}^3 \omega_i^2 Y_i \bX_i \right] = \frac{1}{3}(4 + 0 + 0) = \frac{4}{3}.
$$

Therefore,
$$
\bs\theta_{\text{Reg}} = \frac{\frac{4}{3}}{1} = \frac{4}{3}.
$$

\subsection{Additional discussion on the VIM estimator}
\label{appendix:VIM_discussion}
To compare $\VIM$ and $\Reg$, we again begin with the no-interference setting. In this setting, $\estchat=\E(\frac{1}{n}\sumn \omega_i^2\bX_i\bX_i^\top)^{-1} (\frac{1}{n}\sumn \omega_i^2\bX_iY_i)$, and $\rbhat =  (\frac{1}{n}\sumn \omega_i^2\bX_i\bX_i^\top)^{-1}(\frac{1}{n}\sumn \omega_i^2\bX_iY_i)$. The forms of $\estchat$ and $\rbhat$ under the no-interference setting are nearly identical: the only difference is that $\estchat$ uses the expectation of the Gram matrix, whereas $\rbhat$ relies on its sample counterpart. Under Assumptions \ref{as:interference}--\ref{as:variance} and the no-interference setting, we can show that $\estchat - \rbhat \overset{p}{\to} 0$ by arguments analogous to those in the proof of Proposition \ref{prop:consistrb}. Hence, in the no-interference setting, the adjustment coefficient of $\VIM$ is asymptotically equivalent to that of $\Reg$. Note that this result does not require treatment probabilities to be identical across units.

Moreover, continuing the discussion in Appendix~\ref{appendix:reg_estimator}, we establish that, under regularity conditions and the additional assumption that treatment probabilities are identical across all units, the adjustment coefficients for $\Reg$, $\VIM$, and \citet{Lin2013}’s estimator are asymptotically equivalent.

\begin{proposition}\label{prop:regvimlin}
Let $\bs \theta_{\text{Lin}} = (1-p_1)S_{\bX \bX}^{-1}S_{\bX Y(1)} + p_1S_{\bX \bX}^{-1}S_{\bX Y(0)}$, where $S_{\bX \bX} = \frac{1}{n}\sumn (\bs X_i  - \bar \bX) (\bs X_i  - \bar \bX)^\top$, $S_{\bX Y(q)} = \frac{1}{n}\sumn (\bs X_i - \bar \bX) (Y_i(q) - \bar Y_q)$, $\bar \bX = \frac{1}{n}\sumn \bX_i$ and $\bar Y_q = \frac{1}{n}\sumn Y_i(q)$ for $q = 0, 1$. Under Assumptions \ref{as:interference}--\ref{as:variance}, suppose further that there is no interference and that the treatment probabilities are identical across units. Then $\rbhat - \bs \theta_{\text{Lin}}$, $\estchat - \bs \theta_{\text{Lin}}$, and $\hat{\bs \theta}_{\text{Lin}} - \bs \theta_{\text{Lin}}$ converge to $0$ in probability. Moreover, $\bs \theta_{\text{Lin}}$ has a finite limit, denoted by $\bs \theta_{\text{Lin}}^\ast$. Therefore, $\rbhat$, $\estchat$, and $\hat{\bs \theta}_{\text{Lin}}$ all converge to $\bs \theta_{\text{Lin}}$ in probability.
\end{proposition}

\begin{figure}
  \centering
  \begin{subfigure}[b]{0.24\textwidth}
    \includegraphics[width=\textwidth]{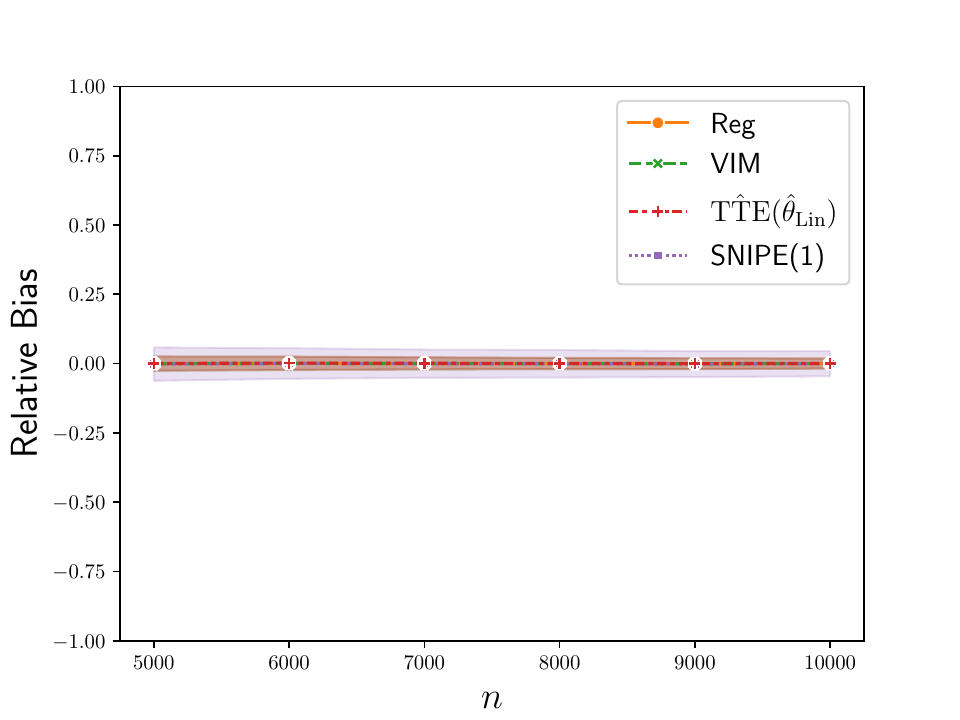}
  \end{subfigure}
  \hfill
  \begin{subfigure}[b]{0.24\textwidth}
    \includegraphics[width=\textwidth]{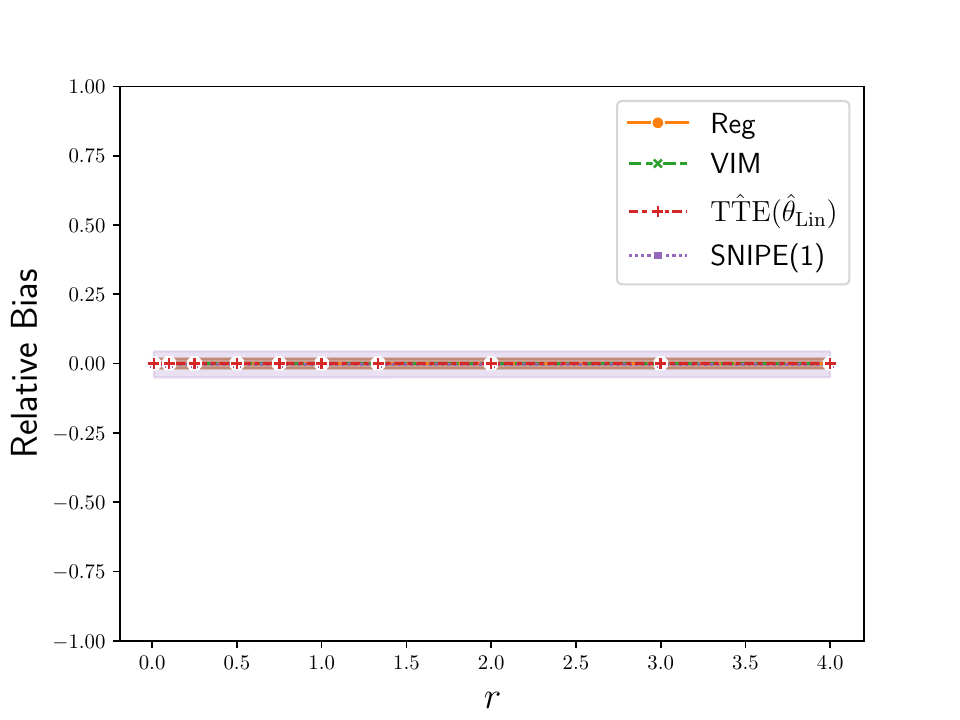}
  \end{subfigure}
  \hfill
  \begin{subfigure}[b]{0.24\textwidth}
    \includegraphics[width=\textwidth]{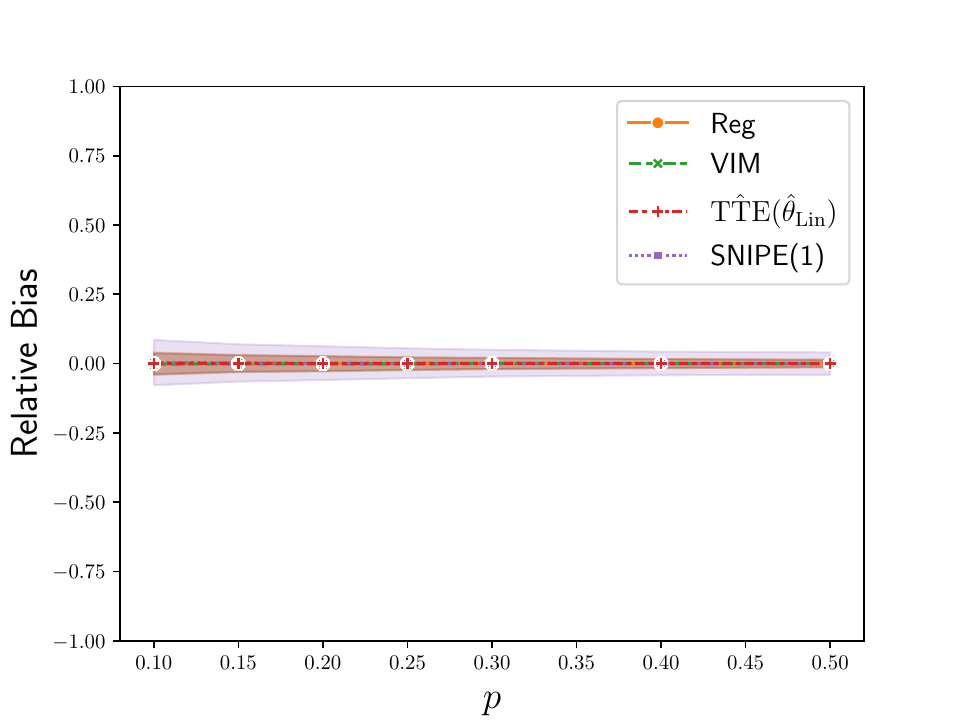}
  \end{subfigure}
    \hfill
  \begin{subfigure}[b]{0.24\textwidth}
    \includegraphics[width=\textwidth]{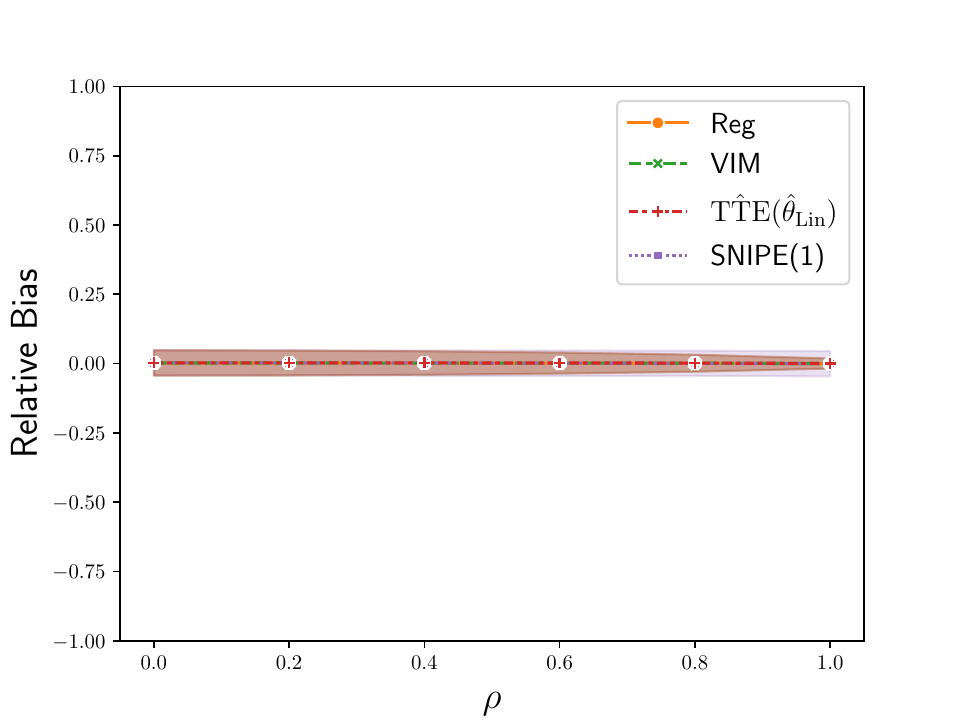}
  \end{subfigure}
    \vspace{0.2cm}

   \begin{subfigure}[b]{0.24\textwidth}
    \includegraphics[width=\textwidth]{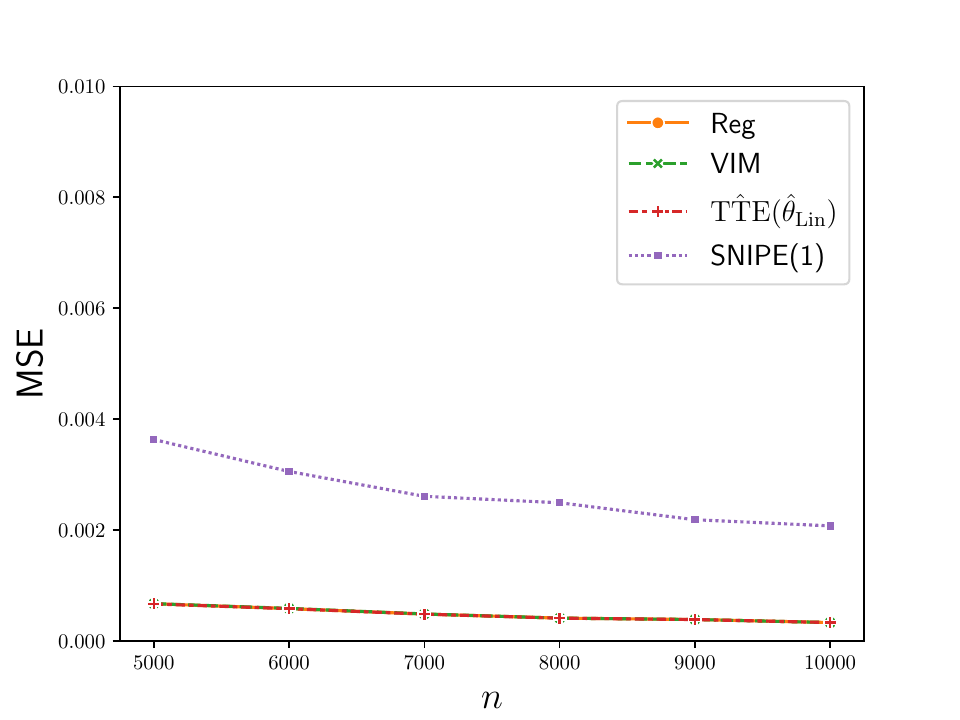}
  \end{subfigure}
  \hfill
  \begin{subfigure}[b]{0.24\textwidth}
    \includegraphics[width=\textwidth]{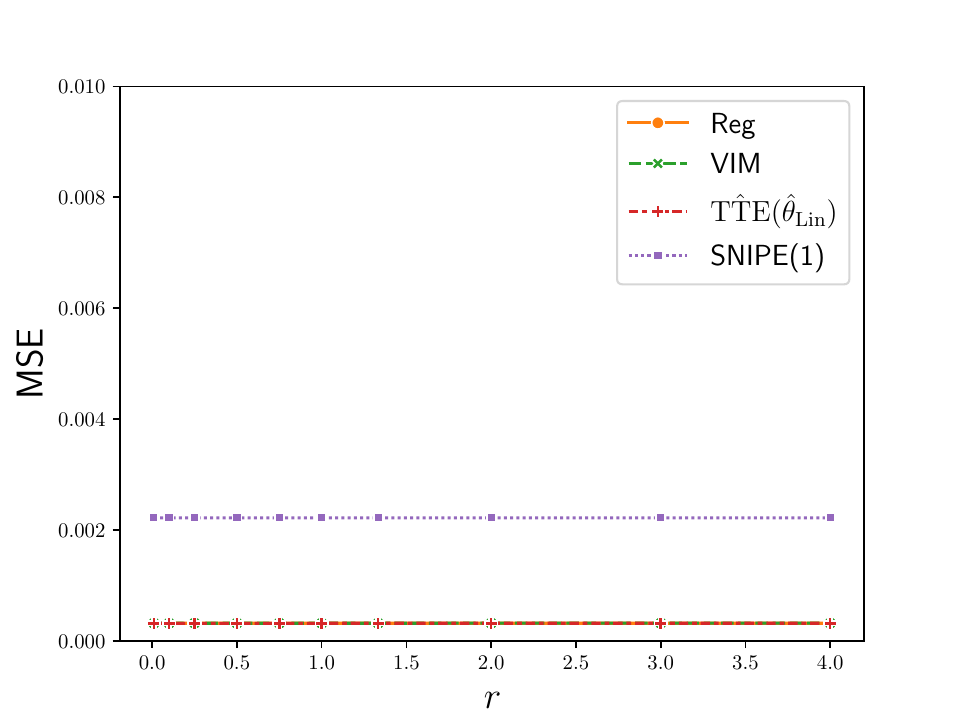}
  \end{subfigure}
  \hfill
  \begin{subfigure}[b]{0.24\textwidth}
    \includegraphics[width=\textwidth]{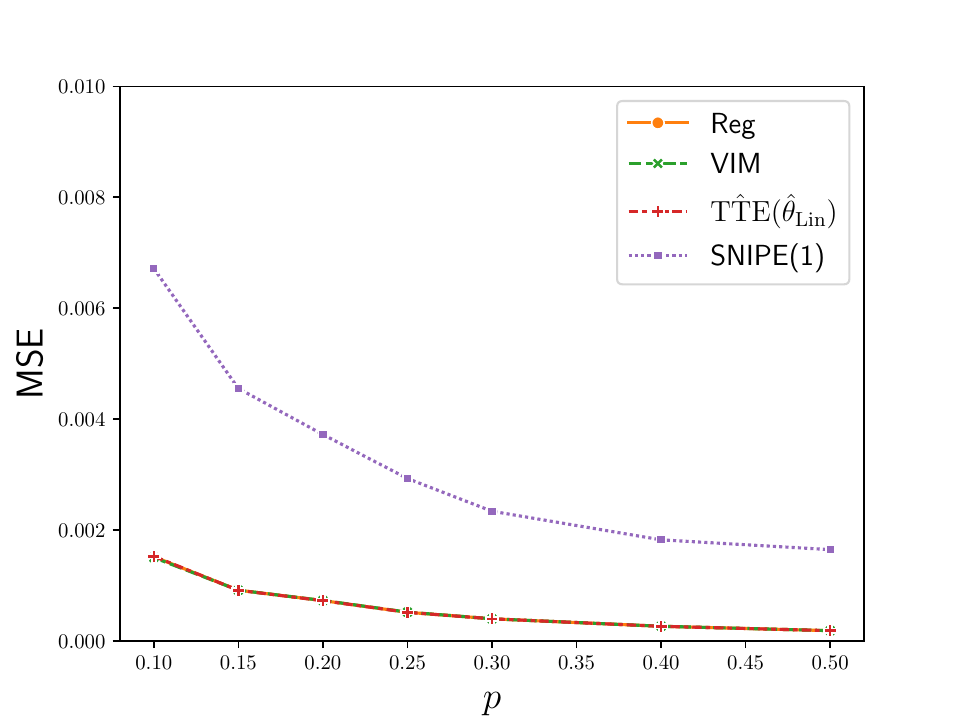}
  \end{subfigure}
  \hfill
  \begin{subfigure}[b]{0.24\textwidth}
    \includegraphics[width=\textwidth]{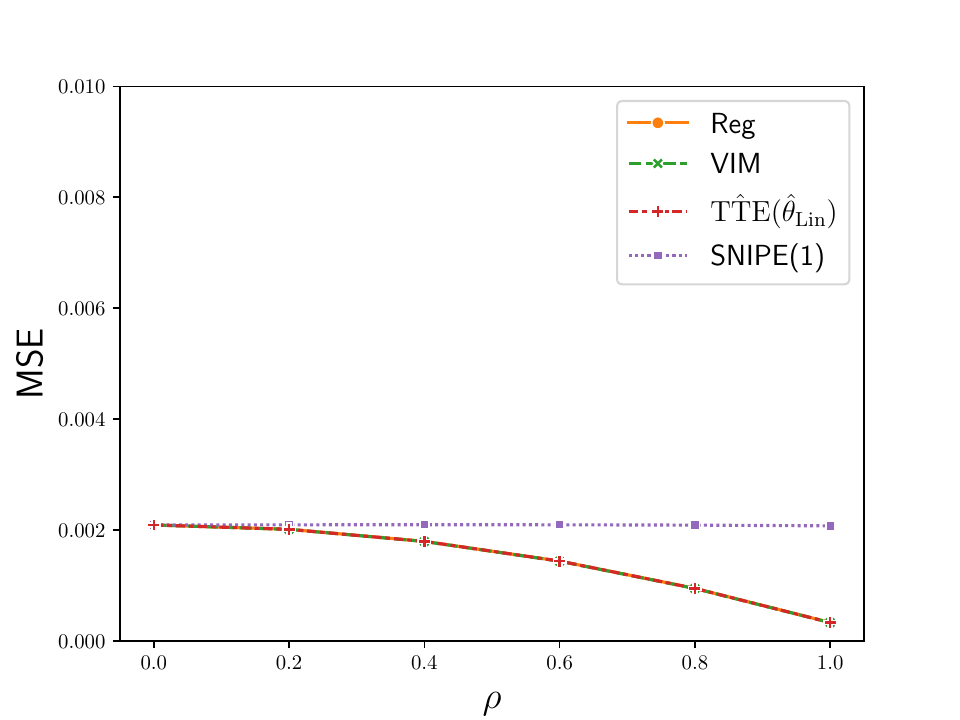}
    \end{subfigure}
  \caption{Relative bias (top row) and mean squared error (MSE; bottom row) of $\Reg$, $\VIM$, $\TTEhat(\hat{\bs \theta}_{\text{Lin}})$, and  $\SNIPE$ under SUTVA with identical treatment probabilities across units. The three covariate-adjusted estimators are asymptotically equivalent, unbiased, and consistently achieve lower MSE than $\SNIPE$.
}
  \label{fig:settingSUTVA}
\end{figure}

We conduct a simulation study to empirically verify that the three adjustment coefficients are asymptotically equivalent under the no-interference setting. Specifically, we generate outcomes and covariates under the no-interference setting with identical treatment probabilities across all units, compute the three adjustment coefficients, and substitute them into the general covariate-adjusted estimator defined in \eqref{equ: ttehat} for comparison. Details of the outcome and covariate generation procedures are provided in Section~\ref{sec:simulation}. By varying the parameters of the data-generating process, we evaluate the resulting relative bias and MSE of each estimator. As shown in Figure~\ref{fig:settingSUTVA}, $\Reg$, $\VIM$, and $\TTEhat(\hat{\bs \theta}_{\text{Lin}})$ are asymptotically equivalent, confirming Proposition~\ref{prop:regvimlin}. Moreover, all three estimators are asymptotically unbiased, and the covariate-adjusted estimators consistently achieve lower MSE than $\SNIPE$.

\subsection{\texorpdfstring{Variance difference between $\TTEhat(\bs 0)$ and $\TTEhat(\rb)$}{Variance difference between tauhat(0) and tauhat(thetaReg)}}
\label{appendix:variance_difference}

When the treatment probability is the same across all units (i.e., $p_i = p$), 
the difference in variances between $\TTEhat(\bs 0)$ and $\TTEhat(\rb)$ is
\begin{align*}
&\Var( \TTEhat(\bs 0)) -  \Var( \TTEhat(\rb)) \nonumber\\
&= -\frac{1}{n^2}\Var\left(\sumn \rb^\top \bX_i \sum_{j \in \mathcal N_i} \frac{Z_j - p_1}{p_1(1-p_1)}\right) + \frac{2}{n^2}\Cov\left(\sumn Y_i\sum_{j \in \mathcal N_i} \frac{Z_j - p_1}{p_1(1-p_1)}, \sumn \rb^\top \bX_i \sum_{j' \in \mathcal N_i} \frac{Z_{j'} - p_1}{p_1(1-p_1)}\right)\\
&= \frac{1}{n^2}\Var\left(\sumn \rb^\top \bX_i \sum_{j \in \mathcal N_i} \frac{Z_j - p_1}{p_1(1-p_1)}\right) \\
&\quad + \frac{2}{n^2}\Cov\left(\sumn \left(Y_i - \rb^\top \bX_i\right)\sum_{j \in \mathcal N_i} \frac{Z_j - p_1}{p_1(1-p_1)}, \sumn \rb^\top \bX_i \sum_{j' \in \mathcal N_i} \frac{Z_{j'} - p_1}{p_1(1-p_1)}\right)\\
&= \frac{1}{n^2}\sumn\Var\left( \rb^\top \bX_i \sum_{j \in \mathcal N_i} \frac{Z_j - p_1}{p_1(1-p_1)}\right) \\
&\quad+ \frac{2}{n^2}\sumn\Cov\left( \left(Y_i - \rb^\top \bX_i\right)\sum_{j \in \mathcal N_i} \frac{Z_j - p_1}{p_1(1-p_1)}, \rb^\top \bX_i \sum_{j' \in \mathcal N_i} \frac{Z_{j'} - p_1}{p_1(1-p_1)}\right)\\
&\quad +\frac{1}{n^2}\sumn\sum_{i': \mathcal N_i \cap \mathcal N_{i'} \neq \varnothing, i'\neq i}\Cov\left( \rb^\top \bX_i \sum_{j \in \mathcal N_i} \frac{Z_j - p_1}{p_1(1-p_1)}, \rb^\top \bX_{i'} \sum_{j' \in \mathcal N_{i'}} \frac{Z_{j'} - p_1}{p_1(1-p_1)}\right) \\
&\quad + \frac{2}{n^2}\sumn\sum_{i': \mathcal N_i \cap \mathcal N_{i'} \neq \varnothing, i'\neq i}\Cov\left(\left(Y_i - \rb^\top \bX_i\right)\sum_{j \in \mathcal N_i} \frac{Z_j - p_1}{p_1(1-p_1)}, \rb^\top \bX_{i'} \sum_{j' \in \mathcal N_{i'}} \frac{Z_{j'} - p_1}{p_1(1-p_1)}\right)\\
&= \frac{1}{n^2}\sumn\Var\left( \rb^\top \bX_i \sum_{j \in \mathcal N_i} \frac{Z_j - p_1}{p_1(1-p_1)}\right) \\
&\quad-\frac{1}{n^2}\sumn\sum_{i': \mathcal N_i \cap \mathcal N_{i'} \neq \varnothing, i'\neq i}\Cov\left( \rb^\top \bX_i \sum_{j \in \mathcal N_i} \frac{Z_j - p_1}{p_1(1-p_1)}, \rb^\top \bX_{i'} \sum_{j' \in \mathcal N_{i'}} \frac{Z_{j'} - p_1}{p_1(1-p_1)}\right) \\
&\quad + \frac{2}{n^2}\sumn\sum_{i': \mathcal N_i \cap \mathcal N_{i'} \neq \varnothing, i'\neq i}\Cov\left(Y_i\sum_{j \in \mathcal N_i} \frac{Z_j - p_1}{p_1(1-p_1)}, \rb^\top \bX_{i'} \sum_{j' \in \mathcal N_{i'}} \frac{Z_{j'} - p_1}{p_1(1-p_1)}\right)\\
&= \frac{1}{p_1(1-p_1)n^2}\sumn\Bigg[ |\mathcal N_i|\rb^\top  \bX_i\bX_i^\top\rb\\
&\quad + \sum_{i': \mathcal N_i \cap \mathcal N_{i'} \neq \varnothing, i'\neq i}  \rb^\top \bX_{i'}\sum_{j \in \mathcal N_i \cap \mathcal N_{i'}} (2(1-2p_1)\alpha_{i, \{j\}} +2(\alpha_{i,\varnothing}+p_1\sum_{j' \in \mathcal N_i}\alpha_{i, \{j'\}}) - \bX_i^\top\rb)\Bigg].
\end{align*}  

\subsection{Construction of covariates}
\label{appendix:construction_covariates}

In our framework, we do not impose strong assumptions on the covariates $\bs X_i$. As shown in Section~\ref{section:theory}, to establish that $\VIM$ has smaller asymptotic variance than $\SNIPE$,  we only require Assumptions~\ref{as:bounded}--\ref{as:variance} on the covariates $\bs X_i$.  These are mild regularity conditions that are typically satisfied for a wide range of choices of $\bs X_i$.

Importantly, we do not assume that the covariates $\bs X_i$ are independent or identically distributed across units; they may be dependent.  They may also depend on the interference network.  The key requirement is that the covariates are independent of the treatment assignment vector $Z$.

Given a set of raw covariates $\bs X^{\operatorname{raw}}$, below we present several possible ways of constructing the covariates $\bs X$.
\begin{coexample}[Raw covariates]
We can directly use the raw covariates for $\Reg$ or $\VIM$: $\bs X = \bs X^{\operatorname{raw}}$. 
For instance, if the outcome $Y$ represents a health outcome and the raw covariates include a patient’s medical history (e.g., chronic conditions, prior hospitalizations), then it is natural to adjust for these covariates directly.
\end{coexample}

\begin{coexample}[Transformation of raw covariates]
We can also construct transformed covariates by applying a non-linear feature map to the raw covariates. Common approaches include polynomial terms, spline bases, and interaction terms, but one can also use kernels, radial basis expansions, or neural network–style transformations such as ReLU features. These transformations are particularly helpful if the outcome $Y$ depends on $\bs X^{\operatorname{raw}}$ in a non-linear manner. 
For example, again in a healthcare setting, age may have a non-linear effect (with risk accelerating at older ages), BMI may interact with blood pressure, and kernel or ReLU features may help capture more complex dependencies. In such cases, using transformed covariates $\bs X = \Phi(\bs X^{\operatorname{raw}})$ can substantially improve adjustment.
\end{coexample}

\begin{coexample}[Network-based covariates]
We can also use network information to construct covariates. 
For instance, one may define $\bs X_i = |\mathcal{N}_i|$, the degree of unit $i$. 
If the outcome concerns how active a user is on a social network platform, then the number of friends is a natural predictor. 
More sophisticated functions of the network can also be used. 
For example, \citet{li2022random} show that adjusting for the first few eigenvectors of the adjacency matrix can substantially reduce the variance of causal effect estimators under neighborhood interference.
\end{coexample}

\begin{coexample}[Network-based and raw covariates]
We can also combine network information with raw covariates. 
A natural approach is to use graph neural networks (GNNs), which iteratively aggregate information from a unit’s neighbors together with its own raw covariates \citep{scarselli2008graph}. 
\end{coexample}

\begin{coexample}[Pre-experiment outcomes]
We may also use outcomes measured prior to the experiment as covariates. 
Such pre-experiment outcomes often have strong predictive power for the post-experiment outcome and can substantially improve precision when used for adjustment.
\end{coexample}

\section{Additional Numerical Results}
\label{appendix:additional_simulation}

\begin{table}[H]
\centering
\small
\setlength{\tabcolsep}{2.2pt}
\renewcommand{\arraystretch}{0.9}
\caption{True treatment effect and average confidence interval lengths for the Erd\H{o}s--R\'enyi design. CI Len (Old) corresponds to Wald-type confidence intervals constructed using the conservative variance estimator of \citet{CortezRodriguezEichhornYu+2023}, whereas CI Len (New) corresponds to those constructed using the proposed variance estimator.}
\begin{minipage}[t]{0.49\textwidth}
\centering
\textbf{Size}\\[0.3em]
\begin{tabular}{lccc}
\hline
Method & $\TTE$ & CI Len (Old) & CI Len (New) \\
\hline
$\mathrm{Reg}_{5000}$ & 14.756 & 549.213 & 14.295 \\
$\mathrm{VIM}_{5000}$ & 14.756 & 549.213 & 14.288 \\
$\mathrm{SNIPE}(1)_{5000}$ & 14.756 & 549.291 & 17.057 \\
$\mathrm{Reg}_{6000}$ & 14.958 & 571.383 & 13.361 \\
$\mathrm{VIM}_{6000}$ & 14.958 & 571.383 & 13.356 \\
$\mathrm{SNIPE}(1)_{6000}$ & 14.958 & 571.447 & 15.871 \\
$\mathrm{Reg}_{7000}$ & 14.850 & 480.574 & 12.424 \\
$\mathrm{VIM}_{7000}$ & 14.850 & 480.574 & 12.419 \\
$\mathrm{SNIPE}(1)_{7000}$ & 14.850 & 480.640 & 14.767 \\
$\mathrm{Reg}_{8000}$ & 15.171 & 393.372 & 11.975 \\
$\mathrm{VIM}_{8000}$ & 15.171 & 393.372 & 11.973 \\
$\mathrm{SNIPE}(1)_{8000}$ & 15.171 & 393.447 & 14.238 \\
$\mathrm{Reg}_{9000}$ & 14.856 & 396.207 & 11.028 \\
$\mathrm{VIM}_{9000}$ & 14.856 & 396.207 & 11.026 \\
$\mathrm{SNIPE}(1)_{9000}$ & 14.856 & 396.271 & 13.149 \\
$\mathrm{Reg}_{10000}$ & 14.971 & 383.016 & 10.551 \\
$\mathrm{VIM}_{10000}$ & 14.971 & 383.016 & 10.549 \\
$\mathrm{SNIPE}(1)_{10000}$ & 14.971 & 383.075 & 12.512 \\
\hline
\end{tabular}
\\[0.8em]
\textbf{Treatment Probability}\\[0.3em]
\begin{tabular}{lccc}
\hline
Method & $\TTE$ & CI Len (Old) & CI Len (New) \\
\hline
$\mathrm{Reg}_{0.1}$ & 14.971 & 2347.190 & 12.715 \\
$\mathrm{VIM}_{0.1}$ & 14.971 & 2347.190 & 12.706 \\
$\mathrm{SNIPE}(1)_{0.1}$ & 14.971 & 2347.200 & 14.428 \\
$\mathrm{Reg}_{0.15}$ & 14.971 & 1392.654 & 11.903 \\
$\mathrm{VIM}_{0.15}$ & 14.971 & 1392.654 & 11.898 \\
$\mathrm{SNIPE}(1)_{0.15}$ & 14.971 & 1392.668 & 13.506 \\
$\mathrm{Reg}_{0.2}$ & 14.971 & 892.997 & 11.390 \\
$\mathrm{VIM}_{0.2}$ & 14.971 & 892.997 & 11.386 \\
$\mathrm{SNIPE}(1)_{0.2}$ & 14.971 & 893.019 & 13.011 \\
$\mathrm{Reg}_{0.25}$ & 14.971 & 618.446 & 11.004 \\
$\mathrm{VIM}_{0.25}$ & 14.971 & 618.446 & 11.001 \\
$\mathrm{SNIPE}(1)_{0.25}$ & 14.971 & 618.479 & 12.698 \\
$\mathrm{Reg}_{0.3}$ & 14.971 & 464.633 & 10.741 \\
$\mathrm{VIM}_{0.3}$ & 14.971 & 464.633 & 10.738 \\
$\mathrm{SNIPE}(1)_{0.3}$ & 14.971 & 464.678 & 12.552 \\
$\mathrm{Reg}_{0.4}$ & 14.971 & 351.368 & 10.463 \\
$\mathrm{VIM}_{0.4}$ & 14.971 & 351.368 & 10.461 \\
$\mathrm{SNIPE}(1)_{0.4}$ & 14.971 & 351.438 & 12.600 \\
$\mathrm{Reg}_{0.5}$ & 14.971 & 417.907 & 10.886 \\
$\mathrm{VIM}_{0.5}$ & 14.971 & 417.907 & 10.884 \\
$\mathrm{SNIPE}(1)_{0.5}$ & 14.971 & 417.979 & 13.397 \\
\hline
\end{tabular}
\end{minipage}
\hfill
\begin{minipage}[t]{0.49\textwidth}
\centering
\textbf{Ratio}\\[0.3em]
\begin{tabular}{lccc}
\hline
Method & $\TTE$ & CI Len (Old) & CI Len (New) \\
\hline
$\mathrm{Reg}_{0.01}$ & 5.052 & 253.140 & 3.396 \\
$\mathrm{VIM}_{0.01}$ & 5.052 & 253.140 & 3.395 \\
$\mathrm{SNIPE}(1)_{0.01}$ & 5.052 & 253.153 & 4.247 \\
$\mathrm{Reg}_{0.1}$ & 5.501 & 257.556 & 3.690 \\
$\mathrm{VIM}_{0.1}$ & 5.501 & 257.556 & 3.690 \\
$\mathrm{SNIPE}(1)_{0.1}$ & 5.501 & 257.570 & 4.596 \\
$\mathrm{Reg}_{0.25}$ & 6.249 & 265.340 & 4.196 \\
$\mathrm{VIM}_{0.25}$ & 6.249 & 265.340 & 4.196 \\
$\mathrm{SNIPE}(1)_{0.25}$ & 6.249 & 265.358 & 5.190 \\
$\mathrm{Reg}_{0.5}$ & 7.495 & 279.385 & 5.067 \\
$\mathrm{VIM}_{0.5}$ & 7.495 & 279.385 & 5.066 \\
$\mathrm{SNIPE}(1)_{0.5}$ & 7.495 & 279.408 & 6.203 \\
$\mathrm{Reg}_{0.75}$ & 8.741 & 294.601 & 5.959 \\
$\mathrm{VIM}_{0.75}$ & 8.741 & 294.601 & 5.959 \\
$\mathrm{SNIPE}(1)_{0.75}$ & 8.741 & 294.630 & 7.235 \\
$\mathrm{Reg}_{1}$ & 9.987 & 310.816 & 6.865 \\
$\mathrm{VIM}_{1}$ & 9.987 & 310.816 & 6.864 \\
$\mathrm{SNIPE}(1)_{1}$ & 9.987 & 310.850 & 8.279 \\
$\mathrm{Reg}_{1.33333}$ & 11.648 & 333.736 & 8.085 \\
$\mathrm{VIM}_{1.33333}$ & 11.648 & 333.736 & 8.084 \\
$\mathrm{SNIPE}(1)_{1.33333}$ & 11.648 & 333.778 & 9.682 \\
$\mathrm{Reg}_{2}$ & 14.971 & 383.016 & 10.551 \\
$\mathrm{VIM}_{2}$ & 14.971 & 383.016 & 10.549 \\
$\mathrm{SNIPE}(1)_{2}$ & 14.971 & 383.075 & 12.512 \\
$\mathrm{Reg}_{3}$ & 19.956 & 462.734 & 14.279 \\
$\mathrm{VIM}_{3}$ & 19.956 & 462.734 & 14.276 \\
$\mathrm{SNIPE}(1)_{3}$ & 19.956 & 462.818 & 16.785 \\
$\mathrm{Reg}_{4}$ & 24.940 & 546.693 & 18.022 \\
$\mathrm{VIM}_{4}$ & 24.940 & 546.693 & 18.018 \\
$\mathrm{SNIPE}(1)_{4}$ & 24.940 & 546.802 & 21.074 \\
\hline
\end{tabular}
\\[0.8em]
\textbf{\% of Covariates Observed}\\[0.3em]
\begin{tabular}{lccc}
\hline
Method & $\TTE$ & CI Len (Old) & CI Len (New) \\
\hline
$\mathrm{Reg}_{0}$ & 15.356 & 378.228 & 12.808 \\
$\mathrm{VIM}_{0}$ & 15.356 & 378.228 & 12.805 \\
$\mathrm{SNIPE}(1)_{0}$ & 15.356 & 378.228 & 12.809 \\
$\mathrm{Reg}_{0.2}$ & 15.334 & 379.579 & 12.724 \\
$\mathrm{VIM}_{0.2}$ & 15.334 & 379.579 & 12.721 \\
$\mathrm{SNIPE}(1)_{0.2}$ & 15.334 & 379.582 & 12.798 \\
$\mathrm{Reg}_{0.4}$ & 15.299 & 381.057 & 12.479 \\
$\mathrm{VIM}_{0.4}$ & 15.299 & 381.057 & 12.476 \\
$\mathrm{SNIPE}(1)_{0.4}$ & 15.299 & 381.067 & 12.770 \\
$\mathrm{Reg}_{0.6}$ & 15.248 & 382.577 & 12.065 \\
$\mathrm{VIM}_{0.6}$ & 15.248 & 382.577 & 12.062 \\
$\mathrm{SNIPE}(1)_{0.6}$ & 15.248 & 382.598 & 12.726 \\
$\mathrm{Reg}_{0.8}$ & 15.172 & 384.037 & 11.459 \\
$\mathrm{VIM}_{0.8}$ & 15.172 & 384.037 & 11.457 \\
$\mathrm{SNIPE}(1)_{0.8}$ & 15.172 & 384.075 & 12.664 \\
$\mathrm{Reg}_{1}$ & 14.971 & 383.016 & 10.551 \\
$\mathrm{VIM}_{1}$ & 14.971 & 383.016 & 10.549 \\
$\mathrm{SNIPE}(1)_{1}$ & 14.971 & 383.075 & 12.512 \\
\hline
\end{tabular}
\end{minipage}
\end{table}

\begin{table}[H]
\centering
\small
\setlength{\tabcolsep}{2.2pt}
\renewcommand{\arraystretch}{0.9}
\caption{True treatment effect and average confidence interval lengths for the soft random geometric graph design. CI Len (Old) corresponds to Wald-type confidence intervals constructed using the conservative variance estimator of \citet{CortezRodriguezEichhornYu+2023}, whereas CI Len (New) corresponds to those constructed using the proposed variance estimator.}
\begin{minipage}[t]{0.49\textwidth}
\centering
\textbf{Size}\\[0.3em]
\begin{tabular}{lccc}
\hline
Method & $\TTE$ & CI Len (Old) & CI Len (New) \\
\hline
$\mathrm{Reg}_{5000}$ & 14.730 & 2.06e+05 & 30.708 \\
$\mathrm{VIM}_{5000}$ & 14.730 & 2.06e+05 & 30.311 \\
$\mathrm{SNIPE}(1)_{5000}$ & 14.730 & 2.06e+05 & 43.715 \\
$\mathrm{Reg}_{6000}$ & 14.540 & 5.44e+06 & 42.449 \\
$\mathrm{VIM}_{6000}$ & 14.540 & 5.44e+06 & 41.644 \\
$\mathrm{SNIPE}(1)_{6000}$ & 14.540 & 5.44e+06 & 63.177 \\
$\mathrm{Reg}_{7000}$ & 16.538 & 2.22e+08 & 51.316 \\
$\mathrm{VIM}_{7000}$ & 16.538 & 2.22e+08 & 50.291 \\
$\mathrm{SNIPE}(1)_{7000}$ & 16.538 & 2.22e+08 & 76.078 \\
$\mathrm{Reg}_{8000}$ & 14.588 & 4.56e+08 & 58.269 \\
$\mathrm{VIM}_{8000}$ & 14.588 & 4.56e+08 & 57.068 \\
$\mathrm{SNIPE}(1)_{8000}$ & 14.588 & 4.56e+08 & 88.129 \\
$\mathrm{Reg}_{9000}$ & 17.144 & 5.08e+09 & 68.446 \\
$\mathrm{VIM}_{9000}$ & 17.144 & 5.08e+09 & 66.919 \\
$\mathrm{SNIPE}(1)_{9000}$ & 17.144 & 5.08e+09 & 103.615 \\
$\mathrm{Reg}_{10000}$ & 15.535 & 1.89e+11 & 75.351 \\
$\mathrm{VIM}_{10000}$ & 15.535 & 1.89e+11 & 73.662 \\
$\mathrm{SNIPE}(1)_{10000}$ & 15.535 & 1.89e+11 & 116.188 \\
\hline
\end{tabular}
\\[0.8em]
\textbf{Treatment Probability}\\[0.3em]
\begin{tabular}{lccc}
\hline
Method & $\TTE$ & CI Len (Old) & CI Len (New) \\
\hline
$\mathrm{Reg}_{0.1}$ & 15.535 & 3.29e+12 & 100.430 \\
$\mathrm{VIM}_{0.1}$ & 15.535 & 3.29e+12 & 100.172 \\
$\mathrm{SNIPE}(1)_{0.1}$ & 15.535 & 3.29e+12 & 117.710 \\
$\mathrm{Reg}_{0.15}$ & 15.535 & 1.28e+12 & 95.453 \\
$\mathrm{VIM}_{0.15}$ & 15.535 & 1.28e+12 & 95.045 \\
$\mathrm{SNIPE}(1)_{0.15}$ & 15.535 & 1.28e+12 & 115.063 \\
$\mathrm{Reg}_{0.2}$ & 15.535 & 6.42e+11 & 89.881 \\
$\mathrm{VIM}_{0.2}$ & 15.535 & 6.42e+11 & 89.239 \\
$\mathrm{SNIPE}(1)_{0.2}$ & 15.535 & 6.42e+11 & 113.236 \\
$\mathrm{Reg}_{0.25}$ & 15.535 & 3.82e+11 & 84.979 \\
$\mathrm{VIM}_{0.25}$ & 15.535 & 3.82e+11 & 84.025 \\
$\mathrm{SNIPE}(1)_{0.25}$ & 15.535 & 3.82e+11 & 113.069 \\
$\mathrm{Reg}_{0.3}$ & 15.535 & 2.53e+11 & 80.852 \\
$\mathrm{VIM}_{0.3}$ & 15.535 & 2.53e+11 & 79.541 \\
$\mathrm{SNIPE}(1)_{0.3}$ & 15.535 & 2.53e+11 & 114.798 \\
$\mathrm{Reg}_{0.4}$ & 15.535 & 1.62e+11 & 71.339 \\
$\mathrm{VIM}_{0.4}$ & 15.535 & 1.62e+11 & 69.316 \\
$\mathrm{SNIPE}(1)_{0.4}$ & 15.535 & 1.62e+11 & 119.369 \\
$\mathrm{Reg}_{0.5}$ & 15.535 & 1.82e+11 & 68.761 \\
$\mathrm{VIM}_{0.5}$ & 15.535 & 1.82e+11 & 66.200 \\
$\mathrm{SNIPE}(1)_{0.5}$ & 15.535 & 1.82e+11 & 130.948 \\
\hline
\end{tabular}
\end{minipage}
\hfill
\begin{minipage}[t]{0.49\textwidth}
\centering
\textbf{Ratio}\\[0.3em]
\begin{tabular}{lccc}
\hline
Method & $\TTE$ & CI Len (Old) & CI Len (New) \\
\hline
$\mathrm{Reg}_{0.01}$ & 5.055 & 1.04e+11 & 8.092 \\
$\mathrm{VIM}_{0.01}$ & 5.055 & 1.04e+11 & 8.084 \\
$\mathrm{SNIPE}(1)_{0.01}$ & 5.055 & 1.04e+11 & 17.572 \\
$\mathrm{Reg}_{0.1}$ & 5.529 & 1.07e+11 & 9.679 \\
$\mathrm{VIM}_{0.1}$ & 5.529 & 1.07e+11 & 9.644 \\
$\mathrm{SNIPE}(1)_{0.1}$ & 5.529 & 1.07e+11 & 21.301 \\
$\mathrm{Reg}_{0.25}$ & 6.319 & 1.13e+11 & 13.637 \\
$\mathrm{VIM}_{0.25}$ & 6.319 & 1.13e+11 & 13.498 \\
$\mathrm{SNIPE}(1)_{0.25}$ & 6.319 & 1.13e+11 & 28.104 \\
$\mathrm{Reg}_{0.5}$ & 7.636 & 1.23e+11 & 21.640 \\
$\mathrm{VIM}_{0.5}$ & 7.636 & 1.23e+11 & 21.276 \\
$\mathrm{SNIPE}(1)_{0.5}$ & 7.636 & 1.23e+11 & 40.160 \\
$\mathrm{Reg}_{0.75}$ & 8.952 & 1.34e+11 & 30.304 \\
$\mathrm{VIM}_{0.75}$ & 8.952 & 1.34e+11 & 29.706 \\
$\mathrm{SNIPE}(1)_{0.75}$ & 8.952 & 1.34e+11 & 52.579 \\
$\mathrm{Reg}_{1}$ & 10.269 & 1.44e+11 & 39.178 \\
$\mathrm{VIM}_{1}$ & 10.269 & 1.44e+11 & 38.326 \\
$\mathrm{SNIPE}(1)_{1}$ & 10.269 & 1.44e+11 & 65.155 \\
$\mathrm{Reg}_{1.33333}$ & 12.024 & 1.59e+11 & 51.153 \\
$\mathrm{VIM}_{1.33333}$ & 12.024 & 1.59e+11 & 49.988 \\
$\mathrm{SNIPE}(1)_{1.33333}$ & 12.024 & 1.59e+11 & 82.043 \\
$\mathrm{Reg}_{2}$ & 15.535 & 1.89e+11 & 75.183 \\
$\mathrm{VIM}_{2}$ & 15.535 & 1.89e+11 & 73.516 \\
$\mathrm{SNIPE}(1)_{2}$ & 15.535 & 1.89e+11 & 116.004 \\
$\mathrm{Reg}_{3}$ & 20.801 & 2.37e+11 & 111.423 \\
$\mathrm{VIM}_{3}$ & 20.801 & 2.37e+11 & 108.809 \\
$\mathrm{SNIPE}(1)_{3}$ & 20.801 & 2.37e+11 & 167.130 \\
$\mathrm{Reg}_{4}$ & 26.067 & 2.85e+11 & 147.784 \\
$\mathrm{VIM}_{4}$ & 26.067 & 2.85e+11 & 144.168 \\
$\mathrm{SNIPE}(1)_{4}$ & 26.067 & 2.85e+11 & 218.338 \\
\hline
\end{tabular}
\\[0.8em]
\textbf{\% of Covariates Observed}\\[0.3em]
\begin{tabular}{lccc}
\hline
Method & $\TTE$ & CI Len (Old) & CI Len (New) \\
\hline
$\mathrm{Reg}_{0}$ & 16.886 & 5.59e+12 & 124.380 \\
$\mathrm{VIM}_{0}$ & 16.886 & 5.59e+12 & 124.326 \\
$\mathrm{SNIPE}(1)_{0}$ & 16.886 & 5.59e+12 & 124.381 \\
$\mathrm{Reg}_{0.2}$ & 16.469 & 1.46e+13 & 125.432 \\
$\mathrm{VIM}_{0.2}$ & 16.469 & 1.46e+13 & 112.903 \\
$\mathrm{SNIPE}(1)_{0.2}$ & 16.469 & 1.46e+13 & 126.810 \\
$\mathrm{Reg}_{0.4}$ & 16.011 & 1.92e+11 & 106.404 \\
$\mathrm{VIM}_{0.4}$ & 16.011 & 1.92e+11 & 83.945 \\
$\mathrm{SNIPE}(1)_{0.4}$ & 16.011 & 1.92e+11 & 111.396 \\
$\mathrm{Reg}_{0.6}$ & 15.358 & 1.70e+09 & 81.654 \\
$\mathrm{VIM}_{0.6}$ & 15.358 & 1.70e+09 & 63.243 \\
$\mathrm{SNIPE}(1)_{0.6}$ & 15.358 & 1.70e+09 & 91.400 \\
$\mathrm{Reg}_{0.8}$ & 15.478 & 7.81e+08 & 64.001 \\
$\mathrm{VIM}_{0.8}$ & 15.478 & 7.81e+08 & 53.153 \\
$\mathrm{SNIPE}(1)_{0.8}$ & 15.478 & 7.81e+08 & 80.176 \\
$\mathrm{Reg}_{1}$ & 15.535 & 1.89e+11 & 75.351 \\
$\mathrm{VIM}_{1}$ & 15.535 & 1.89e+11 & 73.662 \\
$\mathrm{SNIPE}(1)_{1}$ & 15.535 & 1.89e+11 & 116.188 \\
\hline
\end{tabular}
\end{minipage}
\end{table}

\section{Proofs of Theorems and Propositions}\label{sec:main_proof}
Throughout the proofs, we make use of Lemma 3 in \cite{CortezRodriguezEichhornYu+2023}, which states that for any subset $\mathcal{S} \subseteq [n]$, $|g(\mathcal{S})| \leq 1$. We also employ the standard combinatorial inequality $\sum_{k=0}^{\beta}\binom{d}{k} \leq \left(\frac{e d}{\beta}\right)^{\beta}$, which holds for any integers $d \geq \beta \geq 1$.

\subsection{Proof of Proposition \ref{prop:consistrb}}\label{sec:D.1}
First, we show that $\rb$ is bounded from above. To start with, we provide a lower bound and an upper bound for $\E(\omega_i^2)$. For each unit $i$, $i = 1, \ldots,n$
\begin{align*}
    \E(\omega_i^2) &= \E\left(\sum_{\mathcal S \in \mathcal S_i^\beta}\sum_{\mathcal T \in \mathcal S_i^\beta} g(\mathcal S)g(\mathcal T) \prod_{j \in \mathcal S}\frac{Z_{j} - p_j}{p_j(1-p_j)}\prod_{t \in \mathcal T} \frac{Z_{t} - p_t}{p_t(1-p_t)}\right)\\
    &=\sum_{\mathcal S \in \mathcal S_i^\beta} g^2(\mathcal S) \prod_{j \in \mathcal S}\frac{1}{p_j(1-p_j)}. 
\end{align*}
For each unit, the set of neighbors always includes the unit itself and contains at most $d_{\text{in}}$ units. Therefore, the above equation can be bounded by
\begin{align*}
4 \leq \E(\omega_i^2) \leq \left(\frac{ed_{\text{in}}}{\beta p(1-p)}\right)^\beta.  
\end{align*}
To briefly step aside from the main proof, we note that the above result with Assumption \ref{as:invertible} leads to the boundedness of $\rb$. This is because
\begin{align*}
    \|\rb\| &= \left\|\E\left(\frac{1}{n}\sumn \omega_i^2\bX_i\bX_i^\top\right)^{-1}\E\left(\frac{1}{n}\sumn \omega_i^2Y_i\bX_i\right)\right\|\\
    &= \left\|\left(\frac{1}{n}\sumn \E(\omega_i^2)\bX_i\bX_i^\top\right)^{-1}\left(\frac{1}{n}\sumn \E(\omega_i^2)Y_i\bX_i\right)\right\|\\
    &\leq \lambda^{-1}_{\min}\left(\frac{1}{n}\sumn \E(\omega_i^2)\bX_i\bX_i^\top\right)\left\|\frac{1}{n}\sumn \E(\omega_i^2)Y_i\bX_i\right\|\\
    & \leq \frac{1}{4}\left(\frac{ed_{\text{in}}}{\beta p(1-p)}\right)^\beta\lambda^{-1}_{\min}\left(\frac{1}{n}\sumn \bX_i\bX_i^\top\right)\left\|\frac{1}{n}\sumn Y_i\bX_i\right\|\\
    &\leq \frac{1}{4 c_{\lambda_{\min}}}\left(\frac{ed_{\text{in}}}{\beta p(1-p)}\right)^\beta X_{\max} Y_{\max}.
\end{align*}
The above argument shows that $\rb$ is bounded from above, which correspond to Assumption \hyperref[as:variance(i)]{5(i)} is well motivated. Based on Assumption \hyperref[as:variance(i)]{5(i)}, $\rb$ has a finite limit.

Now, to prove that $\rbhat - \rb = o_p(1)$, under Assumption \ref{as:invertible}, it suffices to show that
\begin{align}
   \left\|\frac{1}{n} \sumn \omega_i ^2\bX_i\bX_i^\top - \E\left(\frac{1}{n}\sumn \omega_i^2\bX_i\bX_i^\top\right)\right\| &\overset{p}{\to} 0,\label{equ: consistXX}\\
   \frac{1}{n} \sumn \omega_i^2Y_i\bX_i -\E\left(\frac{1}{n}\sumn \omega_i^2Y_i\bX_i\right) &\overset{p}{\to} 0.\label{equ: consistXY}
\end{align}
Firstly, we demonstrate \eqref{equ: consistXY}.
\begin{align*}
\E\left\| \frac{1}{n} \sumn \omega_i^2Y_i\bX_i -\E\left(\frac{1}{n}\sumn \omega_i^2Y_i\bX_i\right)\right\|^2 = \text{tr} \left[\Var\left(\frac{1}{n} \sumn \omega_i^2Y_i\bX_i\right)\right]  \le d_{\bX} \left\|\Var\left(\frac{1}{n} \sumn \omega_i^2Y_i\bX_i\right)\right\|. 
\end{align*}
Then to bound the above operator norm,
\begin{align*}
&\left\|\Var\left(\frac{1}{n} \sumn \omega_i^2Y_i\bX_i\right)\right\| = \left\|\frac{1}{n^2} \sumn\sum_{i':  \mathcal N_i \cap \mathcal N_{i'} \neq \varnothing}\Cov\left(\omega_{i}^2Y_i, \omega_{i'}^2Y_{i'}\right)\bX_i\bX_{i'}^\top\right\|\\
& = \Bigg\|\frac{1}{n^2} \sumn\sum_{i':  \mathcal N_i \cap \mathcal N_{i'} \neq \varnothing}\sum_{\mathcal S\in \mathcal S_i^\beta}\sum_{\mathcal T\in \mathcal S_i^\beta}\sum_{\mathcal S'\in \mathcal S_{i'}^\beta}\sum_{\mathcal T'\in \mathcal S_{i'}^\beta}\sum_{\mathcal U\in \mathcal S_{i}^\beta}\sum_{\mathcal U'\in \mathcal S_{i'}^\beta}g(\mathcal S)g(\mathcal T)g(\mathcal S')g(\mathcal T')\alpha_{i, \mathcal U}\alpha_{i', \mathcal U'}\\
&\quad \times \Cov\left(\prod_{s \in \mathcal S}\frac{Z_s - p_s}{p_s(1-p_s)}\prod_{t \in \mathcal T}\frac{Z_t - p_t}{p_t(1-p_t)}\prod_{u \in \mathcal U}Z_u, \prod_{s' \in \mathcal S'}\frac{Z_{s'} - p_{s'}}{p_{s'}(1-p_{s'})}\prod_{t' \in \mathcal T'}\frac{Z_{t'} - p_{t'}}{p_{t'}(1-p_{t'})}\prod_{u' \in \mathcal U'}Z_{u'} \right)\bX_i\bX_{i'}^\top\Bigg\|\\
&\leq \frac{X_{\max}^2}{n^2} \sumn\sum_{i':  \mathcal N_i \cap \mathcal N_{i'} \neq \varnothing}\sum_{\mathcal U\in \mathcal S_{i}^\beta}|\alpha_{i, \mathcal U}|\sum_{\mathcal U'\in \mathcal S_{i'}^\beta}  |\alpha_{i', \mathcal U'}|\sum_{\mathcal S\in\mathcal S_i^\beta}\sum_{\mathcal T\in\mathcal S_i^\beta}\sum_{\mathcal S'\in\mathcal S_{i'}^\beta}\sum_{\mathcal T'\in\mathcal S_{i'}^\beta}\\
& \quad \times \left|\Cov\left(\prod_{s \in \mathcal S}\frac{Z_s - p_s}{p_s(1-p_s)}\prod_{t \in \mathcal T}\frac{Z_t - p_t}{p_t(1-p_t)}\prod_{u \in \mathcal U}Z_u, \prod_{s' \in \mathcal S'}\frac{Z_{s'} - p_{s'}}{p_{s'}(1-p_{s'})}\prod_{t' \in \mathcal T'}\frac{Z_{t'} - p_{t'}}{p_{t'}(1-p_{t'})}\prod_{u' \in \mathcal U'}Z_{u'} \right)\right|\\
&\leq \frac{X_{\max}^2}{n^2} \sumn\sum_{i':  \mathcal N_i \cap \mathcal N_{i'} \neq \varnothing}\sum_{\mathcal U\in \mathcal S_{i}^\beta}|\alpha_{i, \mathcal U}|\sum_{\mathcal U'\in \mathcal S_{i'}^\beta}  |\alpha_{i', \mathcal U'}|\sum_{\mathcal S\in\mathcal S_i^\beta}\sum_{\mathcal T\in\mathcal S_i^\beta}\sum_{\mathcal S'\in\mathcal S_{i'}^\beta}\sum_{\mathcal T'\in\mathcal S_{i'}^\beta} \I(\mathcal T' \subseteq (\mathcal S \cup \mathcal S' \cup \mathcal T))\left(\frac{p^3+(1-p^3)}{p^3(1-p)^3)}\right)^{5\beta}\\
&\leq \frac{2^{3\beta}X_{\max}^2}{n^2} \sumn\sum_{i':  \mathcal N_i \cap \mathcal N_{i'} \neq \varnothing}\sum_{\mathcal U\in \mathcal S_{i}^\beta}|\alpha_{i, \mathcal U}|\sum_{\mathcal U'\in \mathcal S_{i'}^\beta}  |\alpha_{i', \mathcal U'}|\sum_{\mathcal S\in\mathcal S_i^\beta}\sum_{\mathcal T\in\mathcal S_i^\beta}\sum_{\mathcal S'\in\mathcal S_{i'}^\beta}\left(\frac{p^3+(1-p)^3}{p^3(1-p)^3}\right)^{5\beta}\\
&\leq \frac{2^{3\beta}d_{\text{in}}d_{\text{out}}Y_{\max}^2 X_{\max}^2}{n} \left(\frac{ed_{\text{in}}}{\beta}\right)^{3\beta} \cdot \left(\frac{p^3+(1-p)^3}{p^3(1-p)^3}\right)^{5\beta}= O\left(\frac{1}{n}\right).
\end{align*}
The last equality is based on Assumption \ref{as:bounded} and the assumption that the maximum of the in- and out-degrees of the graph $d$ is of constant order with respect to $n$. Then
$$
\E\left\| \frac{1}{n} \sumn \omega_i^2Y_i\bX_i -\E\left(\frac{1}{n}\sumn \omega_i^2Y_i\bX_i\right)\right\|^2 = O\left(\frac{1}{n}\right).
$$
Therefore, we have the convergence in probability stated in \eqref{equ: consistXY}. \eqref{equ: consistXX} can be derived using similar procedures hence we omit it here.

\subsection{Proof of Proposition \ref{prop:consistent}}
Firstly, we briefly step aside from the main proof and show that $\estchat$ is bounded from above. Under Assumption \ref{as:invertible}, we have
\begin{align*}
 \|\estchat\| &= \left\|\E\left(\frac{1}{n^2}\sumn\sum_{i':  \mathcal N_i \cap \mathcal N_{i'} \neq \varnothing} \omega_i\omega_{i'} \bX_i\bX_{i'}^\top\right)^{-1}\E\left(\frac{1}{n^2}\sumn\sum_{i':  \mathcal N_i \cap \mathcal N_{i'} \neq \varnothing} \omega_i\omega_{i'} \bX_i Y_{i'}\right)\right\|\\
 &= \left\|\left(\frac{1}{n}\sumn\sum_{i':  \mathcal N_i \cap \mathcal N_{i'} \neq \varnothing} \sum_{\mathcal S \in\mathcal S_i^\beta \cap S_{i'}^\beta } g(\mathcal S)^2\prod_{j \in \mathcal{S}} \frac{1}{p_j(1-p_j)}\bX_i\bX_{i'}^\top\right)^{-1}\right\|\\
 &\quad \times \left\|\left(\frac{1}{n}\sumn\sum_{i':  \mathcal N_i \cap \mathcal N_{i'} \neq \varnothing} \sum_{\mathcal S \in\mathcal S_i^\beta \cap S_{i'}^\beta } g(\mathcal S)^2\prod_{j \in \mathcal{S}} \frac{1}{p_j(1-p_j)} \bX_i Y_{i'}\right)\right\|\\
 &\leq \frac{d_{\text{in}}d_{\text{out}}}{c_{\lambda_{\min}}}\left(\frac{e d_{\text{in}}}{\beta p(1-p)} \right)^\beta X_{\max}Y_{\max}.
\end{align*}
The above argument shows that $\estc$ is bounded from above, which correspond to Assumption \hyperref[as:variance(ii)]{5(ii)}. Based on Assumption \hyperref[as:variance(ii)]{5(ii)}, $\estc$ has a finite limit. 

Then, we show that $\estchat - \estc = o_p(1)$. It suffices to show that $\E(\estchat) = \estc$ and $\Var(\estchat) = o_p(1)$. Firstly, we show that $\E(\estchat) = \estc$. The expectation of $\estchat$ is
\begin{align*}
    \E(\estchat) &= \E\left(\frac{1}{n^2}\sumn\sum_{i':  \mathcal N_i \cap \mathcal N_{i'} \neq \varnothing} \omega_i\omega_{i'} \bX_i\bX_{i'}^\top\right)^{-1} \left[\frac{1}{n^2}\sumn \sum_{i':  \mathcal N_i \cap \mathcal N_{i'} \neq \varnothing} \sum_{\mathcal S\in\mathcal S_i^\beta} \E\left(\omega_i\omega_{i'} \bX_{i'}\bs \prod_{k \in \mathcal{S}} Z_{k}\right)\E\left(\hat \alpha_{i, \mathcal{S}}^{\text{unadj}}\right)\right]\\
    &= \E\left(\frac{1}{n^2}\sumn\sum_{i':  \mathcal N_i \cap \mathcal N_{i'} \neq \varnothing} \omega_i\omega_{i'} \bX_i\bX_{i'}^\top\right)^{-1} \left[\frac{1}{n^2}\sumn \sum_{i':  \mathcal N_i \cap \mathcal N_{i'} \neq \varnothing} \sum_{\mathcal S\in\mathcal S_i^\beta} \E\left(\omega_i\omega_{i'} \bX_{i'}\prod_{k \in \mathcal{S}} Z_{k}\right)\alpha_{i, \mathcal{S}}\right]\\
    &= \E\left(\frac{1}{n^2}\sumn\sum_{i':  \mathcal N_i \cap \mathcal N_{i'} \neq \varnothing} \omega_i\omega_{i'} \bX_i\bX_{i'}^\top\right)^{-1} \left[\frac{1}{n^2}\sumn \sum_{i':  \mathcal N_i \cap \mathcal N_{i'} \neq \varnothing} \sum_{\mathcal S\in\mathcal S_i^\beta} \E\left(\omega_i\omega_{i'} \bX_{i'} \alpha_{i, \mathcal{S}} \prod_{k \in \mathcal{S}} Z_{k}\right)\right]\\
    &=\E\left(\frac{1}{n^2}\sumn\sum_{i':  \mathcal N_i \cap \mathcal N_{i'} \neq \varnothing} \omega_i\omega_{i'} \bX_i\bX_{i'}^\top\right)^{-1} \left[\frac{1}{n^2}\sumn \sum_{i':  \mathcal N_i \cap \mathcal N_{i'} \neq \varnothing}  \E\left(\omega_iY_i\omega_{i'}\bX_{i'}\right)\right]\\
    & =  \E\left(\frac{1}{n^2}\sumn\sum_{i':  \mathcal N_i \cap \mathcal N_{i'} \neq \varnothing} \omega_i\omega_{i'} \bX_i\bX_{i'}^\top\right)^{-1}\E\left(\frac{1}{n^2}\sumn\sum_{i':  \mathcal N_i \cap \mathcal N_{i'} \neq \varnothing} \omega_i\omega_{i'} \bX_iY_{i'}\right) = \estc.
\end{align*}
Next, we show that $\|\Var(\estchat)\|$ is $O(\frac{1}{n})$. 
\begin{align*}
    &\|\Var(\estchat)\| \\&= \Bigg\|\Var\Bigg\{\E\left(\frac{1}{n^2}\sumn\sum_{i':  \mathcal N_i \cap \mathcal N_{i'} \neq \varnothing} \omega_i\omega_{i'} \bX_i\bX_{i'}^\top\right)^{-1} \\
    &\quad \times \left[\frac{1}{n}\sumn \sum_{\mathcal S\in\mathcal S_i^\beta} Y_i \E\left(\frac{1}{n}\sum_{i':  \mathcal N_i \cap \mathcal N_{i'} \neq \varnothing} \omega_{i} \omega_{i'} \bX_{i'} \prod_{k \in \mathcal{S}} Z_{k}\right)\prod_{j \in \mathcal{S}} \frac{-1}{p_j}\sum_{\mathcal{U} \in \mathcal S_i^{\beta}, \mathcal S \subseteq \mathcal{U}}\prod_{l \in \mathcal{U}}\frac{p_l - Z_l}{1-p_l}\right]\Bigg\}\Bigg\|\\
    &\leq \left\|\E\left(\frac{1}{n^2}\sumn \sum_{i':  \mathcal N_i \cap \mathcal N_{i'} \neq \varnothing}\omega_i \omega_{i'}\bX_i\bX_{i'}^\top  \right)^{-1}\right\|^2\\
    &\quad \times \left\|\Var \left[\frac{1}{n}\sumn \sum_{\mathcal S\in\mathcal S_i^\beta} Y_i \E\left(\frac{1}{n}\sum_{i':  \mathcal N_i \cap \mathcal N_{i'} \neq \varnothing} \omega_i\omega_{i'} \bX_{i'}\prod_{k \in \mathcal{S}} Z_{k}\right)\prod_{j \in \mathcal{S}} \frac{-1}{p_j}\sum_{\mathcal{U} \in \mathcal S_i^{\beta}, \mathcal S \subseteq \mathcal{U}}\prod_{l \in \mathcal{U}}\frac{p_l - Z_l}{1-p_l}\right]\right\|.
 \end{align*}
Firstly, we derive the upper bound of the first term. 
\begin{align}\label{equ: consistp1}
&\left\|\E\left(\frac{1}{n^2}\sumn \sum_{i':  \mathcal N_i \cap \mathcal N_{i'} \neq \varnothing}\omega_i \omega_{i'}\bX_i\bX_{i'}^\top  \right)^{-1}\right\|^2\nonumber\\
&=  \left\|\E\left(\frac{1}{n^2}\sumn \sum_{i':  \mathcal N_i \cap \mathcal N_{i'} \neq \varnothing}\sum_{\mathcal S \in\mathcal S_i^\beta} \sum_{\mathcal S' \in\mathcal S_{i'}^\beta}g(\mathcal S)g(\mathcal S')\prod_{j \in \mathcal{S}} \frac{Z_j - p_j}{p_j(1-p_j)}\prod_{j' \in \mathcal S'}\frac{Z_{j'} - p_{j'}}{p_{j'}(1-p_{j'})}\bX_i\bX_{i'}^\top  \right)^{-1}\right\|^2\nonumber\\
&= \left\|\left(\frac{1}{n^2}\sumn \sum_{i':  \mathcal N_i \cap \mathcal N_{i'} \neq \varnothing}\sum_{\mathcal S \in\mathcal S_i^\beta \cap S_{i'}^\beta } g(\mathcal S)^2\prod_{j \in \mathcal{S}} \frac{1}{p_j(1-p_j)}\bX_i\bX_{i'}^\top  \right)^{-1}\right\|^2\nonumber\\
& \leq \frac{n^2}{c_{\lambda_{\min}}^2}.
\end{align}
Then we derive the upper bound of the variance term. 
\begin{align*}  
    &\left\|\Var \left[\frac{1}{n}\sumn \sum_{\mathcal S\in\mathcal S_i^\beta} Y_i \E\left(\frac{1}{n}\sum_{i':  \mathcal N_i \cap \mathcal N_{i'} \neq \varnothing} \omega_i\omega_{i'} \bX_{i'}\prod_{k \in \mathcal{S}} Z_{k}\right)\prod_{j \in \mathcal{S}} \frac{-1}{p_j}\sum_{\mathcal{U} \in \mathcal S_i^{\beta}, \mathcal S \subseteq \mathcal{U}}\prod_{l \in \mathcal{U}}\frac{p_l - Z_l}{1-p_l}\right]\right\|\\
    &= \left\|\Var \left[\frac{1}{n}\sumn \sum_{\mathcal S\in\mathcal S_i^\beta} \sum_{\mathcal S'\in\mathcal S_i^\beta} \alpha_{i, \mathcal S'}\E\left(\frac{1}{n}\sum_{i':  \mathcal N_i \cap \mathcal N_{i'} \neq \varnothing} \omega_i\omega_{i'} \bX_{i'}\prod_{k \in \mathcal{S}} Z_{k}\right)\prod_{j \in \mathcal{S}} \frac{-1}{p_j}\sum_{\mathcal{U} \in \mathcal S_i^{\beta}, \mathcal S \subseteq \mathcal{U}}\prod_{l \in \mathcal{U}}\frac{p_l - Z_l}{1-p_l}\prod_{j' \in \mathcal S'}Z_{j'}\right]\right\|\\
    &\leq \frac{1}{n^2}\sumn \sum_{i': \mathcal N_i \cap\mathcal N_{i'}}\sum_{\mathcal S'\in\mathcal S_i^\beta} |\alpha_{i, \mathcal S'}|\sum_{\mathcal T'\in\mathcal S_{i'}^\beta} |\alpha_{i', \mathcal T'}|\sum_{\mathcal U\in\mathcal S_i^\beta}\sum_{\mathcal U'\in\mathcal S_{i'}^\beta}\sum_{\mathcal S\subseteq\mathcal U}\sum_{\mathcal T\subseteq\mathcal U'}\prod_{j \in \mathcal{S}} \frac{1}{p_j}\prod_{t \in \mathcal{T}} \frac{1}{p_t}\left\|\E\left(\frac{1}{n}\sum_{i':  \mathcal N_i \cap \mathcal N_{i'} \neq \varnothing} \omega_i\omega_{i'} \bX_{i'}\prod_{k \in \mathcal{S}} Z_{k}\right)\right\|\\
    & \quad \times\left\|\E\left(\frac{1}{n}\sum_{i'':  \mathcal N_i' \cap \mathcal N_{i''} \neq \varnothing} \omega_{i'}\omega_{i''} \bX_{i''}\prod_{k' \in \mathcal{T}} Z_{k'}\right)\right\|\left|\Cov\left(\prod_{l \in \mathcal{U}}\frac{p_l - Z_l}{1-p_l}\prod_{j' \in \mathcal S'}Z_{j'}, \prod_{l' \in \mathcal{U}'}\frac{p_{l'} - Z_{l'}}{1-p_{l'}}\prod_{t' \in \mathcal T'}Z_{t'}\right)\right|\\
    &\leq \frac{1}{n^2}\sumn \sum_{i': \mathcal N_i \cap\mathcal N_{i'}}\sum_{\mathcal S'\in\mathcal S_i^\beta} |\alpha_{i, \mathcal S'}|\sum_{\mathcal T'\in\mathcal S_{i'}^\beta} |\alpha_{i', \mathcal T'}|\sum_{\mathcal U\in\mathcal S_i^\beta}\sum_{\mathcal U'\in\mathcal S_{i'}^\beta}\sum_{\mathcal S\subseteq\mathcal U}\sum_{\mathcal T\subseteq\mathcal U'}\left\|\E\left(\frac{1}{n}\sum_{i':  \mathcal N_i \cap \mathcal N_{i'} \neq \varnothing} \omega_i\omega_{i'} \bX_{i'}\prod_{k \in \mathcal{S}} Z_{k}\right)\right\|\\
    & \quad \times\left\|\E\left(\frac{1}{n}\sum_{i'':  \mathcal N_i' \cap \mathcal N_{i''} \neq \varnothing} \omega_{i'}\omega_{i''} \bX_{i''}\prod_{k' \in \mathcal{T}} Z_{k'}\right)\right\|\left|\Cov\left(\prod_{l \in \mathcal{U}}\frac{Z_l - p_l}{p_l(1-p_l)}\prod_{j' \in \mathcal S'}Z_{j'}, \prod_{l' \in \mathcal{U}'}\frac{Z_{l'} - p_{l'}}{p_{l'}(1-p_{l'})}\prod_{t' \in \mathcal T'}Z_{t'}\right)\right|\\
    &\leq \frac{2^{2\beta}}{n^2}\sumn \sum_{i': \mathcal N_i \cap\mathcal N_{i'}}\sum_{\mathcal S'\in\mathcal S_i^\beta} |\alpha_{i, \mathcal S'}|\sum_{\mathcal T'\in\mathcal S_{i'}^\beta} |\alpha_{i', \mathcal T'}|\sum_{\mathcal U\in\mathcal S_i^\beta}\sum_{\mathcal U'\in\mathcal S_{i'}^\beta}\sup_{i, \mathcal S}\left\|\E\left(\frac{1}{n}\sum_{i':  \mathcal N_i \cap \mathcal N_{i'} \neq \varnothing} \omega_i\omega_{i'} \bX_{i'}\prod_{k \in \mathcal{S}} Z_{k}\right)\right\|^2\\
    & \quad \times\sup_{\mathcal{U}, \mathcal S', \mathcal{U}', \mathcal T'}\left|\Cov\left(\prod_{l \in \mathcal{U}}\frac{Z_l - p_l}{p_l(1-p_l)}\prod_{j' \in \mathcal S'}Z_{j'}, \prod_{l' \in \mathcal{U}'}\frac{Z_{l'} - p_{l'}}{p_{l'}(1-p_{l'})}\prod_{t' \in \mathcal T'}Z_{t'}\right)\right|.
\end{align*}
We then derive upper bounds for each component of the above expansion. Firstly, we have the following lemma.
\begin{lemma}\label{lem:expupper} For any unit $i$ and $\mathcal S \in \mathcal S_{i}^\beta$ we have
\begin{align*}
\left\|\E\left(\frac{1}{n}\sum_{i':  \mathcal N_i \cap \mathcal N_{i'} \neq \varnothing} \omega_i\omega_{i'} \bX_{i'}\prod_{k \in \mathcal{S}} Z_{k}\right)\right\| \leq \frac{4d_{\text{in}}d_{\text{out}}}{n} X_{\max}\left( \frac{e d_{\text{in}}}{\beta} \cdot \max \left( \beta^2, \frac{1}{p(1 - p)} \right) \right)^{\beta}.
\end{align*}
\end{lemma}

Secondly, the upper bounds of the covariance part is given by Lemma 4 in \cite{CortezRodriguezEichhornYu+2023}. We summarize it in the following lemma. 
\begin{lemma}[Lemma 4 in \cite{CortezRodriguezEichhornYu+2023}]\label{lem:cortez}
Suppose $\{Z_j\}_{j \in [n]}$ are mutually independent with $Z_j \sim \text{Bernoulli}(p_j)$. Then for any subsets $\mathcal{S}, \mathcal{S}', \mathcal{T}, \mathcal{T}' \in [n]$, the covariance satisfies
\begin{align*}
0 &\leq \operatorname{Cov} \left[ 
\prod_{j \in \mathcal{S}} \frac{Z_j - p_j}{p_j(1 - p_j)} 
\prod_{j' \in \mathcal{S}'} Z_{j'}, \,
\prod_{k \in \mathcal{T}} \frac{Z_k - p_k}{p_k(1 - p_k)} 
\prod_{k' \in \mathcal{T}'} Z_{k'} 
\right] \leq \mathbb{I}(\mathcal{S} \triangle \mathcal{T} \subseteq \mathcal{S}' \cup \mathcal{T}') 
\left( \frac{1}{p(1-p)} \right)^{|\mathcal{S} \cap \mathcal{T}|},
\end{align*}
where $\mathcal{S} \triangle \mathcal{T} = (\mathcal{S} \cup \mathcal{T}) \setminus (\mathcal{S} \cap \mathcal{T})$ denotes the symmetric difference.
\end{lemma}

Therefore, proceeding along similar lines as the proof of Theorem 1 in \cite{CortezRodriguezEichhornYu+2023}, we then have
\begin{align}\label{equ: consistp2}
    &\left\|\Var \left[\frac{1}{n}\sumn \sum_{\mathcal S\subseteq\mathcal S_i^\beta} Y_i \E\left(\frac{1}{n}\sum_{i'=1}^n \omega_i\omega_{i'} \bX_{i'}\prod_{k \in \mathcal{S}} Z_{k}\right)\prod_{j \in \mathcal{S}} \frac{Z_j - p_j}{p_j}\right]\right\|\nonumber\\
    &\qquad \qquad \leq \frac{4^{3+\beta} d_{\text{in}}^3 d_{\text{out}}^3 Y_{\max}^2 X_{\max}^2}{n^3} \left(\frac{e d_{\text{in}}}{\beta}\cdot \max \left( 4\beta^2, \frac{1}{p(1 - p)} \right) \right)^{3\beta}. 
\end{align}
Under (\ref{equ: consistp1}), (\ref{equ: consistp2}), Assumption \ref{as:bounded} and the assumption that maximum degree of the interference network satisfies
$d = O(1), $the variance of $\estchat$ is $O(\frac{1}{n})$. Therefore, $\estchat$ converges to $\estc$ in probability.

\subsection{Proof of Proposition \ref{prop:regvimlin}}
Firstly, we show that $\hat{\bs \theta}_{\text{Lin}} - \bs \theta_{\text{Lin}} \xrightarrow{p}    
0$. We rewrite $\hat {\bs \theta}_1$ as
\begin{align*}
    \hat {\bs \theta}_1 &= \left(\sum_{i: Z_i = 1}(\bs X_i - \hat{\bar \bX}_1)(\bs X_i - \hat{\bar \bX}_1)^\top \right)^{-1}\left(\sum_{i: Z_i = 1}(\bs X_i - \hat{\bar \bX}_1)(\bs Y_i - \hat{\bar Y}_1)\right)\\
    &= \left(\frac{1}{n}\sumn Z_i(\bs X_i - \hat{\bar \bX}_1)(\bs X_i - \hat{\bar \bX}_1)^\top \right)^{-1}\left(\frac{1}{n}\sumn Z_i(\bs X_i - \hat{\bar \bX}_1)(\bs Y_i - \hat{\bar Y}_1)\right),
\end{align*}
where $\hat{\bar \bX}_1 = \frac{1}{\sumn Z_i}\sumn Z_i \bs X_i$ and $\hat{\bar Y}_1 = \frac{1}{\sumn Z_i}\sumn Z_i \bs Y_i$. Since $\frac{1}{n}\sumn Z_i \xrightarrow{p}    
p_1$ and $\frac{1}{n}\sumn Z_i \bX_i - \frac{p_1}{n}\sumn \bX_i \xrightarrow{p}    
0$, by Slutsky’s theorem we have $\hat{\bar \bX}_1 - \bar \bX = o_p(1)$. Therefore, for the first component, $\frac{1}{n}\sumn Z_i(\bs X_i - \hat{\bar \bX}_1)(\bs X_i - \hat{\bar \bX}_1)^\top$, based on Assumption \ref{as:bounded}, we have
\begin{align*}
&\left\|\frac{1}{n}\sumn Z_i(\bs X_i - \hat{\bar \bX}_1)(\bs X_i - \hat{\bar \bX}_1)^\top - \frac{1}{n}\sumn Z_i(\bs X_i - \bar \bX)(\bs X_i - \bar \bX)^\top \right\| \\
&= \left\|\frac{1}{n}(\bar \bX - \hat{\bar \bX}_1)\sumn Z_i(\bs X_i - \hat{\bar \bX}_1)^\top + \frac{1}{n}\left[\sumn Z_i(\bs X_i - \bar \bX)\right](\bar \bX - \hat{\bar \bX}_1)^\top\right\| = o_p(1).
\end{align*}
Then we show that $\frac{1}{n}\sumn Z_i(\bs X_i - \bar \bX)(\bs X_i - \bar \bX)^\top - p_1 S_{\bX\bX}\xrightarrow{p}    
0$.
\begin{align*}
    \E \left(\frac{1}{n}\sumn Z_i(\bs X_i - \bar \bX)(\bs X_i - \bar \bX)^\top\right) &= \frac{p_1}{n}\sumn(\bs X_i - \bar \bX)(\bs X_i - \bar \bX)^\top  = p_1 S_{\bX\bX}.\\
    \left\|\Var \left(\frac{1}{n}\sumn Z_i(\bs X_i - \bar \bX)(\bs X_i - \bar \bX)^\top\right)\right\| &\leq  \frac{p_1(1-p_1)}{n} X^4_{\max}.
\end{align*}
Thus
$$
\E\left\|\frac{1}{n}\sumn Z_i(\bs X_i - \bar \bX)(\bs X_i - \bar \bX)^\top - \E\left(\frac{1}{n}\sumn Z_i(\bs X_i - \bar \bX)(\bs X_i - \bar \bX)^\top\right) \right\|^2 = O(n^{-1}).
$$
Therefore, based on Assumption \ref{as:bounded}, $\|\frac{1}{n}\sumn Z_i(\bs X_i - \bar \bX)(\bs X_i - \bar \bX)^\top - p_1 S_{\bX\bX}\|\xrightarrow{p}    
0$. Thus, $\frac{1}{n}\sumn Z_i(\bs X_i - \hat{\bar \bX}_1)(\bs X_i - \hat{\bar \bX}_1)^\top - p_1 S_{\bX\bX}\xrightarrow{p}    
0$. Similarly, we can show that $\frac{1}{n}\sumn Z_i(\bs X_i - \hat{\bar \bX}_1)(Y_i - \hat{\bar Y}_1 ) - p_1 S_{\bX Y(1)}\xrightarrow{p}    
0$ and therefore $\hat{\bs\theta}_1 - \bs\theta_1 \xrightarrow{p}    
0$. Following similar steps, we can also show $\hat{\bs\theta}_0 - \bs\theta_0 \xrightarrow{p}    
0$ and further show $\hat{\bs\theta}_{\text{Lin}} - \bs\theta_{\text{Lin}} \xrightarrow{p}    
0$.

Next, we show that $\rbhat - \bs\theta_{\text{Lin}} \xrightarrow{p}    
0$. When there is no interference ($\mathcal S_i^\beta = \{\varnothing, \{i\}\}$) and the treatment probabilities are the same across units, the regression-based adjustment coefficient $\rbhat$ can be written as
\begin{align*}
    \rbhat &= \left(\frac{1}{p_1^2}\sum_{i: Z_i = 1}\bs X_i \bs X_i^\top + \frac{1}{(1-p_1)^2}\sum_{i: Z_i = 0}\bs X_i \bs X_i^\top\right)^{-1}\left(\frac{1}{p_1^2}\sum_{i: z_i = 1}\bs X_iY_i + \frac{1}{(1-p_1)^2}\sum_{i: z_i = 0}\bs X_iY_i\right)\\
    &= \left(\frac{1}{n p_1^2}\sumn Z_i\bs X_i \bs X_i^\top + \frac{1}{n(1-p_1)^2}\sumn(1-Z_i)\bs X_i \bs X_i^\top\right)^{-1}\left(\frac{1}{n p_1^2}\sumn Z_i\bs X_iY_i + \frac{1}{n(1-p_1)^2}\sumn (1-Z_i)\bs X_iY_i\right).
\end{align*}
Next, we show that$
\left\|\frac{1}{n p_1^2}\sum_{i} Z_i\bs X_i\bs X_i^\top - \frac{1}{p_1} S_{\bX\bX}\right\| \xrightarrow{p}    
0$.
\begin{align*}
    \E \left(\frac{1}{n p_1^2}\sumn Z_i\bs X_i \bs X_i^\top\right) = \frac{1}{n p_1}\sumn\bs X_i \bs X_i^\top  = \frac{1}{p_1} S_{\bX\bX}.\\
    \left\|\Var \left(\frac{1}{n p_1^2}\sumn Z_i\bs X_i \bs X_i^\top\right)\right\| \leq  \frac{1-p_1}{np_1^3} X^4_{\max}. 
\end{align*}
Based on Assumption \ref{as:bounded}, $\left\|\Var \left(\frac{1}{n p_1^2}\sumn Z_i\bs X_i \bs X_i^\top\right)\right\| = o(1)$. Therefore, $
\frac{1}{n p_1^2}\sum_{i} Z_i\bs X_i\bs X_i^\top - \frac{1}{p_1} S_{\bX\bX} \xrightarrow{p}    
0$. Similarly, we can show that
\begin{align*}
&\left\|\frac{1}{n(1- p_1)^2}\sum_{i} (1-Z_i)\bs X_i \bs X_i^\top- \frac{1}{1-p_1} S_{\bX\bX} \right\|\overset{p}{\to}    
0,\\
&\frac{1}{n p_1^2}\sum_{i} Z_i\bs X_i Y_i - \frac{1}{p_1} S_{\bX Y(1)}\xrightarrow{p}    
0,\\
&\left\|\frac{1}{n (1-p_1)^2}\sum_{i} (1-Z_i)\bs X_iY_i - \frac{1}{1-p_1} S_{\bX Y(0)}\right\|\xrightarrow{p}    
0.
\end{align*}
Therefore, $\rbhat - \bs\theta_{\text{Lin}} \xrightarrow{p}    
0$. Finally, as mentioned in Section \ref{sec:VIM-SNIPE}, $\estchat$ is asymptotically equivalent to $\rbhat$ when there is no interference. Therefore, $\estchat - \bs\theta_{\text{Lin}} \xrightarrow{p}    
0$.

\subsection{Proof of Theorem \ref{thm:vimvar}}\label{sec:proof_thm1}
By definition, the differences between the variances of the SNIPE estimator $\TTEhat(\bs 0)$ and the oracle estimator $\TTEhat(\estc)$ is
\begin{align*}
n\Var\left(\TTEhat(\bs 0)\right) - n\Var\left(\TTEhat(\bs \estc)\right) = \frac{1}{n}\E\left[\left(\sumn  \omega_i \estc^\top \bX_i \right)^2\right],
\end{align*}
which is non-negative and has a finite limit as shown in the proof of Proposition \ref{prop:consistent}. This variance difference is the lowest among the entire class of covariate-adjusted estimators parameterized by $\bs \theta$ by definition. Secondly, we prove that the variance of $n\Var\left(\TTEhat(\bs \estchat)\right)$ converges to the limit of $n\Var(\TTEhat(\bs \estc))$. To establish this result, first we let
 \begin{align*}
     Q_n(\bs\theta) &= \frac{1}{n}\E\left[\left(\sumn  \omega_i \bs \theta^\top \bX_i \right)^2\right] - \frac{2}{n} \sumn \sum_{i':  \mathcal N_i \cap \mathcal N_{i'} \neq \varnothing}\sum_{\mathcal S\in\mathcal S_i^\beta} \alpha_{i, \mathcal{S}}\E\left[\omega_i \omega_{i'} \bs \theta^\top \bX_{i'} \prod_{k \in \mathcal{S}}Z_{k}\right],\\
     \widetilde Q_n(\bs \theta) &= \frac{1}{n}\E\left[\left(\sumn  \omega_i \bs \theta^\top \bX_i \right)^2\right] - \frac{2}{n} \sumn \sum_{i':  \mathcal N_i \cap \mathcal N_{i'} \neq \varnothing}\sum_{\mathcal S\in\mathcal S_i^\beta} \hat\alpha^{\text{unadj}}_{i, \mathcal{S}}\E\left[\omega_i \omega_{i'} \bs \theta^\top \bX_{i'} \prod_{k \in \mathcal{S}}Z_{k}\right].
 \end{align*}
 Expanding $\Var\left(\TTEhat(\bs \estchat)\right)$, we have
 \begin{align*}
     n\Var\left(\TTEhat(\bs \estchat)\right) &= n\E\left(\TTEhat(\bs \estchat)^2\right) - n\left[\E\left(\TTEhat( \estchat)\right)\right]^2\\
     &= n\E\left(\TTEhat(\estchat)^2\right) - n\left[\E\left(\TTEhat( \estc) + \left(\estc - \estchat\right )^\top\frac{1}{n}\sum_{i=1}^n \omega_i \bX_i \right)\right]^2.
\end{align*}
For the second term,
\begin{align*}
     &n\left|\E\left(\left(\estc - \estchat\right )^\top\frac{1}{n}\sum_{i=1}^n \omega_i \bX_i \right)\right|^2\leq n\left(\E\left\|\estc - \estchat\right\|^2\right)
     \left(\E\left\|\frac{1}{n}\sum_{i=1}^n \omega_i \bX_i\right\|^2\right)\\
     &= n\left(\E\left\|\estc - \estchat\right\|^2\right)
     \left(\frac{1}{n^2}\sum_{i=1}^n \sum_{i': \mathcal N_i \cap \mathcal N_{i'} \neq \varnothing}^n \sum_{\mathcal S \in \mathcal S_i^\beta \cap \mathcal S_{i'}^\beta} g^2(S) \prod_{j \in \mathcal S}\frac{1}{p_j(1-p_j)} \bX_i^\top\bX_{i'}\right)\\
     &= n\left(\text{tr}\left(\Var\left(\estc - \estchat\right)\right)\right)
     \left(\frac{1}{n^2}\sum_{i=1}^n \sum_{i': \mathcal N_i \cap \mathcal N_{i'} \neq \varnothing}^n \sum_{\mathcal S \in \mathcal S_i^\beta \cap \mathcal S_{i'}^\beta} g^2(S) \prod_{j \in \mathcal S}\frac{1}{p_j(1-p_j)} \bX_i^\top\bX_{i'}\right)\\
     &\leq n\left(d_{\bX}\left\|\Var\left(\estc - \estchat\right)\right\|\right)
     \left(\frac{1}{n^2}\sum_{i=1}^n \sum_{i': \mathcal N_i \cap \mathcal N_{i'} \neq \varnothing}^n \sum_{\mathcal S \in \mathcal S_i^\beta \cap \mathcal S_{i'}^\beta} g^2(S) \prod_{j \in \mathcal S}\frac{1}{p_j(1-p_j)} \bX_i^\top\bX_{i'}\right)\\
     &=O\left(n \times \frac{1}{n} \times \frac{d^2}{n} \left(\frac{e d}{\beta p(1-p)}\right)^\beta X_{\max}^2\right) = O\left(\frac{1}{n}\right). 
 \end{align*}
 Therefore $n\Var\left(\TTEhat(\bs \estchat)\right) = n\E\left(\TTEhat(\bs \estchat)^2\right) + o(1)$. Then it suffices to show that $Q_n(\estchat) - Q_n(\estc)\to 0$. Under Assumption \ref{as:bounded}, there exists a compact space $\bs\Theta_{\text{VIM}}$ containing both $\estchat$ and $\estc$. We proceed with the proof in the following three steps:
\begin{enumerate}[label=(\roman*)]
\item $\sup_{\bs \theta \in \bs\Theta_{\text{VIM}}}|\widetilde Q_n(\bs \theta) - Q_n(\bs \theta)| = o_p(1)$;  
\item $Q_n(\estchat) \le Q_n(\estc) + o_p(1)$;
\item $Q_n(\estchat) - Q_n(\estc) \to  0.$
 \end{enumerate}
First, we show the uniform convergence. By definition, 
\begin{align*}
 \sup_{\bs \theta \in \bs\Theta_{\text{VIM}}}|\widetilde Q_n(\bs \theta) - Q_n(\bs \theta)| &=  \sup_{\bs \theta \in \bs\Theta_{\text{VIM}}}\left|\frac{2}{n} \sumn \sum_{i':  \mathcal N_i \cap \mathcal N_{i'} \neq \varnothing}\sum_{\mathcal S\subseteq\mathcal S_i^\beta} (\alpha_{i, \mathcal{S}} -\hat\alpha^{\text{unadj}}_{i, \mathcal{S}})\E\left(\omega_i \omega_{i'} \bs \theta^\top \bX_{i'} \prod_{k \in \mathcal{S}}Z_{k}\right)\right|\\
 & \leq \sup_{\bs \theta \in \bs\Theta_{\text{VIM}}} \|\bs \theta\| \left\|2 \sumn \sum_{\mathcal S\subseteq\mathcal S_i^\beta} (\alpha_{i, \mathcal{S}} -\hat\alpha^{\text{unadj}}_{i, \mathcal{S}})\E\left(\frac{1}{n}\sum_{i':  \mathcal N_i \cap \mathcal N_{i'} \neq \varnothing}\omega_i \omega_{i'} \bX_{i'} \prod_{k \in \mathcal{S}}Z_{k}\right)\right\|.
\end{align*}
Under Assumption \ref{as:bounded} and the assumption that the maximum degree of the interference network satisfies
$d = O(1),$ $\sup_{\bs \theta \in \bs\Theta_{\text{VIM}}} \|\bs \theta\|$ is bounded from above. Next, we prove that the second component is $o_p(1)$. For simplicity, we let $\E_{i, \mathcal S} = \E\left(\frac{1}{n}\sum_{i':  \mathcal N_i \cap \mathcal N_{i'} \neq \varnothing}\omega_i \omega_{i'} \bX_{i'} \prod_{k \in \mathcal{S}}Z_{k}\right)$. It is easy to see that
\begin{align*}
\E\left(2 \sumn \sum_{\mathcal S\in\mathcal S_i^\beta} (\alpha_{i, \mathcal{S}} -\hat\alpha^{\text{unadj}}_{i, \mathcal{S}})\E_{i, \mathcal S}\right)=0    
\end{align*}
because of the unbiasedness of $\hat\alpha^{\text{unadj}}_{i, \mathcal{S}}$. Next, we show that its variance vanishes as $n \to \infty$.
\begin{align*}
 &\Var\left(\left\|2 \sumn \sum_{\mathcal S\in\mathcal S_i^\beta} (\alpha_{i, \mathcal{S}} -\hat\alpha^{\text{unadj}}_{i, \mathcal{S}})\E_{i, \mathcal S}\right\|\right)\\
 &\leq4 \sumn \sum_{i':  \mathcal N_i \cap \mathcal N_{i'} \neq \varnothing} \sum_{\mathcal S\in\mathcal S_i^\beta}  \sum_{\mathcal S'\in\mathcal S_{i'}^\beta} \| \E_{i, \mathcal S}\|^2 \left|\Cov(\hat\alpha^{\text{unadj}}_{i, \mathcal{S}}, \hat\alpha^{\text{unadj}}_{i', \mathcal{S'}})\right|\\
 &= 4 \sumn \sum_{i':  \mathcal N_i \cap \mathcal N_{i'} \neq \varnothing} \sum_{\mathcal S\in\mathcal S_i^\beta}  \sum_{\mathcal S'\in\mathcal S_{i'}^\beta} \| \E_{i, \mathcal S}\|^2 \\
 &\quad\times\left|\Cov\left(\sum_{\mathcal T \in \mathcal S_{i}^\beta} \alpha_{i, \mathcal T'} \prod_{t \in T}Z_t \prod_{j \in \mathcal{S}} \frac{-1}{p_j}\sum_{\mathcal{U} \in \mathcal S_i^{\beta}, \mathcal S \subseteq \mathcal{U}}\prod_{l \in \mathcal{U}}\frac{p_l - Z_l}{1-p_l}, \sum_{\mathcal T' \in \mathcal S_{i'}^\beta} \alpha_{i', \mathcal T} \prod_{t' \in T'}Z_{t'} \prod_{j' \in \mathcal{S'}} \frac{-1}{p_{j'}}\sum_{\mathcal{U'} \in \mathcal S_{i'}^{\beta}, \mathcal S' \subseteq \mathcal{U'}}\prod_{l' \in \mathcal{U'}}\frac{p_{l'} - Z_{l'}}{1-p_{l'}}\right)\right|\\
 & \leq 4 \sumn \sum_{i':  \mathcal N_i \cap \mathcal N_{i'} \neq \varnothing} \sum_{\mathcal S\in\mathcal S_i^\beta}  \sum_{\mathcal S'\in\mathcal S_{i'}^\beta} \| \E_{i, \mathcal S}\|^2 \sum_{\mathcal T \in \mathcal S_{i}^\beta} |\alpha_{i, \mathcal T'}| \sum_{\mathcal T' \in \mathcal S_{i'}^\beta} |\alpha_{i', \mathcal T}|\sum_{\mathcal{U} \in \mathcal S_i^{\beta}}\sum_{\mathcal{U'} \in \mathcal S_{i'}^{\beta}}\\
 &\quad\times\left|\Cov\left(\prod_{l \in \mathcal{U}}\frac{p_l - Z_l}{p_l(1-p_l)}\prod_{t \in T}Z_t,   \prod_{l' \in \mathcal{U'}}\frac{p_{l'} - Z_{l'}}{p_{l'}(1-p_{l'})}\prod_{t' \in T'}Z_{t'}\right)\right|.
\end{align*}
Then based on Lemma \ref{lem:expupper} and \ref{lem:cortez} we have
\begin{align*}
 &\Var\left(\left\|2 \sumn \sum_{\mathcal S\in\mathcal S_i^\beta} (\alpha_{i, \mathcal{S}} -\hat\alpha^{\text{unadj}}_{i, \mathcal{S}})\E\left(\frac{1}{n}\sum_{i':  \mathcal N_i \cap \mathcal N_{i'} \neq \varnothing}\omega_i \omega_{i'} \bX_{i'} \prod_{k \in \mathcal{S}}Z_{k}\right)\right\|\right)\\ 
 &\leq \frac{4^{4} d_{\text{in}}^3 d_{\text{out}}^3 Y_{\max}^2 X_{\max}^2}{n} \left(\frac{e d_{\text{in}}}{\beta}\cdot \max \left( 4\beta^2, \frac{1}{p(1 - p)} \right) \right)^{3\beta} = O\left(\frac{1}{n}\right).
\end{align*}
The last equality is based on Assumption \ref{as:bounded} and the assumption that the maximum degree of the interference network satisfies
$d = O(1)$. Therefore, $\sup_{\bs \theta \in \bs\Theta_{\text{VIM}}}|\widetilde Q_n(\bs \theta) - Q_n(\bs \theta)| = o_p(1)$. Based on the uniform convergence, we have
\begin{align*}
    |\widetilde Q_n(\estchat) - Q_n(\estchat)| = o_p(1), \quad
|\widetilde Q_n(\estc) - Q_n(\estc)| = o_p(1).
\end{align*}
Since by definition $\estchat$ minimizes $\widetilde Q_n$, we have $\widetilde Q_n(\estchat) \leq \widetilde Q_n(\estc).$ Therefore, it gives
\begin{align*}
Q_n(\estchat)
= \widetilde Q_n(\estchat) + o_p(1)
\le \widetilde Q_n(\estc) + o_p(1)
= Q_n(\estc) + o_p(1).    
\end{align*}
This implies $Q_n(\estchat) \le Q_n(\estc) + o_p(1).$ By definition, we have $Q_n(\estc) \le Q_n(\estchat)$, then
\begin{align*}
 Q_n(\estchat) - Q_n(\estc) \to  0.    
\end{align*}

\subsection{Proof of Theorem \ref{thm:ubound}}
Recall that under Assumption \ref{as:model}, $Y_i
= \alpha_{i,\varnothing}
+ \sum_{S \in \mathcal S_i^\beta,\, S \neq \varnothing} \alpha_{i,S}
\prod_{j \in \mathcal S} Z_j.$ Observe that
\[
Y_i - \bs \theta^\top \bs X_i
= \bigl(\alpha_{i,\varnothing} - \bs \theta^\top \bs X_i\bigr)
+ \sum_{S \in \mathcal S_i^\beta,\, S \neq \varnothing} \alpha_{i,S}
\prod_{j \in \mathcal S} Z_j .
\]
Thus, subtracting $\bs \theta^\top \bs X_i$ only modifies the intercept term
$\alpha_{i,\varnothing}$, while leaving all higher-order interaction terms
$\alpha_{i,S}$ unchanged. Applying the same argument in the proof of Theorem 1 in \cite{CortezRodriguezEichhornYu+2023} with $\alpha_{i,\varnothing}$ replaced by
$\alpha_{i,\varnothing} - \bs \theta^\top \bs X_i$ yields

\begin{align*}
\Var\left(\widehat{\mathrm{TTE}}(\theta)\right) \leq \frac{4d_\text{in}d_\text{out}}{n}\left(\max_{i \in [n]} |\alpha_{i, \mathcal \varnothing} - \bs \theta^\top \bX_i| + \sum_{\substack{\mathcal S \in \mathcal S_{i}^\beta\\
\mathcal S \neq \varnothing}}|\alpha_{i, \mathcal S}|\right)^2 \left(
\frac{ed_\text{in}}{\beta}\max\left\{4\beta^2,\frac{1}{p(1-p)}\right\}\right)^\beta. 
\end{align*}
\subsection{Proof of Theorem \ref{thm:clt}}
Firstly, we rewrite $\sqrt{n}\TTEhat(\hat{\bs\theta})$ as
\begin{align*}
\sqrt{n}(\TTEhat(\hat{\bs\theta}) - \tau) = \sqrt{n}(\hat{\tau}(\bs\theta) - \tau) - (\hat{\bs\theta} - \bs\theta)^\top \sqrt{n}\frac{1}{n}\sumn \omega_i\bX_i.   
\end{align*}
\begin{lemma} \label{lem:cltfix}
 Under Assumptions \ref{as:interference} - \ref{as:bounded} and \ref{as:variance}, and the assumption that $d$ is bounded, $\sqrt{n}(\TTEhat(\bs\theta) - \tau) $ converges in distribution to $\mathcal{N}(0, V(\bs \theta^\ast))$.   
\end{lemma}
The asymptotic normality of $\sqrt{n}\frac{1}{n}\sumn \omega_i\bX_i$ follows from similar arguments to the proof of Lemma \ref{lem:cltfix} by viewing $\bX_i$'s as outcomes. Therefore, by Assumption \ref{as:thetahat} and Slutsky’s Theorem, we have
\begin{align*}
    (\hat{\bs \theta} - \bs\theta)^\top \sqrt{n}\frac{1}{n}\sumn \omega_i\bX_i = o_p(1).
\end{align*}
Then combining Lemma \ref{lem:cltfix}, we have $\sqrt{n}(\hat{\tau}(\hat{\bs\theta}) - \tau) $ converges in distribution to $\mathcal{N}(0,V(\bs \theta^\ast))$.

\subsection{Proof of Theorem \ref{thm:var_est}}
Recall that the reported variance estimator is
\[
\hat V(\hat{\bs \theta})=\hat V(\bs 0) - n\hat\Delta(\hat{\bs \theta}).
\]
By Appendix~\ref{sec:proof_thm1}, where $\widetilde Q_n(\bs\theta)= -n \hat\Delta(\bs\theta)$ and
$Q_n(\bs\theta)= -n\Delta_n(\bs\theta)$, we have
\begin{equation}\label{eq:Delta_hat_consistency_thm4}
\hat\Delta(\hat{\bs\theta})-\Delta(\bs\theta^\ast)=o_p(1).
\end{equation}
Therefore, it suffices to show that
\begin{equation}\label{eq:V0_consistency_goal_thm4}
\hat V(\bs 0)-\widetilde V(\bs 0)=o_p(1),
\end{equation}
because then
\[
\hat V(\hat{\bs \theta})-\widetilde V(\bs\theta^\ast)
=
\{\hat V(\bs 0)-\widetilde V(\bs 0)\}
-
\{\hat\Delta(\hat{\bs\theta})-\Delta(\bs\theta^\ast)\}
=o_p(1)
\]
by \eqref{eq:Delta_hat_consistency_thm4} and \eqref{eq:V0_consistency_goal_thm4},
where $\widetilde V(\bs\theta^\ast)=\widetilde V(\bs 0)-\Delta(\bs\theta^\ast)$.

\medskip

We now prove \eqref{eq:V0_consistency_goal_thm4}. Define the index set of dependent
pairs
\[
\mathcal E_n=\{(i,i'):\ \mathcal N_i\cap \mathcal N_{i'}\neq\varnothing\}.
\]
The population target $\widetilde V(\bs 0)$ is defined by replacing
$(\hat\alpha^{\text{unadj}}_{i, \mathcal S}\hat\alpha^{\text{unadj}}_{i', \mathcal S'},\hat\gamma^{\text{unadj}})$ in $\hat V(\bs 0)$ by $(\E(\hat\alpha^{\text{unadj}}_{i, \mathcal S}\hat\alpha^{\text{unadj}}_{i', \mathcal S'}),\gamma^{\text{unadj}})$:
\begin{align*}
\widetilde V(\bs 0)
&= \frac{2}{n}\sum_{(i,i')\in\mathcal E_n}\,
\sum_{\mathcal S \in \mathcal S_i^\beta} \sum_{\mathcal S' \in \mathcal S_{i'}^\beta}
\E(\hat\alpha^{\text{unadj}}_{i, \mathcal S}\hat\alpha^{\text{unadj}}_{i', \mathcal S'})
+ \frac{2}{n}\sum_{(i,i')\in\mathcal E_n}\E(\hat\alpha_{i, \varnothing}^{\text{unadj}}
\hat\alpha_{i', \varnothing}^{\text{unadj}}) \\
&\quad - \frac{2}{n}\sum_{(i,i')\in\mathcal E_n}\sum_{\mathcal T \in \mathcal T_{ii'}^{\beta}}
\gamma_{ii', \mathcal T}^{\text{unadj}}
- \frac{2}{n}\sum_{(i,i')\in\mathcal E_n}\gamma_{ii', \varnothing }^{\text{unadj}}.
\end{align*}
Hence, $\hat V(\bs 0)-\widetilde V(\bs 0)$ can be decomposed into four terms:
\[
\hat V(\bs 0)-\widetilde V(\bs 0)=A_{n,1}+A_{n,2}-A_{n,3}-A_{n,4},
\]
where
\begin{align*}
A_{n,1}
&=
\frac{2}{n}\sum_{(i,i')\in\mathcal E_n}
\sum_{\mathcal S \in \mathcal S_i^\beta} \sum_{\mathcal S' \in \mathcal S_{i'}^\beta}
\Bigl(
\hat\alpha_{i, \mathcal S}^{\text{unadj}}\hat\alpha_{i', \mathcal S'}^{\text{unadj}}
-\E(\hat\alpha_{i, \mathcal S}^{\text{unadj}}\hat\alpha_{i', \mathcal S'}^{\text{unadj}})
\Bigr),\\
A_{n,2}
&=
\frac{2}{n}\sum_{(i,i')\in\mathcal E_n}
\Bigl(
\hat\alpha_{i, \varnothing}^{\text{unadj}}\hat\alpha_{i', \varnothing}^{\text{unadj}}
-\E(\hat\alpha_{i, \varnothing}^{\text{unadj}}\hat\alpha_{i', \varnothing}^{\text{unadj}})
\Bigr),\\
A_{n,3}
&=
\frac{2}{n}\sum_{(i,i')\in\mathcal E_n}\sum_{\mathcal T \in \mathcal T_{ii'}^{\beta}}
\Bigl(\hat\gamma_{ii', \mathcal T}^{\text{unadj}}-\gamma_{ii', \mathcal T}^{\text{unadj}}\Bigr),\\
A_{n,4}
&=
\frac{2}{n}\sum_{(i,i')\in\mathcal E_n}
\Bigl(\hat\gamma_{ii', \varnothing }^{\text{unadj}}-\gamma_{ii', \varnothing }^{\text{unadj}}\Bigr).
\end{align*}
We show that $A_{n,k}=o_p(1)$ for each $k\in\{1,2,3,4\}$ by verifying that
$\E(A_{n,k})=0$ and $\Var(A_{n,k})\to 0$.

\medskip

Under the bounded-degree condition $d=O(1)$, the set of dependent pairs satisfies
$|\mathcal E_n|=O(n)$. Moreover, for any index tuple appearing in the sums below
(e.g., $(i,i',\mathcal S,\mathcal S')$ or $(i,i',\mathcal T)$), the random variable
$\hat\alpha_{i,\mathcal S}^{\text{unadj}}$ (resp.\ $\hat\gamma_{ii',\mathcal T}^{\text{unadj}}$)
depends only on the treatment assignments in a bounded-size neighborhood determined
by $\mathcal N_i$ (resp.\ $\mathcal N_i\cup\mathcal N_{i'}$) and the indices in
$\mathcal S$ (resp.\ $\mathcal T$). Consequently, for each fixed summand, there
exist at most $O(1)$ other summands with which it can have nonzero covariance.
We use this observation repeatedly below; it is the same sparsity technique used
throughout Appendix~\ref{sec:main_proof}.

\medskip

\noindent\textbf{The $\gamma$-terms.}
By construction of the pseudo-inverse estimators,
$\hat\gamma_{ii',\mathcal T}^{\text{unadj}}$ is unbiased for
$\gamma_{ii',\mathcal T}^{\text{unadj}}$ (and similarly for $\varnothing$), hence
$\E(A_{n,3})=\E(A_{n,4})=0$. Consider $A_{n,3}$; the argument for $A_{n,4}$ is identical.
Using the covariance expansion,
\begin{align*}
\Var(A_{n,3})
&=
\frac{4}{n^2}
\sum_{(i,i')\in\mathcal E_n}\sum_{(j,j')\in\mathcal E_n}
\sum_{\mathcal T\in\mathcal T_{ii'}^\beta}\sum_{\mathcal U\in\mathcal T_{jj'}^\beta}
\Cov\!\Bigl(\hat\gamma_{ii',\mathcal T}^{\text{unadj}},\hat\gamma_{jj',\mathcal U}^{\text{unadj}}\Bigr).
\end{align*}
By the sparsity counting bound above, for each fixed $(i,i',\mathcal T)$ there are
only $O(1)$ choices of $(j,j',\mathcal U)$ giving nonzero covariance, and
$|\mathcal E_n|=O(n)$.
Under Assumption~\ref{as:bounded}, the covariance terms are uniformly bounded in
absolute value. Since $|\mathcal T_{ii'}^\beta|$ is uniformly bounded under $d=O(1)$,
it follows that $\Var(A_{n,3})=O(1/n)$ and hence $A_{n,3}=o_p(1)$ by Chebyshev's
inequality. The same argument yields $A_{n,4}=o_p(1)$.

\medskip

\medskip

\noindent\textbf{The $\alpha$-terms.}
We treat $A_{n,1}$; the proof for $A_{n,2}$ is identical. By definition,
\[
A_{n,1}
=
\frac{2}{n}\sum_{(i,i')\in\mathcal E_n}
\sum_{\mathcal S \in \mathcal S_i^\beta} \sum_{\mathcal S' \in \mathcal S_{i'}^\beta}
\Bigl(
\hat\alpha_{i, \mathcal S}^{\text{unadj}}\hat\alpha_{i', \mathcal S'}^{\text{unadj}}
-\E(\hat\alpha_{i, \mathcal S}^{\text{unadj}}\hat\alpha_{i', \mathcal S'}^{\text{unadj}})
\Bigr),
\]
so $\E(A_{n,1})=0$.

We now control its variance. By covariance expansion,
\begin{align*}
\Var(A_{n,1})
&=
\frac{4}{n^2}
\sum_{(i,i')\in\mathcal E_n}\sum_{(j,j')\in\mathcal E_n}
\sum_{\mathcal S\in\mathcal S_i^\beta}\sum_{\mathcal S'\in\mathcal S_{i'}^\beta}
\sum_{\mathcal U\in\mathcal S_j^\beta}\sum_{\mathcal U'\in\mathcal S_{j'}^\beta} \\
&\quad \times
\Cov\!\Bigl(
\hat\alpha_{i,\mathcal S}^{\text{unadj}}\hat\alpha_{i',\mathcal S'}^{\text{unadj}},
\hat\alpha_{j,\mathcal U}^{\text{unadj}}\hat\alpha_{j',\mathcal U'}^{\text{unadj}}
\Bigr).
\end{align*}
Indeed, subtracting expectations does not change covariance.

Under Assumption~\ref{as:bounded}, the outcomes are uniformly bounded and the treatment
probabilities are uniformly bounded away from $0$ and $1$. Since $d=O(1)$ and $\beta$ is fixed,
the sets $\mathcal S_i^\beta$ are uniformly bounded in size. Therefore,
$\hat\alpha_{i,\mathcal S}^{\text{unadj}}$ is uniformly bounded in absolute value, and hence
\[
\sup_{i,i',\mathcal S,\mathcal S'}
\Var\!\left(
\hat\alpha_{i,\mathcal S}^{\text{unadj}}\hat\alpha_{i',\mathcal S'}^{\text{unadj}}
\right)
\le C
\]
for some constant $C<\infty$. By Cauchy--Schwarz,
\[
\left|
\Cov\!\Bigl(
\hat\alpha_{i,\mathcal S}^{\text{unadj}}\hat\alpha_{i',\mathcal S'}^{\text{unadj}},
\hat\alpha_{j,\mathcal U}^{\text{unadj}}\hat\alpha_{j',\mathcal U'}^{\text{unadj}}
\Bigr)
\right|
\le C.
\]

Moreover, each product
$\hat\alpha_{i,\mathcal S}^{\text{unadj}}\hat\alpha_{i',\mathcal S'}^{\text{unadj}}$
depends only on the treatment assignments in a bounded-size region determined by
$\mathcal N_i\cup\mathcal N_{i'}$. Under Assumption~\ref{as:interference} and $d=O(1)$,
for each fixed $(i,i',\mathcal S,\mathcal S')$, there are only $O(1)$ choices of
$(j,j',\mathcal U,\mathcal U')$ for which the corresponding dependence regions overlap,
and hence only $O(1)$ choices giving nonzero covariance.

Since $|\mathcal E_n|=O(n)$ and the numbers of admissible
$\mathcal S,\mathcal S'$ are uniformly bounded, the total number of summands indexed by
$(i,i',\mathcal S,\mathcal S')$ is $O(n)$. Therefore,
\[
\Var(A_{n,1})=O\!\left(\frac{1}{n}\right),
\]
and thus $A_{n,1}=o_p(1)$ by Chebyshev's inequality.

The same argument yields $A_{n,2}=o_p(1)$.

\medskip

Combining $A_{n,k}=o_p(1)$ for $k=1,2,3,4$ yields
$\hat V(\bs 0)-\tilde V(\bs 0)=o_p(1)$, proving \eqref{eq:V0_consistency_goal_thm4}.
Together with \eqref{eq:Delta_hat_consistency_thm4}, we conclude that
\[
\hat V(\hat{\bs\theta})-\tilde V(\bs\theta^\ast)=o_p(1).
\]

Finally, we show conservativeness. By construction of $\tilde V(\bs 0)$ in
Section~\ref{sec:var_estimator}, it upper-bounds the asymptotic variance of the
unadjusted estimator, i.e.\ $\tilde V(\bs 0)\ge V(\bs 0)$, where $V(\bs 0)$
denotes the asymptotic variance evaluated at $\bs\theta=\bs 0$.
Since $\tilde V(\bs\theta^\ast)=\tilde V(\bs 0)-\Delta(\bs\theta^\ast)$ and
$V(\bs\theta^\ast)=V(\bs 0)-\Delta(\bs\theta^\ast)$ by definition of $\Delta(\cdot)$,
we have $\tilde V(\bs\theta^\ast)\ge V(\bs\theta^\ast)$.
Therefore, $\hat V(\hat{\bs\theta})$ is asymptotically conservative for $V(\bs\theta^\ast)$.

\section{Proofs of Lemmas}

\subsection{Proof of Lemma \ref{lem:unbias}}
\label{appendix:proof_lemma_unbiased}
Under Assumptions \ref{as:interference}--\ref{as:model}, we compute the expectation of $\hat{\alpha}_{i, \mathcal S}^{\text{unadj}}$ as
\begin{align*}
    \E(\hat{\alpha}_{i, \mathcal S}^{\text{unadj}}) &= \E\left(Y_i\prod_{j \in \mathcal{S}} \frac{-1}{p_j}\sum_{\mathcal{U} \in \mathcal S_i^{\beta}, \mathcal S \subseteq \mathcal{U}}\prod_{l \in \mathcal{U}}\frac{p_l - Z_l}{1-p_l} \right)\\
    &=\E\left( \sum_{\mathcal T \in \mathcal S_{i}^\beta} \alpha_{i, \mathcal T} \prod_{t \in \mathcal T} Z_t \prod_{j \in \mathcal{S}} \frac{-1}{p_j}\sum_{\mathcal{U} \in \mathcal S_i^{\beta}, \mathcal S \subseteq \mathcal{U}}\prod_{l \in \mathcal{U}}\frac{p_l - Z_l}{1-p_l} \right)\\
    &= \prod_{j \in \mathcal{S}} \frac{-1}{p_j} \sum_{\mathcal T \in \mathcal S_{i}^\beta} \alpha_{i, \mathcal T} \sum_{\mathcal{U} \in \mathcal S_i^{\beta}} \I(\mathcal S \subseteq \mathcal U \subseteq \mathcal T)\E\left(\prod_{t \in \mathcal T} Z_t\prod_{l \in \mathcal{U}}\frac{p_l - Z_l}{1-p_l}\right)\\
    &= \sum_{\mathcal T \in \mathcal S_{i}^\beta, \mathcal S \subseteq \mathcal T} \alpha_{i, \mathcal T} \prod_{t \in \mathcal T - \mathcal S} p_{t}\sum_{\mathcal{U} \in \mathcal S_i^{\beta}}\I(\mathcal S \subseteq \mathcal U \subseteq \mathcal T) (-1)^{|\mathcal U - \mathcal S|}\\
    &=\alpha_{i, \mathcal S}.
\end{align*}
The last equality holds because
$$
\sum_{\mathcal U \in \mathcal S_i^{\beta}} \I(\mathcal S \subseteq \mathcal U \subseteq \mathcal T)(-1)^{|\mathcal U - \mathcal S|} = 0,
$$
whenever $\mathcal T \neq \mathcal S$.

\subsection{Proof of Lemma \ref{lem:expupper}}
\begin{align*}
&\left\|\E\left[\frac{1}{n}\sum_{i':  \mathcal N_i \cap \mathcal N_{i'} \neq \varnothing} \omega_i\omega_{i'} \bX_{i'}\prod_{k \in \mathcal{S}} Z_{k}\right]\right\|\\
&=
 \left\|\E\left[\frac{1}{n}\sum_{i':  \mathcal N_i \cap \mathcal N_{i'} \neq \varnothing} \bX_{i'}\sum_{\mathcal S' \in\mathcal S_{i'}^\beta}g(\mathcal S')\prod_{j\in\mathcal S'}\frac{Z_{j}-p_{j}}{p_{j}(1-p_{j})}\sum_{\mathcal T \in\mathcal S_{i}^\beta}g(\mathcal T)\prod_{t\in\mathcal T}\frac{Z_t-p_t}{p_t(1-p_t)}\prod_{k \in \mathcal{S}} Z_{k}\right]\right\|\\
 & \leq \frac{1}{n}\sum_{i':  \mathcal N_i \cap \mathcal N_{i'} \neq \varnothing} \left\|\bX_{i'}\right\|\sum_{\mathcal S' \in\mathcal S_{i'}^\beta}\left|g(\mathcal S')\right|\sum_{\mathcal T \in\mathcal S_{i}^\beta}\left|g(\mathcal T)\right|\left|\E\left[\prod_{j\in\mathcal S'}\frac{Z_{j}-p_{j}}{p_{j}(1-p_{j})}\prod_{t\in\mathcal T}\frac{Z_t-p_t}{p_t(1-p_t)}\prod_{k \in \mathcal{S}} Z_{k}\right]\right|\\
 &\leq \frac{1}{n}\sum_{i':  \mathcal N_i \cap \mathcal N_{i'} \neq \varnothing} X_{\max}\sum_{\mathcal S' \in\mathcal S_{i'}^\beta}\sum_{\mathcal T \in\mathcal S_{i}^\beta}\left|\E\left[\prod_{j\in\mathcal S'}\frac{Z_{j}-p_{j}}{p_{j}(1-p_{j})}\prod_{t\in\mathcal T}\frac{Z_t-p_t}{p_t(1-p_t)}\prod_{k \in \mathcal{S}} Z_{k}\right]\right|\\
 &\leq \frac{1}{n}\sum_{i':  \mathcal N_i \cap \mathcal N_{i'} \neq \varnothing} X_{\max}\sum_{\mathcal S' \in\mathcal S_{i'}^\beta}\sum_{\mathcal T \in\mathcal S_{i}^\beta} \I\left(\left(\mathcal S' \cup \mathcal T\right)\setminus \left(\mathcal S' \cap \mathcal T\right) \subseteq \mathcal S\right)\left(\frac{1}{p(1 - p)}\right)^{|\mathcal S' \cap \mathcal T|}\\
 &\leq \frac{4d_{\text{in}}d_{\text{out}}}{n}X_{\max}\left( \frac{e d_{\text{in}}}{\beta} \cdot \max \left( \beta^2, \frac{1}{p(1 - p)} \right) \right)^{\beta}.
\end{align*}

\subsection{Proof of Lemma \ref{lem:cltfix}}
 Our proof follows similar arguments as proof of Theorem 3 in \cite{CortezRodriguezEichhornYu+2023}. Let 
\begin{align*}
    &R_i := \frac{1}{n}\left[\omega_iY_i - \omega_i\bs\theta^\top \bX_i - \E\left(\omega_i Y_i  - \omega_i\bs\theta^\top\bX_i\right)\right],\\
    &\nu^2 := \Var\left(\sum_{i=1}^n R_i\right), \quad Q:= \frac{1}{\nu} \left(\TTEhat(\bs\theta) - \tau\right),
\end{align*}
where $\tau = \E\left(\omega_i  Y_i \right) $ by the unbiasedness results in \cite{CortezRodriguezEichhornYu+2023}. Since $\E\left(\omega_i\right) = 0$ by construction, $\tau = \E\left(\omega_i  Y_i\right) - 0= \E\left(\omega_i  Y_i  - \omega_i\bs\theta^\top \bX_i\right)$. Next, we have the following upper bound
\begin{align*}
    |\omega_i| = \left|\sum_{\mathcal S\in\mathcal S_i^\beta, \mathcal S \neq \varnothing}g(\mathcal S)\prod_{j\in\mathcal S}\frac{Z_j-p_j}{p_j(1-p_j)}\right| = \left|\sum_{\mathcal S \in \mathcal S_i^\beta, \mathcal S \neq \varnothing} \frac{1}{p^{|\mathcal S|}}\right| \leq \left(\frac{d}{p}\right)^\beta.
\end{align*}
Therefore
\begin{align*}
    \left|R_i\right| &\leq Y_{\max}\left(\frac{d}{ p}\right)^\beta + Y_{\max} + \|\bs \theta\|_2 X_{\max}\left(\frac{d}{ p}\right)^\beta \\
    & \leq \left(Y_{\max} + \|\bs \theta\|_2 X_{\max} \right)\left(\frac{d}{ p}\right)^\beta + \left[Y_{\max} + \|\bs \theta\|_2 X_{\max} \right].
\end{align*}
Following analogous steps in \cite{CortezRodriguezEichhornYu+2023}, based on Assumption \ref{as:variance}--\ref{as:no_degeneracy}, we have
\begin{align*}
    d_W(Q, \zeta) = O\left(\frac{(Y_{\max} + \|\bs \theta\|_2 X_{\max})^3 d^{3\beta + 4}}{n^{1/2} p^{3\beta}} + \frac{(Y_{\max} + \|\bs \theta\|_2 X_{\max})^2 d^{2\beta + 3}}{n^{1/2} p^{2\beta}} \right),
\end{align*}
where $\zeta$ is a standard normal random variable. Based on Assumption \ref{as:bounded} and the assumption that $d$ is $O\left(1\right)$, the Wasserstein distance between $Q$ and
$\zeta$ goes to $0$ as $n \to \infty$. Next, we calculate $n\nu^2$.
\begin{align*}
n\nu^2 &= \Var\left(\sqrt{n}\TTEhat(\bs \theta)\right) =  \frac{1}{n}\Var\left(\sumn \omega_i\left(Y_i  - \bs \theta^\top \bX_i \right )\right) \nonumber \\
&= \frac{1}{n}\Var\left(\sumn \omega_i Y_i\right)+\frac{1}{n}\E\left[\left(\sumn  \omega_i \bs \theta^\top \bX_i \right)^2\right] - \frac{2}{n} \sumn \sum_{i':  \mathcal N_i \cap \mathcal N_{i'} \neq \varnothing}\bs \theta^\top\E\left[ \omega_i\omega_{i'}  \bX_{i'} Y_i\right]\\
&=\frac{1}{n}\Var\left(\sumn \omega_i Y_i\right)+\frac{1}{n}\bs \theta^\top \bX^\top \bs M \bX \bs \theta - \frac{2}{n} \sumn \sum_{i':  \mathcal N_i \cap \mathcal N_{i'} \neq \varnothing}\bs \theta^\top\E\left[\omega_i \omega_{i'}  \bX_{i'} Y_i\right] = V_n(\bs \theta).
\end{align*}
Since $\TTEhat(\bs\theta) = Q\nu + \tau$, under Assumption \ref{as:variance}--\ref{as:no_degeneracy}, the distribution of $\sqrt{n}(\TTEhat(\bs\theta) - \tau)$ converges to $\mathcal{N}(0,V(\bs \theta^\ast))$.   

\end{appendix}

\end{document}